\documentclass{aa}  
\usepackage{graphicx,changepage}
\usepackage{txfonts}
\usepackage{subfig}
\usepackage{natbib}
\usepackage{comment}

\DeclareGraphicsExtensions{.pdf,.png,.jpg,.ps}
\bibpunct{(}{)}{;}{a}{}{,} 
\usepackage{hyperref}
\begin{document} 

\authorrunning{Castignani et al.}
\titlerunning{Molecular gas and star formation activity in LIRGs in intermediate-redshift clusters}

\title{Molecular gas and star formation activity in luminous infrared
     galaxies in clusters at intermediate redshifts}

\author{G. Castignani
          \inst{1}\fnmsep\thanks{e-mail: gianluca.castignani@epfl.ch}
          \and
          P. Jablonka\inst{1,2}
          \and
          F. Combes\inst{3,4}
          \and      
          C.~P. Haines\inst{5}
          \and
          T. Rawle\inst{6}  
          \and
          M. Jauzac\inst{7,8,9}
          \and          
          E. Egami\inst{10}
           \and
          M. Krips\inst{11}
          \and
          D. Sp\'{e}rone-Longin\inst{1}
          \and
          M. Arnaud\inst{12}
          \and
          S. Garc\'{i}a-Burillo\inst{13}
          \and
          E. Schinnerer\inst{14}
          \and
          F. Bigiel\inst{15}
          }

   \institute{Laboratoire d'astrophysique, \'{E}cole Polytechnique F\'{e}d\'{e}rale de Lausanne (EPFL), Observatoire de Sauverny, 1290 Versoix, Switzerland
              \and   
      Observatoire de Paris, GEPI, CNRS, Sorbonn\'{e} University, PSL Research Universty, 75014 Paris, France
                    \and
              Observatoire de Paris, LERMA, CNRS, Sorbonn\'{e} University, PSL Research Universty, 75014 Paris, France
              \and
              Coll\`{e}ge de France, 11 Place Marcelin Berthelot, 75231 Paris, France
              \and
              Universidad de Atacama, Instituto de Astronom\'{i}a y Ciencias Planetarias de Atacama,  Av. Copayapu 485, Copiap\'{o}, Regi\'{o}n de Atacama, Chile
               \and
               ESA, Science Operations Department, STScI, Baltimore, MD 21218, USA
              \and
              Centre for Extragalactic Astronomy, Durham University, South Road, Durham DH1 3LE, UK
              \and
              Institute for Computational Cosmology, Durham University, South Road, Durham DH1 3LE, UK
              \and 
              Astrophysics and Cosmology Research Unit, School of Mathematical Sciences, University of KwaZulu-Natal, Durban 4041, South Africa              
              \and
          Steward Observatory, University of Arizona, 933 N. Cherry Ave., Tucson, AZ 85721, USA
          \and
          IRAM, Domaine Universitaire, 300 rue de la Piscine, 38406 Saint-Martin-d'H\`{e}res, France
          \and
          Laboratoire AIM, IRFU/Service d'Astrophysique -CEA - CNRS, Universit\'{e} Paris Diderot, Gif-sur-Yvette, France
          \and
          Observatorio  Astron\'{o}mico  Nacional  (OAN-IGN)-Observatorio  de Madrid, Alfonso XII, 3, 28014 Madrid, Spain
          \and
          MPI for Astronomy, K\"{o}nigstuhl 17, D-69117, Heidelberg, Germany
          \and
           Argelander-Institut f\"{u}r Astronomie, Universit\"{a}t Bonn, Auf dem H\"{u}gel 71, 53121 Bonn, Germany
           \vspace{0.3cm}
                   }

          \date{Received November 26, 2019; Accepted May 17 2020}
 
   \abstract{We investigate the role of dense  megaparsec-scale environments in processing molecular gas of cluster galaxies as they fall into the cluster cores. We selected a sample of $\sim20$ luminous infrared galaxies (LIRGs) belonging to intermediate-redshift clusters, mainly from the {\textit{Herschel} Lensing Survey (HLS) and the Local Cluster Substructure Survey (LoCuSS).} These galaxies include MACS~J0717.5+3745 at $z=0.546$ and {Abell~697, 963, 1763, and 2219} at $z=0.2-0.3$. We performed spectral energy distribution modeling from the  far-infrared to ultraviolet of the LIRGs, which span  cluster-centric distances within $r/r_{200}\simeq0.2-1.6$. We observed the LIRGs in CO(1$\rightarrow$0) or CO(2$\rightarrow$1) with the  Plateau de Bure interferometer and its successor NOEMA, as part of five observational programs carried out between 2012 and 2017. We compared the molecular gas to stellar mass ratio $M(H_2)/M_\star$, star formation rate (SFR), and depletion time ($\tau_{\rm dep}$) of the LIRGs with those of a compilation of cluster and field star-forming galaxies from the literature. The targeted LIRGs have SFR, $M(H_2)/M_\star$, and $\tau_{\rm dep}$ that are consistent with those of both main-sequence (MS) field galaxies and star-forming galaxies from the comparison sample. However we find that the depletion time, normalized to the MS value, { tentatively} increases with increasing  $r/r_{200}$, with a significance of $2.8\sigma$, which is ultimately due to a deficit of cluster-core LIRGs with $\tau_{\rm dep}\gtrsim\tau_{\rm dep,MS}$. We suggest that a rapid exhaustion of the molecular gas reservoirs occurs in the cluster LIRGs and is indeed effective in suppressing their star formation and  ultimately {quenching} them. This mechanism may explain the exponential decrease of the fraction of cluster LIRGs with cosmic time. The compression of the gas in LIRGs, possibly induced by intra-cluster medium shocks, may be responsible for the short timescales that are observed in a large fraction of cluster-core LIRGs. Some of our LIRGs may also belong to a population of infalling filament galaxies.}



\maketitle
%

\section{Introduction}\label{sec:introduction}
Galaxy clusters are the most massive gravitationally bound structures in the Universe, which originate from density fluctuations of the primordial density field \citep[see][for a review]{KravtsovBorgani2012}. They are also excellent laboratories for studying galaxy evolution.  Following the pioneering morphology versus density relation from \citet{Dressler1980}, numerous studies have shown that the  environment is a key factor in governing the star formation rates
(SFRs) of galaxies \citep[e.g,][]{Baldry2006,Peng2010}.


Environmental processes can remove gas through i) tidal heating and stripping  occurring in gravitational interactions and mergers between galaxies \citep{Merritt1983,Moore1998}, ii) ram-pressure stripping due to a passage through the hot intra-cluster gas \citep{Gunn_Gott1972, Roediger_Henssler2005,Jachym2014}, or iii) starvation, that is, the suppression of cold gas accretion from the cosmic web \citep{Larson1980,Balogh2000,vandeWoort2017}. However, their relative efficiencies and timescales are still debated \citep[][]{Haines2015,Fillingham2015,Balogh2016,Wagner2017}. 

Environmental quenching has an overall efficiency that increases with cosmic time. \citet{Nantais2017} found indeed that the fraction of quenched galaxies increases from $\sim40\%$ at $z=1.6$ to $\sim90\%$ at $z\lesssim1.1$. 
The efficiency of each quenching process also peaks at  different cluster-centric distances \citep{DeLucia2010,Moran2007}, which ultimately results in the suppressed star formation observed in cluster cores, at least at low redshifts. For example, \citet{Pintos-Castro2019} recently showed that the fraction of star-forming galaxies in the cluster cores at $z\sim0.4$ can be as low as $\sim10\%$ for galaxies with stellar masses of $\log(M_\star/M_\odot)=10.5-11.2$.

There is also ample evidence that the SFR is suppressed at distances up to $\sim5$ virial radii from   the cluster center \citep{Finn2010,Saintonge2008,Lewis2002,Gomez2003,Bahe2013}. Galaxies {are also observed being preprocessed} by dense environments before they fall into the cluster itself \citep{Poggianti1999,Cortese2006,Bianconi2018,Sarron2019,Vulcani2019}.


All these studies strongly suggest that dense Mpc-scale environments play an important role in quenching star formation. Nevertheless, the non-negligible fraction of star-forming systems in clusters may imply a much longer time between {the accretion of the galaxy into the cluster} and star formation shut-off \citep[e.g.,][]{McGee2011,Haines2015,Cantale2016} than that, $\sim500$~Myr , inferred from the abundance of post-starburst galaxies \citep[][]{Dressler2013}.

Molecular gas is also known to be depleted in dense environments, at least in the local Universe \citep{Casoli1998,Lavezzi_Dickey1998,Vollmer2008,Scott2013}. The molecular gas content is indeed correlated with the SFR \citep{Bigiel2008,Schruba2011,Leroy2013}. Nevertheless, the role of dense Mpc-scale environments in processing molecular gas as cluster galaxies from the outskirts of the clusters fall into the cluster cores is substantially unknown.
To determine the effect of the Mpc-scale environment on galaxy evolution, it is necessary to understand how galaxies and their gas properties are altered as they move through the cosmic web and enter the densest regions. To this aim in the present work we study a population of luminous infrared galaxies (LIRGs). These galaxies belong to massive clusters at intermediate redshifts and homogeneously span a broad range of cluster-centric distances, from the cores out to the cluster outskirts.  The present work is part of a wider search for CO in distant cluster galaxies \citep{Jablonka2013,Castignani2018,Castignani2019}. Indeed, recent advancements of millimeter wavelength interferometers such as { the NOrthern Extended Millimeter Array (NOEMA) and the Atacama Large Millimeter/submillimeter Array (ALMA)} now allow unprecedented studies of molecules in distant galaxies.

Throughout this work we adopt a flat $\Lambda \rm CDM$ cosmology with matter density $\Omega_{\rm m} = 0.30$, dark energy density $\Omega_{\Lambda} = 0.70$, and Hubble constant $h=H_0/100\, \rm km\,s^{-1}\,Mpc^{-1} = 0.70$ \citep[see however,][]{PlanckCollaborationVI2018,Riess2019}. The paper is structured as follows. In Sect.~\ref{sec:sample} we describe both cluster and galaxy samples; in {Sect.~\ref{sec:SFR_Mstar} we derive stellar mass and SFR estimates 
of the LIRGs using a multiwavelength spectral energy distribution (SED) modeling;} in Sect.~\ref{sec:data_reduction} we describe the molecular gas observations and data reduction; in Sect.~\ref{sec:comparison_samples} we introduce the comparison samples;
in Sects.~\ref{sec:results} and \ref{sec:discussion} we present and discuss the results, respectively; in Sect.~\ref{sec:conclusions} we draw our conclusions. In the Appendix~\ref{app:NOEMAdetections_images} we report the CO observations and the optical images of the targeted LIRGs.



\section{Samples}\label{sec:sample}

\subsection{Galaxy clusters}\label{sec:cluster_sample}

The LIRGs of this study are spatially distributed from the centers
to the infall regions of five massive intermediate-redshift galaxy clusters: Abell~963 ($z=0.204$), Abell~2219 ($z=0.226$), Abell~1763 ($z=0.232$), Abell~697 ($z=0.282$),   
and MACS~J0717.5+3745 ($z=0.546$), selected from the Local Cluster Substructure Survey (LoCuSS) and the \textit{Herschel} Lensing Survey (HLS).

The Local Cluster Substructure
Survey is a multiwavelength survey of a sample of X-ray galaxy clusters at $0.15\leq z\leq0.3$ drawn from the ROSAT All-Sky Survey cluster catalogs
\citep{Haines2009a}. In addition to the ultraviolet (UV) to near-infrared (NIR) imaging, each
cluster was observed across a $25 \arcmin \times 25 \arcmin$ field of view at 24~$\mu$m with the \textit{Spitzer Space Telescope}. 
{Each cluster was also observed with {\it Herschel} PACS and SPIRE over the same $25 \arcmin \times 25 \arcmin$ field of view, within the LoCuSS {\it Herschel} Key Programme.}

The HLS is a deep \textit{Herschel} PACS
(100 and 160 $\mu$m) and SPIRE (250, 350, and 500 $\mu$m) imaging program of
$\sim40$ massive galaxy clusters, which were selected as the most X-ray luminous
clusters from the ROSAT X-ray all-sky survey \citep{Egami2010}.  The majority of
HLS clusters are also in the LoCuSS cluster sample. A wealth of spectroscopic
information is also available for the LoCuSS and HLS clusters
\citep[e.g.,][]{Ma2008,Richard2010,Haines2013,Ebeling2014}, which has enabled the selection of our targets.

\begin{table*}[htb]
\begin{center}
\begin{tabular}{lccccccc}
\hline\hline
Cluster ID & R.A. (J2000) & Dec. (J2000) & $z_{spec}$ & $\log(M_{200}/M_\odot)$ & $r_{200}$ &  $c_{200}$ & $\sigma_{\rm{cluster}}$ \\
 & (hh:mm:ss.ss) & (dd:mm:ss.ss) & & & (Mpc) & & (km/s)   \\ 
 ~~~~~(1) & (2) & (3) & (4) & (5) & (6) & (7)  & (8) \\
 \hline
Abell 963  &  10:17:03.65  &  +39:02:49.63  &  0.204  &  14.99  &  1.92  &  4.8  &  1119  \\
Abell 2219  &  16:40:22.54  &  +46:42:21.60  &  0.226  &  15.05  &  1.98  &  4.7  &  1332  \\
Abell 1763  &  13:35:18.07  &  +40:59:57.16  &  0.232  &  14.95  &  1.84  &  4.8  &  1358  \\
Abell 697  &  08:42:57.58  &  +36:21:59.54  &  0.282  &  15.25  &  2.28  &  4.4  &  1268  \\
Cl~1416+4446  &  14:16:28.08  &  +44:46:37.92  &  0.397  &  14.24  &  1.00  &  5.1  &  750  \\
Cl~0926+1242  &  09:26:36.60  &  +12:42:58.97  &  0.489  &  14.62  &  1.30  &  4.5  &  810  \\
MACS~J0717.5+3745  &  07:17:30.93  &  +37:45:29.74  &  0.546  &  15.46  &  2.41  &  3.6  &  1660  \\ 
\hline
 \end{tabular}
\end{center}
 \caption{Cluster properties: (1) cluster ID; (2-3) R.A. and Dec. J2000 coordinates of the cluster center; (4) spectroscopic redshift of the cluster; (5-7) cluster mass ($M_{200}$), radius ($r_{200}$), and concentration ($c_{200}$); (8) velocity dispersion.   } 
\label{tab:cluster_properties}
\end{table*}

Abell 963 is classified as a relaxed cluster based on the joint {\it Hubble Space Telescope} ({\it HST}) strong-lensing and X-ray analysis by \citet{Smith2005}. The \textit{XMM-Newton} maps reveal significant substructures on large $\sim$Mpc scales; three infalling groups were identified by \cite{Haines2018}.

Abell 2219 is one of the hottest and brightest X-ray luminous clusters. It is a
merging system with infalls of clumps aligned with a filament in the foreground \citep{Boschin2004}. One of our targets, A2219-1, is projected in the northwest part of the central X-ray emission close to the shock front \citep{Canning2017}.

Abell 1763 is an X-ray cluster connected to Abell 1770 by filaments, which were revealed by the combination of \textit{Spitzer} and ancillary optical data \citep{Fadda2008}. Our cluster targets are located within the virial radius of Abell 1763, however, well outside the cluster-core region where the intra-cluster medium X-ray emission detected by \textit{Chandra} and the BCG radio emission observed with the VLA coexist \citep{Douglass2018}.

In the case of Abell 697, a recent merger seems to be favored by X-ray morphology and a weak-lensing analysis \citep{Cibirka2018}. The importance of this merger though is difficult to assess because its axis is close to the line of sight
\citep{Girardi2006}. In any case it seems weaker than in Abell 1763 and Abell 2219 at comparable redshifts.

MACS~J0717.5+3745 is the most distant of our targeted clusters. It is also one of the most dynamically active and massive galaxy clusters known to date at $z>0.5$, making it an optimal laboratory to catch environmental transformations of galaxies  in the act.  \citet{Ma2009} showed that its core is an active triple merger. On a larger scale, as illustrated in Fig.~\ref{fig:M0717field}, the cluster was shown to be part of a filamentary structure that extends over $\sim4.5$~Mpc \citep{Ebeling2004,Jauzac2012,Jauzac2018,Ellien2019}.

To increase  the sample size, we include two other intermediate-redshift clusters, Cl~1416+4446 ($z=0.397$) and Cl~0926+1242 ($z=0.489$),  in our analysis. For these intermediate-redshift clusters, the galaxy molecular gas reservoirs of three LIRGs was investigated by \citet{Jablonka2013}.


In the following, the cluster masses, $M_\Delta$, are defined in the classical way, that is, as the mass enclosed within $r_\Delta$, the radius that encompasses a matter density $\Delta$ times larger than the critical density. The virial masses ($M_{200}$) of Abell~2219, MACS~J0717.5+3745, and Cl~1416+4446
have been inferred from weak-lensing analysis 
\citep{Okabe2010,Israel2012,Medezinski2013}.  For these clusters we derived the $r_{200}$ radius from $M_{200}$ and estimated the concentration $c_{200}$
at $r_{200}$ from the concentration  versus cluster mass relation by
\citet{Duffy2008}. For the other clusters $M_{500}$ cluster masses inferred from
X-ray analysis were reported by \citet{Jablonka2013}, \citet{Haines2015}, and
references therein. We therefore iteratively used the relations by
\citet{Duffy2008} and \citet{Hu_Kravtsov2003} between the cluster mass
($M_\Delta$), concentration, and redshift to simultaneously estimate $M_{200}$,
$r_{200}$, and the concentration $c_{200}$.

Table~\ref{tab:cluster_properties} summarizes the cluster properties including
the cluster mass, radius, and concentration outlined above.  We also report the
cluster center coordinates and velocity dispersions, which are obtained from the
LoCuSS
webpage\footnote{\url{http://herschel.as.arizona.edu/locuss/locuss_clusterlist.php}},
as well as from previous studies \citep{Ebeling2007,Jablonka2013,Haines2015}.

\subsection{Luminous infrared galaxies}\label{sec:sample_LIRG}
We selected our targets{ primarily} on the basis of their infrared (IR)  luminosity, that is, $\log(L_{\rm{IR}}/L_\odot)\geq11.2$, to maximize the likelihood for a CO detection within eight hours of NOEMA integration time. { The LIRGs were searched for in and around well-studied, intermediate-redshift clusters (see Sect.~\ref{sec:cluster_sample}) up to two virial radii $\sim2\,r_{200}$. }
Our sample encompasses 17 LIRGs with accurate spectroscopic redshifts, 7 in the $z\sim0.2$ LoCuSS clusters and 10 in the HLS cluster MACS~J0717.5+3745  at $z\sim0.5$.

{In Fig.~\ref{fig:HSTimages} we show {\it HST}-ACS images of the 10 HLS~MACS~J0717.5+3745 cluster galaxies {(F818W filter)} as well as the source A2219-1 {(F850LP filter)}. In Fig.~\ref{fig:SUBARUimages} we report Subaru images {(\textsf{i}$^+$, \textsf{I$_{\textsf c}$} bands)} for the remaining 6 LIRGs in the Abell~697, 963, 1763, and 2219 clusters.
{High-resolution ($\sim0.2$~kpc) {\it HST} images show, almost invariably, a presence of disturbed morphologies and clumpy substructures, likely associated with star-forming regions.} 
{From the Subaru images of the remaining LIRGs, at a resolution of 0.2~arcsec~=~(0.7-0.8)~kpc,} we perceive  a central component (e.g., bulge, bar), from the more extended disk component, a few arcsec in size.}



\subsubsection{LIRGs in and around LoCuSS clusters}
The targets in the LoCuSS clusters were initially selected on the grounds of existing \textit{Herschel} PACS and/or SPIRE fluxes in addition to \textit{Spitzer} 24 $\mu$m fluxes.  Abell 1763, Abell 2219, and Abell 963 {host  two LIRGs each}. In this study, we followed up these clusters for CO emission. Abell~697 has three LIRGs. So far we targeted only the brightest and closest LIRG to the cluster center for its outstanding property, namely A697-1. Indeed it is the most luminous cluster LIRG, nearly an ultra luminous infrared galaxy (ULIRG), over the full sample of 30 LoCuSS systems {considered for follow up in CO}, once the brightest cluster galaxies (BCGs) are excluded. {This exclusion does not alter our results, since none out of the BCGs of the 6 LoCuSS/HLS clusters in our sample were detected by {\it Herschel} in the far-infrared (FIR) 100-500~$\mu$m range \citep{Rawle2012}.}
From the optical Subaru images, A697-1 appears to be a major merger; its IR emission peaks midway between the 2 galaxies. 

\subsubsection{LIRGs in and around MACS~J0717.5+3745}
\begin{figure}\centering
\captionsetup[subfigure]{labelformat=empty}
\subfloat[]{\includegraphics[trim={0cm 0cm 0cm 0cm},clip,
page=1,width=0.5\textwidth,clip=true]{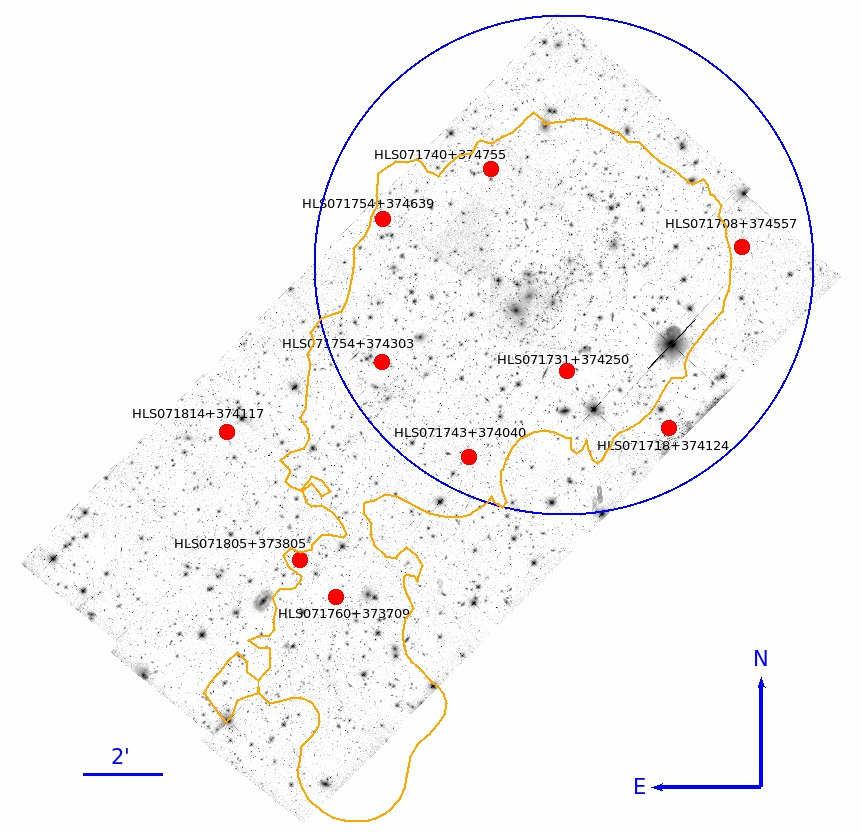}}
\caption{{\it HST}-ACS mosaic (F818W filter) of MACS~J0717.5+3745. Our targets are highlighted as red dots. The solid blue circle has a radius equal to $r_{200}$, while the solid orange contour represents the weak-lensing mass reconstruction by \citet{Jauzac2012} (at a 3$\sigma$ level) showing the cluster and its associated large-scale filament extending in the southeast direction.}
\label{fig:M0717field}
\end{figure}

The LIRGs of MACS~J0717.5+3745 were selected after requiring detection in at least three \textit{Herschel} bands. There are $\sim$18 such systems in and around MACS~J0717.5+3745. From those we chose targets which could sample both the region inside the cluster virial radius and its filamentary structure, as shown in Fig.~\ref{fig:M0717field}. { The 10 target LIRGs of MACS~J0717.5+3745 are either within the virial radius $\sim\,r_{200}$ or belong to the extended filamentary structure toward the southwest. HLS071814+374117 is located in the outskirts of the cluster and is formally just outside the filamentary structure. Because this LIRG is located at $r\sim1.5~r_{200}$ from the cluster center, the source fulfills the selection requirement $r\lesssim2\,r_{200}$ and it has been thus included in our sample of LIRGs.}

\begin{table*}[htb]
\begin{center}
\begin{tabular}{llllccr}
\hline\hline
Cluster ID & Galaxy ID & R.A. (J2000) & Dec. (J2000) & $z_{spec}$ &  $r/r_{200}$ & $v/\sigma_{\rm{cluster}}$ \\
 & & (hh:mm:ss.ss) & (dd:mm:ss.ss) & & &     \\ 
 ~~~~~~(1) & ~~~~~~(2) & (3) & (4) & (5) & (6) & (7)  \\
 \hline
Abell 963  &  A963-1          &  10:17:07.46  &  +39:04:25.04  &  0.211  &  0.19  &  1.56  \\
           &  A963-2          &  10:16:18.06  &  +39:06:13.33  &  0.208  &  1.01  &  0.85  \\
           & J101628.2+390932 &  10:16:28.2    & +39:09:32     & 0.211   &  1.03  & 1.43 \\
  & & & & & &  \\

Abell 2219  &  A2219-1  &  16:40:13.03  &  +46:43:06.31  &  0.230  &  0.20  &  0.84  \\
  &  A2219-2  &  16:39:39.05  &  +46:43:57.40  &  0.233  &  0.86  &  1.30  \\
  & & & & & &  \\

Abell 1763  &  A1763-1  &  13:35:21.13  &  +41:02:29.23  &  0.232  &  0.31  &  $-$0.02  \\
  &  A1763-2  &  13:35:52.91  &  +41:02:08.89  &  0.226  &  0.82  &  $-$1.11  \\ 
  & & & & & &  \\

Abell 697  &  A697-1  &  08:42:47.14  &  +36:29:02.67  &  0.274  &  0.81  &  $-$1.46  \\
  & & & & & &  \\

Cl 1416+4446  &  GAL1416+4446  &  14:16:19.52  &  +44:43:57.12  &  0.396  &  0.99  &  $-$0.17  \\
  & & & & & &  \\

Cl 0926+1242  &  GAL0926+1242-A  &  09:26:32.24  &  +12:42:13.04  &  0.489  &  0.37  &  $-$0.10  \\
  &  GAL0926+1242-B  &  09:26:31.99  &  +12:42:12.59  &  0.489  &  0.38  &  $-$0.10  \\
  & & & & & &  \\

MACS~J0717.5+3745  &  HLS071708+374557  &  07:17:08.30  &  +37:45:56.86  &  0.542  &  0.71  &  $-$0.40  \\
  &  HLS071718+374124  &  07:17:17.69  &  +37:41:24.39  &  0.576  &  0.79  &  3.53  \\
  &  HLS071731+374250  &  07:17:30.66  &  +37:42:49.66  &  0.537  &  0.42  &  $-$0.98  \\
  &  HLS071740+374755  &  07:17:40.33  &  +37:47:54.89  &  0.563  &  0.49  &  2.01  \\
  &  HLS071743+374040  &  07:17:43.02  &  +37:40:40.12  &  0.544  &  0.85  &  $-$0.21  \\
  &  HLS071754+374303  &  07:17:54.17  &  +37:43:03.33  &  0.544  &  0.82  &  $-$0.21  \\
  &  HLS071754+374639  &  07:17:53.97  &  +37:46:39.11  &  0.545  &  0.75  &  $-$0.09  \\
  &  HLS071760+373709  &  07:17:59.98  &  +37:37:08.80  &  0.553  &  1.62  &  0.84  \\
  &  HLS071805+373805  &  07:18:04.51  &  +37:38:04.81  &  0.555  &  1.59  &  1.07  \\
  &  HLS071814+374117  &  07:18:13.82  &  +37:41:17.42  &  0.542  &  1.50  &  $-$0.47  \\
  \hline
 \end{tabular}
\caption{Properties of our targets: (1) cluster ID; (2) galaxy ID; (3-4) R.A. and Dec. J2000 coordinates; (5) spectroscopic redshift;  (6) projected cluster-centric distance in units of $r_{200}$; (7) line-of-sight velocity relative to the cluster redshift normalized to the cluster velocity dispersion.}
\label{tab:galaxy_properties_general}
\end{center}
\end{table*}

\begin{table*}[htb]
\begin{center}
\begin{tabular}{lccccccc}
\hline\hline
 Galaxy ID &  $\log(M_\star/M_\odot)$ & $\log(L_{\rm dust}/L_\odot)$ & $\log(M_{\rm dust}/M_\odot)$  & SFR &  sSFR  & sSFR$_{\rm MS}$ & $\chi^2$ \\
   &  &  &   &  ($M_\odot/{\rm yr}$) & (Gyr$^{-1}$) & (Gyr$^{-1}$) \\ 
 ~~~~~~(1) & (2) & (3) & (4) & (5) & (6) & (7) & (8)  \\
 \hline 
A1763-1  &  $10.51^{+0.10}_{-0.17}$  &  $11.44^{+0.01}_{-0.12}$  &  $7.96^{+0.09}_{-0.06}$  &  $19.4^{+0.5}_{-9.3}$  &  $0.60^{+0.15}_{-0.17}$  &  0.10 & 3.69 \\
A1763-2  &  $11.03^{+0.10}_{-0.01}$  &  $11.39^{+0.10}_{-0.01}$  &  $8.11^{+0.07}_{-0.06}$  &  $3.6^{+3.1}_{-0.04}$  &  $0.03^{+0.02}_{-0.01}$  &  0.05 & 4.33 \\
A2219-1  &  $10.24^{+0.01}_{-0.15}$  &  $11.36^{+0.10}_{-0.11}$  &  $7.77^{+0.08}_{-0.05}$  &  $21.2^{+0.2}_{-15.3}$  &  $1.19^{+0.15}_{-0.72}$  &  0.13 & 1.56 \\
A2219-2  &  $10.32^{+0.33}_{-0.10}$  &  $11.25^{+0.01}_{-0.06}$  &  $7.66^{+0.16}_{-0.13}$  &  $1.7^{+5.5}_{-0.02}$  &  $0.08^{+0.09}_{-0.02}$  &  0.12 & 2.37 \\
A697-1  &  $10.82^{+0.10}_{-0.01}$  &  $11.83^{+0.10}_{-0.10}$  &  $8.28^{+0.11}_{-0.10}$  &  $7.5^{+0.1}_{-0.1}$  &  $0.12^{+0.03}_{-0.02}$  &  0.08 & 2.93 \\
A963-1  &  $10.36^{+0.10}_{-0.10}$  &  $11.38^{+0.10}_{-0.01}$  &  $8.02^{+0.13}_{-0.09}$  &  $10.5^{+2.7}_{-0.1}$  &  $0.47^{+0.12}_{-0.10}$  &  0.10 & 2.24  \\
A963-2  &  $10.76^{+0.10}_{-0.01}$  &  $11.29^{+0.10}_{-0.10}$  &  $7.84^{+0.06}_{-0.05}$  &  $8.3^{+0.1}_{-1.7}$  &  $0.15^{+0.04}_{-0.03}$  &  0.07 & 1.12  \\
J101628.2+390932 & $10.35^{+0.15}_{-0.17}$   &  $11.11^{+0.04}_{-0.02}$   & $8.06^{+0.23}_{-0.15}$   & $10.3^{+1.7}_{-1.5}$   &  $0.47^{+0.28}_{-0.21}$  & 0.11 & 2.78 \\ 
GAL0926+1242-A  &  $10.38^{+0.10}_{-0.08}$  &  $11.13^{+0.07}_{-0.06}$  &  $8.14^{+0.45}_{-0.39}$  &  $10.3^{+1.9}_{-1.4}$  &  $0.42^{+0.17}_{-0.12}$  &  0.25 &  1.32 \\
GAL0926+1242-B  &  $9.69^{+0.08}_{-0.02}$   &  $11.15^{+0.19}_{-0.02}$  &  $8.38^{+0.49}_{-0.38}$  &  $15.2^{+0.5}_{-0.3}$  &  $3.35^{+0.87}_{-0.69}$  &  0.46 & 21.9 \\
GAL1416+4446    &  $11.05^{+0.11}_{-0.13}$    &  $11.54^{+0.07}_{-0.10}$  &  $8.66^{+0.33}_{-0.42}$  &  $15.6^{+3.8}_{-3.2}$  &  $0.13^{+0.08}_{-0.04}$  &  0.10 & 1.08 \\
HLS071708+374557  &  $10.41^{+0.25}_{-0.09}$  &  $11.75^{+0.18}_{-0.23}$  &  $8.48^{+0.24}_{-0.17}$  &  $43.4^{+27.0}_{-16.3}$  &  $1.68^{+0.98}_{-0.93}$  &  0.28 & 1.12 \\
HLS071718+374124  &  $10.70^{+0.11}_{-0.08}$  &  $11.37^{+0.13}_{-0.12}$  &  $8.44^{+0.28}_{-0.21}$  &  $17.1^{+5.2}_{-4.3}$  &  $0.33^{+0.14}_{-0.12}$  &  0.23 & 0.26 \\
HLS071731+374250  &  $10.77^{+0.08}_{-0.07}$  &  $11.74^{+0.06}_{-0.06}$  &  $8.18^{+0.08}_{-0.06}$  &  $34.8^{+9.0}_{-7.5}$  &  $0.60^{+0.25}_{-0.17}$  &  0.20 & 0.95 \\
HLS071740+374755  &  $11.12^{+0.10}_{-0.10}$  &  $11.66^{+0.06}_{-0.08}$  &  $8.20^{+0.10}_{-0.07}$  &  $31.0^{+6.3}_{-4.9}$  &  $0.24^{+0.10}_{-0.07}$  &  0.16 & 0.70 \\
HLS071743+374040  &  $10.54^{+0.15}_{-0.05}$  &  $11.63^{+0.12}_{-0.14}$  &  $9.00^{+0.10}_{-0.10}$  &  $36.9^{+13.5}_{-13.3}$  &  $1.06^{+0.44}_{-0.59}$  &  0.25 & 0.47 \\
HLS071754+374303  &  $10.92^{+0.12}_{-0.06}$  &  $11.76^{+0.16}_{-0.12}$  &  $8.88^{+0.12}_{-0.23}$  &  $47.5^{+19.6}_{-11.1}$  &  $0.60^{+0.25}_{-0.26}$  &  0.18 & 0.34 \\
HLS071754+374639  &  $10.91^{+0.09}_{-0.10}$  &  $11.30^{+0.12}_{-0.14}$  &  $8.40^{+0.32}_{-0.28}$  &  $15.9^{+4.8}_{-4.4}$  &  $0.19^{+0.08}_{-0.06}$  &  0.18 & 0.18 \\
HLS071760+373709  &  $10.68^{+0.18}_{-0.11}$  &  $11.34^{+0.21}_{-0.19}$  &  $8.03^{+0.32}_{-0.20}$  &  $20.7^{+14.1}_{-6.9}$  &  $0.47^{+0.28}_{-0.21}$  &  0.22 &  0.62 \\
HLS071805+373805  &  $10.55^{+0.10}_{-0.01}$  &  $11.57^{+0.10}_{-0.10}$  &  $8.15^{+0.36}_{-0.31}$  &  $4.6^{+1.2}_{-0.1}$  &  $0.13^{+0.03}_{-0.03}$  &  0.25 & 0.14 \\
HLS071814+374117  &  $10.94^{+0.08}_{-0.07}$  &  $11.71^{+0.11}_{-0.11}$  &  $8.34^{+0.09}_{-0.07}$  &  $33.7^{+11.2}_{-8.4}$  &  $0.38^{+0.22}_{-0.11}$  &  0.17 & 2.32 \\
\hline
\end{tabular}
\end{center}
 \caption{Results of the SED modeling: (1) Galaxy ID; (2) stellar mass; (3) total stellar luminosity absorbed by dust; (4) total dust mass; (5) SFR; (6) specific SFR; (7) specific SFR from \citet{Speagle2014} for MS field galaxies of redshift and stellar mass equal to those of the corresponding LIRG; { (8) best fit $\chi^2$ value.}  }
\label{tab:galaxies_SEDproperties}
\end{table*}

\begin{figure*}[!ht]\centering
\captionsetup[subfigure]{labelformat=empty}
\subfloat[A1763-1]{\includegraphics[trim={2.5cm 18.5cm 2.5cm 3cm},clip,page=1,width=0.5\textwidth,clip=true]{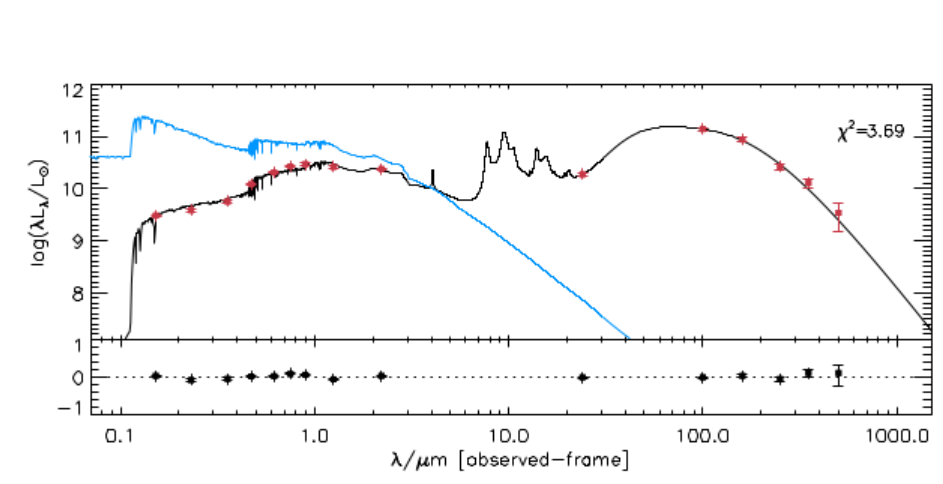}}
\subfloat[A2219-1]{\includegraphics[trim={2.5cm 18.5cm 2.5cm 3cm},clip, page=1,width=0.5\textwidth,clip=true]{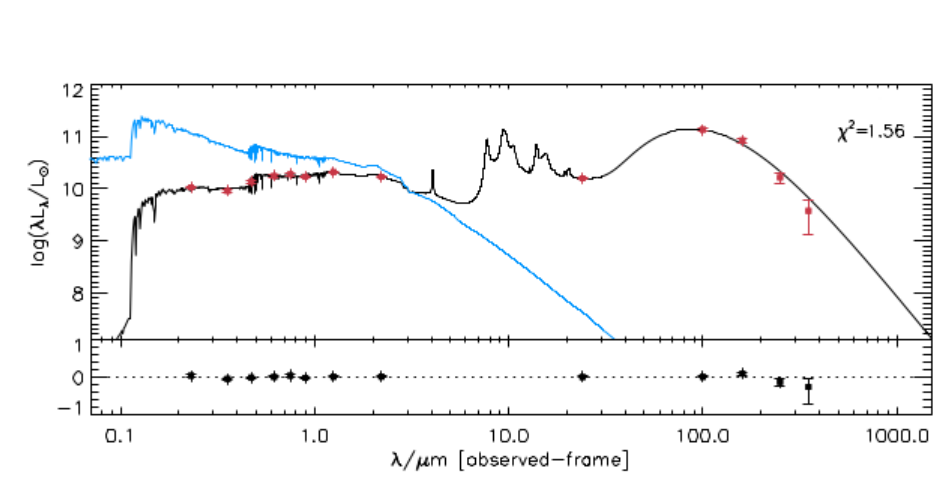}}\\
\subfloat[GAL0926+1242-B]{\includegraphics[trim={2.5cm 18.5cm 2.5cm 3cm},clip,page=1,width=0.5\textwidth,clip=true]{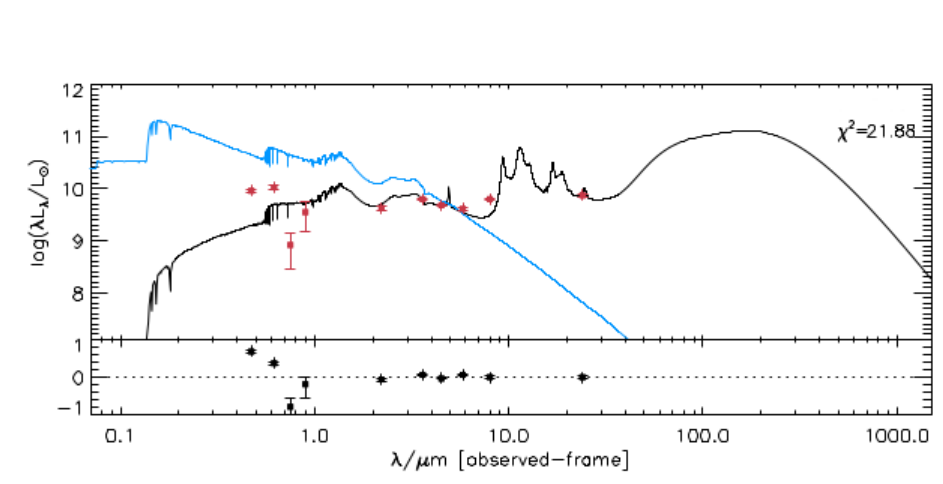}}
\subfloat[HLS071708+374557]{\includegraphics[trim={2.5cm 18.5cm 2.5cm 3cm},clip,page=1,width=0.5\textwidth,clip=true]{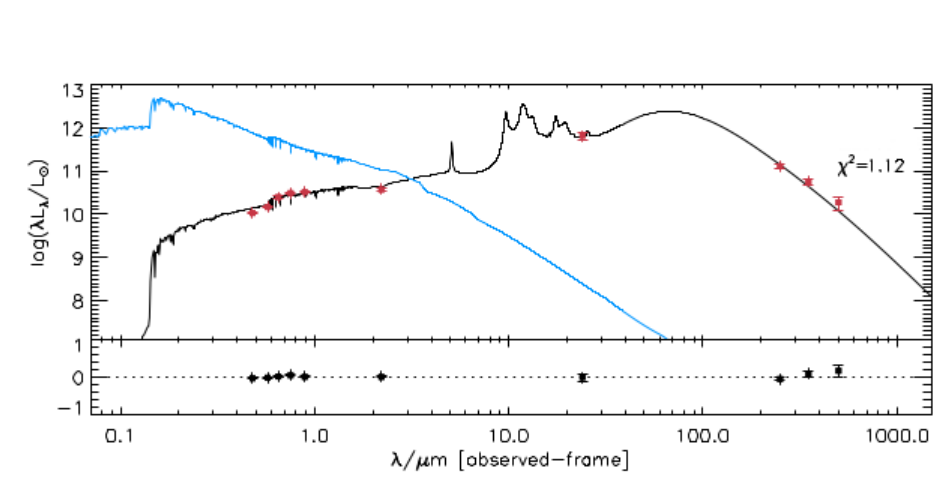}}
\caption{Examples of FIR-to-UV MAGPHYS SED fits (solid black curve) for some of our targets.  The galaxy IDs are indicated at the bottom of each panel. The photometric data points, corrected for Galactic extinction, are shown as red points.  The solid blue curves show the SED fits in the absence of dust. {The bottom panels show the residuals.}}\label{fig:SED}
\end{figure*}


\subsubsection{Additional cluster LIRGs and the final sample}
{In the following we discuss the properties of additional LIRGs, with observations in CO from the literature, which have been included in our sample.} We consider for our analysis 3 LIRGs from the clusters Cl~1416+4446 and Cl~0926+1242, which were detected in CO(2$\rightarrow$1) or CO(1$\rightarrow$0) by \citet{Jablonka2013}.  These 3 LIRGs have IR luminosities, inferred from the SED modeling, in the range $\log(L_{\rm dust}/L_\odot)=11.1-11.5$, which are similar to those of the other 17 new LIRGs considered in this work.

Moreover, in his doctoral dissertation, \citet{Cybulski2016_thesis} reported {25} CO detections among Abell~963 members. {By performing accurate SED modeling (see Sect.~\ref{sec:SFR_Mstar}) in the FIR to UV wavelengths, we found only 1 source among the 25 with an estimated FIR luminosity $>10^{11}~L_\odot$,  typical of LIRGs, which has not been already included in our sample of LIRGs.
Therefore we included this source, namely  J101628.2+390932 at $z=0.211$, in our analysis.}

{We also note that \citet{Cybulski2016} reported a CO detection for an additional cluster galaxy with an estimated FIR luminosity $>10^{11}~L_\odot$. It is the Abell~2192 source J162644.6+422530 at $z=0.189$, with a FIR luminosity $\log(L_{\rm IR}/L_\odot)=11.13$ reported by the authors. However, it is located in the outskirts of the cluster and has not been included in our sample of LIRGs. The source has indeed a projected cluster-centric distance of 2.8~Mpc, corresponding to $r/r_{200} = 1.9$ \citep{Verheijen2007}.}

Including the LIRGs of the above-mentioned publications yields a total sample of 21
LIRGs that are considered in this
work. Table~\ref{tab:galaxy_properties_general} lists the coordinates and cluster-centric distances of the LIRGs.

The sample of LIRGs considered in this work is the largest sample of  LIRGs in and around clusters at intermediate redshifts with CO detections. 
Our selection implies that each of the seven clusters considered in this work has at least one LIRG within the cluster virial radius to address any possible radial variation. 

{According to this constraint, we did not include in our sample the five intermediate-$z$ LIRGs detected in CO(1$\rightarrow$0) by \citet{Geach2011}. We checked that they are indeed located well outside the virial radius, in  the  outskirts  of the $z\sim0.4$ cluster Cl~0024+16: they are all at projected cluster-centric distances between $(1-3)\times~r_{200}$ and  three out of the five  also have line-of-sight velocities $>3\sigma_{\rm cluster}$, relative to the cluster redshift. }

\section{Star formation rates and stellar masses}\label{sec:SFR_Mstar}

We  consistently calculated SFRs and both stellar and dust masses of
our new sample of LIRGs, based on FIR-to-UV photometry, wherever available. Stellar
masses and SFRs have been derived with the Multi-wavelength
Analysis of Galaxy Physical Properties (MAGPHYS) package \citep{daCunha2008},
which enables self-consistent SED fits of both stellar and dust components. The
calculations were performed with a \citet{Chabrier2003} initial mass function
(IMF) and are shown in
Table~\ref{tab:galaxies_SEDproperties}. Fig.~\ref{fig:SED} provides examples
of SED fits. The LoCuSS galaxies have UV-to-FIR SEDs, which were obtained by combining \textit{GALEX},
SDSS, UKIRT, \textit{Spitzer}, and \textit{Herschel}
photometry. For the HLS sources belonging to MACS~J0717.5+3745 we used
CFHT \textsf{U} and \textsf{K$_{\textsf s}$} imaging, Subaru \textsf{B}, \textsf{V}, \textsf{R},
\textsf{i}, and \textsf{z}-band photometry, in addition to \textit{Spitzer} and \textit{Herschel}
photometry.
The SED fits of GAL0926+1242-A, GAL0926+1242-B, and GAL1416+4446 were done with \textsf{griz},  \textsf{K$_{\textsf s }$} in the
NIR, \textit{Spitzer} IRAC  (3.6, 4.5, 5.8, 8.0~$\mu$m), and \textit{Spitzer} MIPS (24 $ \mu$m) fluxes.

{ Our SED fits are overall good, with $\chi^2$ values on the order of unity (Table~\ref{tab:galaxies_SEDproperties}). GAL0926+1242-B is an exception; this source has a higher $\chi^2=21.9$, mainly because of large \textsf{i} and \textsf{z}-band flux uncertainties (Fig.~\ref{fig:SED}). Because of these uncertainties, it  might be possible that the stellar mass estimate of GAL0926+1242-B {is biased toward a lower value.} However, removing this source from our sample would not change our conclusions. An output of MAGPHYS is also the dust luminosity $L_{\rm dust}$, that is,  the total stellar luminosity absorbed by dust in the stellar birth clouds and in the ambient interstellar medium, which is then reradiated. Therefore, $L_{\rm dust}$ is a physically motivated analog of the total observed IR luminosity $L_{\rm IR}$.  From the best-fit results reported in Table~\ref{tab:galaxies_SEDproperties} the dust luminosities are found in the range $\log(L_{\rm dust}/L_\odot)\simeq11.1-11.8$, and are thus consistent with the fact that the galaxies are LIRGs, for which it holds $\log(L_{\rm IR}/L_\odot)\simeq11-12$.}


\section{Molecular gas}\label{sec:data_reduction}

\subsection{Observations}
The 17 new LIRGs in our sample were observed in CO with the IRAM Plateau de Bure interferometer \citep[PdBI;][]{Guilloteau1992,Cox2011} and its successor { NOEMA} \citep[][]{Schuster2014}. Observations were carried out between 2012 and 2017 as part of five programs: W036, X2D, S14BU, S17BI, and W17CX   (PI:~Jablonka). The summary of these observations is presented in Table \ref{tab:observinglog}.

\begin{table*}[htb]
\begin{center}
\begin{tabular}{lllccl}
\hline\hline
Program ID      & Date                 &  Source                                    &  Configuration & Number of antennas   & Resolution \\
                &                      &                                            &                &              &   \\
\hline
W036            &  Sep-Nov 2012        &                                            &                &              &  \\
                &                      &  A963-1                                    &    C \& D      &    6             & 2.9 \arcsec\ (10.0~kpc) \\
                &                      &  A963-2                                    &     D          &    6             & 2.6 \arcsec\ (8.8~kpc) \\
                &                      &  HLS071814+374117                          &     D          &    5             & 2.8 \arcsec\ (17.8~kpc) \\
                &                      &  HLS071708+374557                          &     D          &    4             & 3.8 \arcsec\ (24.2~kpc) \\

X2D             &  Jun -  Oct 2013     &                                            &                &              &    \\                           
                &                      & A2219-1                                    &     D          &   5          & 4.7 \arcsec\ (17.3~kpc) \\ 
                &                      & A2219-2                                    &     D          &   5          & 4.9 \arcsec\ (18.2~kpc) \\   
                &                      & A1763-1                                    &     D          &   5          & 5.8 \arcsec\ (21.4~kpc) \\ 
                &                      & A1763-2                                    &     D          &   6          & 5.9 \arcsec\ (24.1~kpc) \\ 
                &                      & A697-1                                     &     D          &   6          & 5.7 \arcsec\ (23.8~kpc) \\ 

S14BU           &  Aug - Sep 2014      &                                            &                &              &  \\                              
                &                      & HLS071754+374303                           &     D          &   5          & 2.8  \arcsec\ (17.9~kpc) \\ 
                &                      & HLS071731+374250                           &     D          &   5          & 2.8  \arcsec\ (17.7~kpc) \\ 

S17BI \& W17CX  &   Aug - Dec 2017     &                                            &                &              &  \\     
                &                       & HLS071718+374124                          &     D          &   8,9,10     & 2.6  \arcsec\  (17.1~kpc)   \\ 
                &                      & HLS071743+374040                           &     D          &   7          & 3.0  \arcsec\  (19.1~kpc)  \\
                &                      & HLS071760+373709                           &     D          &   7          & 3.0  \arcsec\  (19.3~kpc)  \\
                &                      & HLS071805+373805                           &     D          &   7          & 3.0  \arcsec\  (19.3~kpc)  \\
                &                      & HLS071754+374639                           &     D          &   7,8        & 3.1  \arcsec\ (19.8~kpc)   \\
                &                      & HLS071740+374755                           &     D          &   0          & 2.6  \arcsec\ (16.9~kpc)   \\

\hline
\end{tabular}
\caption{Summary of the {PdBI / NOEMA} observations}
\label{tab:observinglog}
\end{center}
\end{table*}

At a resolution of a few arcsec, all our targets were unresolved by the observations. 
Data reduction was performed using the GILDAS package.\footnote{https://www.iram.fr/IRAMFR/GILDAS/} Data were calibrated using the standard pipeline adopting a natural weighting scheme to maximize sensitivity. 

\subsection{CO fluxes}

The CO fluxes from our NOEMA observations were calculated following the procedure described in \citet{Castignani2018}, which provides further details.  Each spectrum is fitted using a $\chi^2$ minimization procedure with a
best-fit model given as the sum of a polynomial of degree one and a Gaussian to
account for both the baseline and the CO emission line of each target.  The
significance of the CO detection is assessed by a Monte Carlo simulation of
$N=1,000$ spectra per target, which are then fitted with the same $\chi^2$
minimization procedure described above. The velocity integrated $S_{\rm
  CO(J\rightarrow J-1)}\Delta\varv$ flux, in units of Jy~km~s$^{-1}$, is calculated by
integrating the Gaussian model to the simulated spectrum.  For each source we
then used the median and the 68.27\% confidence region of the flux
distribution to get the velocity integrated flux and its 1$\sigma$ uncertainty.

Velocity integrated fluxes were then converted into velocity integrated luminosities $L^\prime_{\rm CO(J\rightarrow J-1)}$, in units of K~km~s$^{-1}$~pc$^2$, using Eq.~(3) of \citet{Solomon_VandenBout2005}, that is,
\begin{equation}
\label{eq:LpCO}
 L^{\prime}_{\rm CO(J\rightarrow J-1)}=3.25\times10^7\,S_{\rm CO(J\rightarrow J-1)}\,\Delta\varv\,\nu_{\rm obs}^{-2}\,D_L^2\,(1+z)^{-3}\,,
\end{equation}
where $\nu_{\rm obs}$, in GHz,  is the observer-frame frequency of the CO(J$\rightarrow$J-1) transition, $D_L$ is the luminosity distance in Mpc, and $z$ the redshift.

{Our NOEMA observations yield 15 CO detections out of 17 targets. For HLS071760+373709 and HLS071805+373805, in the outskirts of MACS~J0717.5+3745, we did not find any CO emission associated with the 2 LIRGs, after carefully inspecting their data cubes. Therefore, for these 2 LIRGs we set 3$\sigma$ upper limits $\lesssim0.5~{\rm Jy}~{\rm km}~{\rm s}^{-1}$ for the CO(2$\rightarrow$1) flux, at 300~km/s resolution in velocity. Throughout this work we remove these 2 sources when estimating statistical quantities of our sample that are related to CO.}
{Concerning the other 4 sources in our sample, from the literature,} the $S_{\rm CO(J\rightarrow J-1)}$ fluxes of GAL0926+1242-A, GAL0926+1242-B, and GAL1416+4446  are from \citet{Jablonka2013}, {while that of J101628.2+390932 is from \citet{Cybulski2016_thesis}}, which we refer to for further details. Table~\ref{tab:galaxy_properties_mol_gas}
summarizes the results.

Figs.~\ref{fig:HSTimages} and \ref{fig:SUBARUimages} show the  {\it HST} and Subaru images of our targets, together with the CO contours. Figure~\ref{fig:NOEMAdetections} shows the CO intensity maps and the corresponding spectra.
{The CO emission peak of A697-1 is slightly shifted toward its southern companion, which is consistent with a similar offset found in the IR map (see also Sect.~\ref{sec:sample_LIRG}).}

Interestingly, \citet{Cybulski2016} observed 5 galaxies belonging to Abell~963
in CO(1$\rightarrow$0) with the Large Millimeter Telescope (LMT). Their sample includes A963-2, for which the authors report a velocity integrated flux of $S_{\rm CO(1\rightarrow0)}\Delta \varv=(1.771\pm0.348)$~Jy~km~s$^{-1}$,
consistent with that independently estimated in this work, $S_{\rm
  CO(1\rightarrow0)}\Delta \varv=(1.92^{+0.13}_{-0.12}) $~Jy~km~s$^{-1}$.

\citet{Cybulski2016_thesis} also includes A963-1, with an estimated $S_{\rm
  CO(1\rightarrow0)}\Delta \varv=(2.290\pm0.437)$~Jy~km~s$^{-1}$. For this source,
we obtained a { lower} flux $S_{\rm CO(1\rightarrow0)}\Delta
\varv=(0.85^{+0.12}_{-0.11})$~Jy~km~s$^{-1}$, at higher signal-to-noise ratio,
S/N=7.7 versus S/N=5.2 reported by \citet{Cybulski2016_thesis}. We suggest that the flux discrepancy {may be due to the underestimated uncertainty for the LMT observations: our interferometric data likely have a more stable spectral baseline compared to the single-dish LMT data.}
We note that S/N$\simeq$5 detections are somehow uncertain, even with
ALMA \citep{Walter2016,Decarli2016}.

\subsection{Molecular gas masses}\label{sec:molecular_gas_masses}

We estimated the total molecular gas masses of our targets as
$M({\rm H_2})=\alpha_{\rm CO}L^{\prime}_{\rm
  CO(1\rightarrow0)}=\alpha_{\rm CO}L^{\prime}_{{\rm CO}(J\rightarrow
  J-1)}/r_{J1}$, where $r_{J1}= L^{\prime}_{{\rm CO}(J\rightarrow
  J-1)}/L^{\prime}_{{\rm CO}(1\rightarrow0)}$ is the excitation ratio. In
this work the CO(2$\rightarrow$1) and CO(1$\rightarrow$0) transitions
are considered and we assumed $r_{21}=0.85$, typical of
submillimeter galaxies \citep[][for a review]{Carilli_Walter2013}.

As reported in Table~\ref{tab:galaxies_SEDproperties} our targets span a broad
range of SFRs, from $\sim0.5\times{\rm SFR}_{\rm MS}$ up to $\sim10\times{\rm
  SFR}_{\rm MS}$, where ${\rm SFR}_{\rm MS}$ is the SFR at the main sequence (MS) estimated following the \citet{Speagle2014} prescription.

By only assuming a Galactic CO-to-H$_2$ conversion factor $X_{\rm CO}\simeq2\times10^{20}$~cm$^{-1}$/(K km/s), that is,  $\alpha_{\rm CO}=4.36~M_\odot\,({\rm K~km~s}^{-1}~{\rm pc}^2)^{-1}$, typical of MS galaxies  that commonly have ${\rm SFR}\lesssim3\times{\rm SFR}_{\rm MS}$ \citep{Solomon1997,Bolatto2013}, we  could potentially overestimate the molecular gas masses of our most active sources, that is, well above the MS \citep[see, e.g.,][for a discussion]{Noble2017,Castignani2018}. A value of $\alpha_{\rm CO}\simeq0.8~M_\odot\,({\rm K~km~s}^{-1}~{\rm pc}^2)^{-1}$ is typically assumed for highly star-forming sources such the ULIRGs \citep[][for a review]{Bolatto2013}. Since our sample includes sources with SFR both within  and above the MS, we adopt the following heuristic prescription for the CO-to-H$_2$ conversion factor, unless specified otherwise:
\begin{equation}
\label{eq:alphaCO}
\frac{\alpha_{\rm CO}(x)}{M_\odot\,({\rm K~km~s}^{-1}~{\rm pc}^2)^{-1}} = 
    \begin{cases}  3.56\exp(-\frac{x}{9})+0.8 & \text{if}\; x\geq0\\  
    4.36 & \text{otherwise,}\\     
    \end{cases}
\end{equation}
where $x=\frac{\rm SFR}{{\rm SFR}_{\rm MS}}-1$. { The adopted $\alpha_{\rm CO}$ conversion factor can be linearized for $x\gtrsim0$ as $\alpha_{\rm CO} = 4.76 -0.4\frac{\rm SFR}{{\rm SFR}_{\rm MS}}+o(\frac{\rm SFR}{{\rm SFR}_{\rm MS}}-1)$.} The $\alpha_{\rm CO}{\rm(SFR)}$ prescription was chosen in such a way that $\alpha_{\rm CO}$ corresponds to the Galactic value for ${\rm SFR}\lesssim {\rm SFR}_{\rm MS}$, while it asymptotically declines, with no discontinuity, down to $\alpha_{\rm   CO}=0.8~M_\odot\,({\rm K~km~s}^{-1}~{\rm pc}^2)^{-1}$ for ${\rm SFR}\gg {\rm   SFR}_{\rm MS}$.

{ Our SFR-dependent prescription is heuristic and adopted in this work to partially circumvent the problem of choosing appropriate values of $\alpha_{\rm CO}$ for both MS and more star-forming LIRGs. According to Eq.~\ref{eq:alphaCO} galaxies with $\frac{\rm SFR}{{\rm SFR}_{\rm MS}}=1$, 3, and 10 have $\frac{\alpha_{\rm CO}}{M_\odot\,({\rm K~km~s}^{-1}~{\rm pc}^2)^{-1}}=4.36$, 3.65, and 2.11, respectively.}
Similarly, for the targeted LIRGs, the resulting $\alpha_{\rm CO}$ varies within the range $(2.16-4.36)~M_\odot\,({\rm K~km~s}^{-1}~{\rm pc}^2)^{-1}$, { safely} above $\alpha_{\rm CO}\simeq1~M_\odot\,({\rm K~km~s}^{-1}~{\rm pc}^2)^{-1}$, which is indeed typical of ULIRGs. {The impact of the choice of $\alpha_{\rm CO}$ is discussed in Sect.~\ref{sec:results_molecular_gas_tdep}.}

\subsection{Gas fraction and depletion timescale}

We calculated the gas depletion timescale, $\tau_{\rm dep}=M_{{\rm H}_2}/{\rm SFR}$, associated with the consumption of the molecular gas and the molecular gas to stellar mass ratios, $M({\rm H}_2)/M_\star$, from the galaxy stellar masses and the SFRs from our SED fits and the galaxy molecular gas mass estimates,
which we introduced in previous sections. 

The molecular gas properties of J101628.2+390932 \citep{Cybulski2016_thesis},
$M({\rm H}_2)$ and $\tau_{\rm dep}$, have been estimated analogously to the
other sources in our sample. To this aim, for this source we adopted
$S_{\rm CO(1\rightarrow0)}\Delta \varv=(1.250\pm0.313)$~Jy~km~s$^{-1}$
\citep{Cybulski2016_thesis} and $\alpha_{\rm CO}=3.25~M_\odot\,({\rm
  K~km~s}^{-1}~{\rm pc}^2)^{-1}$, using Eq.~(\ref{eq:alphaCO}).

For comparison we also estimated the depletion time $\tau_{\rm dep, MS}$ and the
molecular gas fraction $\big(\frac{M({\rm H_2})}{M_\star}\big)_{\rm MS}$ for the MS
field galaxies with redshifts and stellar masses corresponding to those of our
targets, following the \citet{Tacconi2018} prescription.

\begin{table*}
\begin{adjustwidth}{-0.9cm}{}
\centering
\begin{tabular}{lccccccccccccc}
\hline\hline
 Galaxy ID &  $z_{spec}$ & {J$\rightarrow$J-1}  & $\nu_{\rm obs}$ & $S_{\rm CO(J\rightarrow J-1)}\Delta \varv$   &  $M({\rm H_2})$ & $\alpha_{\rm CO}$ & $\tau_{\rm dep}$ & $\frac{M({\rm H_2})}{M_\star}$  & $\tau_{\rm dep, MS}$ & $\big(\frac{M({\rm H_2})}{M_\star}\big)_{\rm MS}$  \\
   &  & & (GHz) &  (Jy~km~s$^{-1}$)  & ($10^{10}~M_\odot$) & $\big(\frac{M_\odot}{{\rm K~km~s}^{-1}{\rm pc}^2}\big)$ & ($10^9$~yr) &  & ($10^9$~yr) & \\ 
 ~~~~~~(1) & (2) & (3) & (4) & (5) & (6) & (7) & (8) & (9) & (10) & (11) \\
 \hline
A963-1  &  0.211  &  1$\rightarrow$0  &  95.163  &  $0.85^{+0.11}_{-0.11}$  &  $0.60^{+0.08}_{-0.07}$  &  3.24  &  $0.57^{+0.07}_{-0.16}$  &  $0.26^{+0.06}_{-0.08}$  &  $1.06^{+0.13}_{-0.12}$  &  $0.11^{+0.21}_{-0.07}$  \\
A963-2  &  0.208  &  1$\rightarrow$0  &  95.415  &  $1.92^{+0.12}_{-0.13}$  &  $1.61^{+0.10}_{-0.11}$  &  3.94  &  $1.94^{+0.42}_{-0.13}$  &  $0.28^{+0.02}_{-0.07}$  &  $1.15^{+0.15}_{-0.13}$  &  $0.08^{+0.15}_{-0.05}$  \\
A2219-1  &  0.230  &  1$\rightarrow$0  &  93.694  &  $0.92^{+0.07}_{-0.08}$  &  $0.52^{+0.04}_{-0.04}$  &  2.16  &  $0.25^{+0.18}_{-0.02}$  &  $0.30^{+0.09}_{-0.03}$  &  $1.03^{+0.13}_{-0.12}$  &  $0.13^{+0.24}_{-0.08}$  \\
A2219-2  &  0.233  &  1$\rightarrow$0  &  93.504  &  $0.36^{+0.06}_{-0.06}$  &  $0.42^{+0.07}_{-0.07}$  &  4.36  &  $2.49^{+0.40}_{-2.01}$  &  $0.20^{+0.05}_{-0.12}$  &  $1.04^{+0.13}_{-0.12}$  &  $0.12^{+0.22}_{-0.08}$  \\
A1763-1  &  0.232  &  1$\rightarrow$0  &  93.549  &  $1.45^{+0.19}_{-0.21}$  &  $1.08^{+0.14}_{-0.16}$  &  2.80  &  $0.56^{+0.28}_{-0.08}$  &  $0.34^{+0.12}_{-0.10}$  &  $1.08^{+0.14}_{-0.12}$  &  $0.10^{+0.19}_{-0.07}$  \\
A1763-2  &  0.226  &  1$\rightarrow$0  &  94.015  &  $1.59^{+0.15}_{-0.12}$  &  $1.75^{+0.16}_{-0.14}$  &  4.36  &  $4.91^{+0.46}_{-4.35}$  &  $0.16^{+0.02}_{-0.04}$  &  $1.21^{+0.17}_{-0.15}$  &  $0.07^{+0.12}_{-0.04}$  \\
A697-1  &  0.274  &  1$\rightarrow$0  &  90.466  &  $0.70^{+0.08}_{-0.08}$  &  $1.11^{+0.12}_{-0.13}$  &  4.21  &  $1.47^{+0.16}_{-0.17}$  &  $0.17^{+0.02}_{-0.05}$  &  $1.13^{+0.15}_{-0.13}$  &  $0.09^{+0.17}_{-0.06}$  \\
GAL1416+4446  &  0.396  &  1$\rightarrow$0  &  82.549  &  1.0$\pm$0.1  &  $3.41^{+0.34}_{-0.34}$  &  4.22  &  $2.19^{+0.50}_{-0.58}$  &  $0.30^{+0.08}_{-0.09}$  &  $1.12^{+0.17}_{-0.14}$  &  $0.11^{+0.17}_{-0.07}$  \\GAL0926+1242-A  &  0.489  &  2$\rightarrow$1  &  154.869  &  0.6$\pm$0.1  &  $0.90^{+0.15}_{-0.15}$  &  4.08  &  $0.87^{+0.19}_{-0.22}$  &  $0.37^{+0.09}_{-0.12}$  &  $0.94^{+0.13}_{-0.11}$  &  $0.24^{+0.35}_{-0.14}$  \\
GAL0926+1242-B  &  0.489  &  2$\rightarrow$1  &  154.869  &  0.5$\pm$0.1  &  $0.49^{+0.10}_{-0.10}$  &  2.67  &  $0.32^{+0.06}_{-0.07}$  &  $1.01^{+0.21}_{-0.29}$  &  $0.81^{+0.14}_{-0.12}$  &  $0.41^{+0.61}_{-0.25}$  \\
HLS071708+374557  &  0.542  &  2$\rightarrow$1  &  149.467  &  $1.11^{+0.13}_{-0.14}$  &  $1.42^{+0.16}_{-0.18}$  &  2.82  &  $0.33^{+0.13}_{-0.21}$  &  $0.55^{+0.12}_{-0.44}$  &  $0.92^{+0.13}_{-0.11}$  &  $0.26^{+0.37}_{-0.15}$  \\
HLS071718+374124  &  0.576  &  2$\rightarrow$1  &  146.28  &  $0.46^{+0.07}_{-0.10}$  &  $1.00^{+0.16}_{-0.22}$  &  4.18  &  $0.59^{+0.17}_{-0.22}$  &  $0.20^{+0.05}_{-0.07}$  &  $0.97^{+0.14}_{-0.12}$  &  $0.22^{+0.31}_{-0.13}$  \\
HLS071731+374250  &  0.537  &  2$\rightarrow$1  &  149.953  &  $1.83^{+0.27}_{-0.31}$  &  $2.99^{+0.44}_{-0.50}$  &  3.66  &  $0.86^{+0.22}_{-0.26}$  &  $0.51^{+0.11}_{-0.13}$  &  $1.00^{+0.14}_{-0.12}$  &  $0.19^{+0.27}_{-0.11}$  \\
HLS071740+374755  &  0.563  &  2$\rightarrow$1  &  147.497  &  $1.99^{+0.06}_{-0.06}$  &  $4.07^{+0.13}_{-0.12}$  &  4.16  &  $1.31^{+0.21}_{-0.27}$  &  $0.31^{+0.06}_{-0.08}$  &  $1.06^{+0.17}_{-0.15}$  &  $0.15^{+0.21}_{-0.09}$  \\
HLS071743+374040  &  0.544  &  2$\rightarrow$1  &  149.312  &  $1.37^{+0.12}_{-0.11}$  &  $2.04^{+0.17}_{-0.17}$  &  3.26  &  $0.55^{+0.21}_{-0.21}$  &  $0.59^{+0.08}_{-0.25}$  &  $0.95^{+0.13}_{-0.11}$  &  $0.24^{+0.33}_{-0.14}$  \\
HLS071754+374303  &  0.544  &  2$\rightarrow$1  &  149.312  &  $2.14^{+0.21}_{-0.21}$  &  $3.51^{+0.34}_{-0.34}$  &  3.58  &  $0.74^{+0.19}_{-0.31}$  &  $0.42^{+0.07}_{-0.14}$  &  $1.03^{+0.15}_{-0.13}$  &  $0.17^{+0.25}_{-0.10}$  \\
HLS071754+374639  &  0.545  &  2$\rightarrow$1  &  149.216  &  $0.58^{+0.08}_{-0.09}$  &  $1.16^{+0.17}_{-0.17}$  &  4.32  &  $0.73^{+0.23}_{-0.24}$  &  $0.14^{+0.04}_{-0.04}$  &  $1.02^{+0.15}_{-0.13}$  &  $0.18^{+0.25}_{-0.10}$  \\
HLS071760+373709  &  0.553  &  2$\rightarrow$1  &  148.447  &  $<0.47$  &  $<0.90$  &  4.01  &  $<0.43$  &  $<0.19$  &  $0.97^{+0.14}_{-0.12}$  &  $0.21^{+0.30}_{-0.13}$  \\
HLS071805+373805  &  0.555  &  2$\rightarrow$1  &  148.256  &  $<0.42$  &  $<0.87$  &  4.36  &  $<1.89$  &  $<0.24$  &  $0.95^{+0.13}_{-0.11}$  &  $0.24^{+0.34}_{-0.14}$  \\
HLS071814+374117  &  0.542  &  2$\rightarrow$1  &  149.525  &  $2.53^{+0.36}_{-0.37}$  &  $4.48^{+0.65}_{-0.65}$  &  3.90  &  $1.33^{+0.38}_{-0.48}$  &  $0.51^{+0.11}_{-0.13}$  &  $1.03^{+0.15}_{-0.13}$  &  $0.17^{+0.24}_{-0.10}$  \\
 \hline
\end{tabular}
\caption{Molecular gas properties: (1) galaxy name;  (2) spectroscopic redshift of the galaxy; (3-4) CO(J$\rightarrow$J-1) transition and observer frame frequency; (5) CO(J$\rightarrow$J-1) velocity integrated flux; (6) molecular gas mass; (7) CO-to-H$_2$ conversion factor; (8) depletion timescale  $\tau_{\rm dep}=M({\rm H_2})/{\rm SFR}$; (9) molecular gas to stellar mass ratio; (10-11) depletion timescale and molecular gas to stellar mass ratio for MS field galaxies \citep{Tacconi2018}. {For HLS071760+373709 and HLS071805+373805, the quantities in columns (5,6,8,9) are $3\sigma$ upper limits.}}
\label{tab:galaxy_properties_mol_gas}
\end{adjustwidth}
\end{table*}

\section{Comparison samples}\label{sec:comparison_samples}


In order to place the properties of our LIRGs into context, we compiled
published datasets of field and cluster {star-forming galaxies}, both in the local and distant
Universe, that have been observed in CO.
In particular, as outlined below, the distant star-forming galaxies that  are used for our comparison include a homogeneous compilation of sources out to $z\simeq1.6$, largely in the field, or at most in the cluster outskirts \citep{Geach2011}.\\



As in \citet{Jablonka2013} for our comparison  we include field and
cluster galaxies in the local Universe, which have been observed in CO and also benefit from IR
luminosities and stellar mass estimates
\citep{Kenney_Young1989,Casoli1991,Young1995,Boselli1997,Lavezzi_Dickey1998,Lavezzi1999,Helfer2003,Gao_Solomon2004,Kuno2007,Saintonge2011,  Garcia-Burillo2012,Scott2013}. 
In line with our sample we restrict the comparison to galaxies with stellar masses
$\log(M_\star/M_\odot)>9.5$ and FIR luminosities in the range $\log(L_{\rm IR}/L_\odot)=9-12$.  Since we are mainly interested in looking for a possible cosmological evolution of galaxy properties, we also remove local sources \emph{stricto sensu}, that is, we only consider galaxies with $z>0.01$.
{This selection yields a total of 154 sources, all at $z<0.054$:  79 field galaxies, with $\log(M_\star/M_\odot)=10.56^{+0.21}_{-0.13}$ and $\log(L_{\rm IR}/L_\odot)=10.28^{+0.47}_{-0.36}$, as well as 75 cluster galaxies, with $\log(M_\star/M_\odot)=10.28^{+0.42}_{-0.11}$ and $\log(L_{\rm
  IR}/L_\odot)=10.34^{+0.20}_{-0.26}$.\footnote{For both stellar mass and IR luminosity, we report the median value and the $68.27\%$ confidence region (1$\sigma$) uncertainties.} }\\





{ We consider the 61  star-forming galaxies at $0.15 < z < 0.35$ from the {\it Herschel} Astrophysical Terahertz Large Area Survey (H-ATLAS) with ALMA CO(1$\rightarrow$0) detections  reported by \citet{Villanueva2017}. They have total IR luminosities and stellar masses in the range $\log(L_{\rm IR}/L_\odot)\simeq10.1-11.9$ and $\log(M/M_\star)\simeq9.7-11.3$, respectively, which were estimated  using SED fits with MAGPHYS, similar to what has been done for our work. We did not consider 6 sources with only upper limits in CO(1$\rightarrow$0) that are part of the full sample of 67 galaxies reported in \citet{Villanueva2017}.}

For the studies quoted above, the galaxy IR luminosities were
converted into SFRs using the \citet{Kennicutt1998} relation, rescaled for a
\citet{Chabrier2003} IMF, on which MAGPHYS SED fits of this work relies. Namely,
following the prescription by \citet{daCunha2010} we adopted the
relation
\begin{equation}
\label{eq:LIR_to_SFR}
 \frac{\rm SFR}{\rm M_\odot/yr}=1.075\times10^{-10}\frac{L_{\rm IR}}{L_\odot}\,.   
\end{equation}

We further include the following samples.\\

$\bullet$ The five LIRGs detected in CO(1$\rightarrow$0) by \citet{Geach2011} in the outskirts ($r/r_{200} \sim$ 1 -
3) of the rich cluster Cl~0024+16 ($z=0.395$). {The authors report 7.7$\mu$m-based SFRs in the range $\sim(30-60)~M_\odot/{\rm yr}$ and stellar masses $M_\star\sim10^{11}~M_\odot$.}\\

$\bullet$ { The sample of 20 LIRGs at $0.2<z<0.7$ detected in CO(3$\rightarrow$2) with ALMA observations by \citet{Lee2017}. The sources fall within the equatorial COSMOS survey \citep{Scoville2007} and are bright submillimeter galaxies observed with {\it Herschel}. These have IR luminosities $\log(L/L_{\rm IR})\simeq11.1-11.6$ and SFR~$\simeq(10-37)~M_\odot$/yr. Estimates for the SFR were obtained by the authors using both IR and UV luminosities using data from \textit{Spitzer}, \textit{Herschel}, and \textit{GALEX}.}\\

$\bullet$  The subsample of 27 star-forming galaxies,
SFR$=(3.4-106)~M_\odot$/yr, at $z=0.05-0.3$, with CARMA observations in CO(1$\rightarrow$0) by \citet{Bauermeister2013}. Our selection excludes their 4 $z\simeq0.5$ sources with  upper limits only in CO(3$\rightarrow$2). {We adopt  stellar  masses  and  SFRs  reported by the authors that correspond to those of the seventh release of SDSS provided  by  the  Max-Planck-Institute  for Astrophysics-John  Hopkins  University  (MPA-JHU)  group.}\footnote{\url{http://www.mpa-garching.mpg.de/SDSS}}\\


$\bullet$ The 8 $z=0.1-0.2$ star-forming galaxies detected in
CO(1$\rightarrow$0) by \citet{Morokuma-Matsui2015} with the Nobeyama Radio Observatory (NRO). We adopt the SFRs, SFR$\simeq10~M_\odot$/yr, reported by the authors, which correspond to those of the
tenth release of the
SDSS\footnote{\url{http://skyserver.sdss.org/dr10/en/home.aspx}} and were
derived using at least five emission lines \citep{Brinchmann2004}.\\

$\bullet$ The recent PHIBSS2 observations from \citet{Freundlich2019}. Their sample includes 60 CO(2$\rightarrow$1) detections of $0.5<z<0.8$ star-forming galaxies, with SFR$=(28-630)~M_\odot$/yr, inferred from UV and IR fluxes, and corrected for { extinction}.\\


$\bullet$ { \citet{Tacconi2013} present} 52 CO(3$\rightarrow$2) detections of star-forming galaxies in two redshift slices centered at $z\sim1.2$ and 2.2. The observations are part of the Plateau de Bure high-$z$ Blue Sequence Survey (PHIBSS) and were done with the PdBI. For our comparison we consider the subsample of 38 detections at $1.0\lesssim z\lesssim1.5$ with SFR$=(28-630)~M_\odot$/yr, {estimated by the authors and based either on the sum of the observed UV and IR luminosities, or on an extinction-corrected H$\alpha$ luminosity. In both cases the SFRs provided by the authors were corrected for a \citet{Chabrier2003} IMF, hence consistent with our estimates.} \\

$\bullet$ The six BzK field galaxies detected in CO by \citet{Daddi2010},at $1.4<z<1.6$ with estimated IR luminosities $L_{\rm IR}=
(0.6-4.0)\times10^{12}~L_\odot$, typical of ULIRGs. The SFRs, (62-400)~M$_\odot$/yr, were calculated
with a \citet{Chabrier2003} IMF, hence consistent with our estimates.
Likewise \citet{Tacconi2013} we include these six sources in the PHIBSS sample.

\section{Results}\label{sec:results}

\subsection{Stellar masses and star formation rates}\label{sec:results_Mstar_SFR}
\begin{figure*}[!ht]\centering
\subfloat[]{\includegraphics[trim={0cm 0cm 0cm 0cm},clip,width=0.5\textwidth,clip=true]{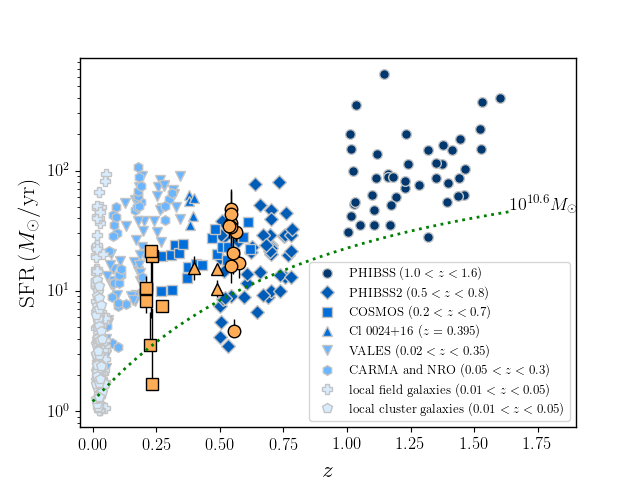}\hspace{0.3cm}}
\subfloat[]{\includegraphics[trim={0cm 0cm 0cm 0cm},clip,width=0.5\textwidth,clip=true]{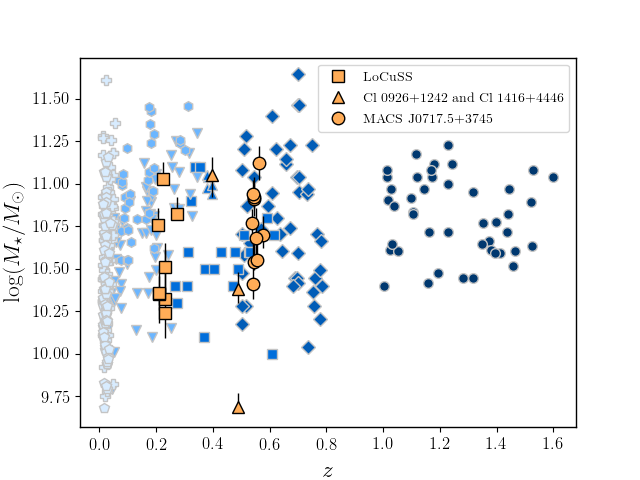}\hspace{0.3cm}}\\
\subfloat[]{\includegraphics[trim={0cm 0cm 0cm 0cm},clip,width=0.5\textwidth,clip=true]{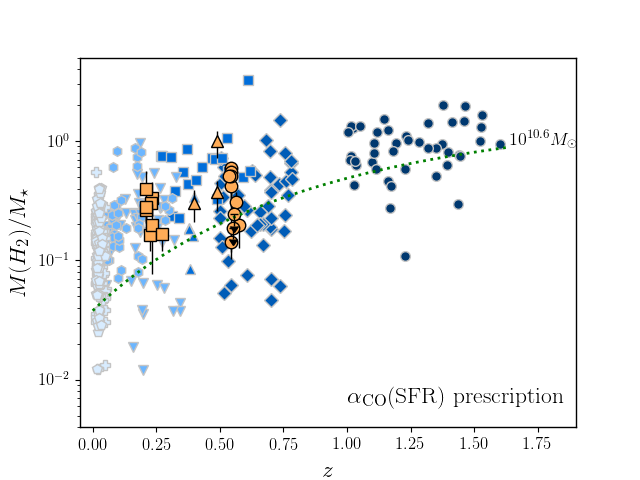}}
\subfloat[]{\includegraphics[trim={0cm 0cm 0cm 0cm},clip,page=5,width=0.5\textwidth,clip=true]{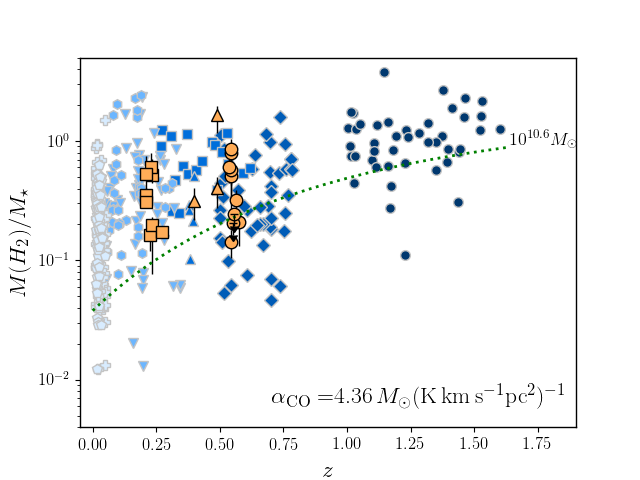}} \\
\caption{{(a) SFR {\it vs.} $z$ and (b) $M_\star$ {\it vs.} $z$ scatter plots.} The green dotted line {in panel~(a)} shows the empirical SFR values by \citet{Speagle2014} for MS field galaxies (SFR$_{\rm MS}$) with stellar mass   $\log(M_\star/M_\odot)=10.6$, which corresponds to the mean stellar mass for the LIRGs in our sample. 
(c,d) Molecular gas to stellar mass ratio as function of redshift; the two plots differ in terms of the adopted $\alpha_{\rm CO}$, as shown at the { bottom right} of the panels. The green dotted lines show the empirical values found by \citet{Tacconi2018} for MS field galaxies with a stellar mass   $\log(M_\star/M_\odot)=10.6$. {The color code for the data points is given in panels (a,b).}}
\label{fig:our_targets_2}
\end{figure*}

\begin{figure*}[!ht]\centering
\subfloat[]{\includegraphics[trim={0cm 0cm 0cm 0cm},clip,width=0.5\textwidth,clip=true]{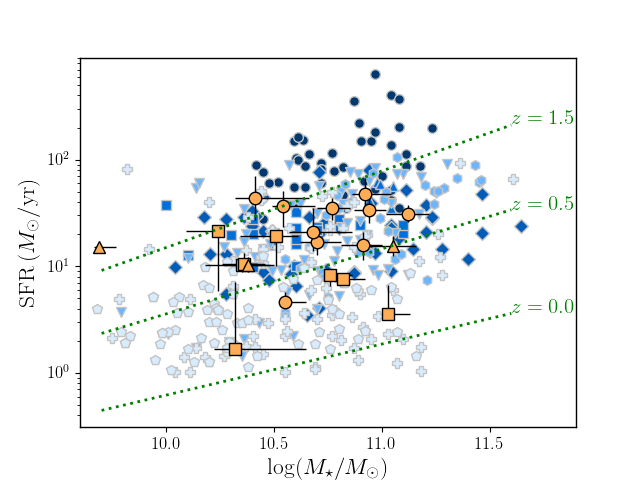}}
\subfloat[]{\includegraphics[trim={0cm 0cm 0cm 0cm},clip,width=0.5\textwidth,clip=true]{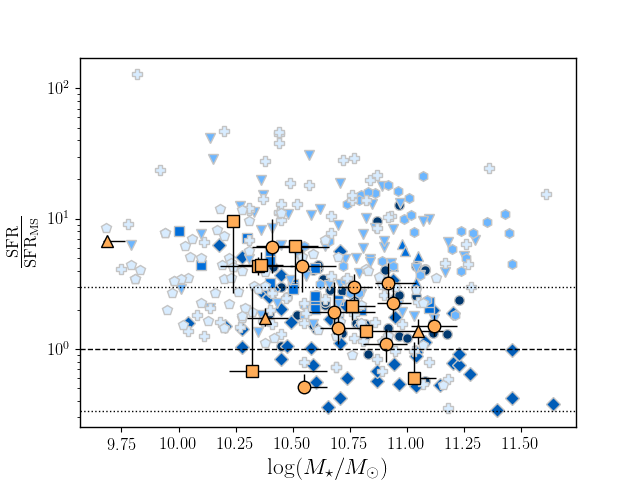}}\\
\subfloat[]{\includegraphics[trim={0cm 0cm 0cm 0cm},clip,width=0.5\textwidth,clip=true]{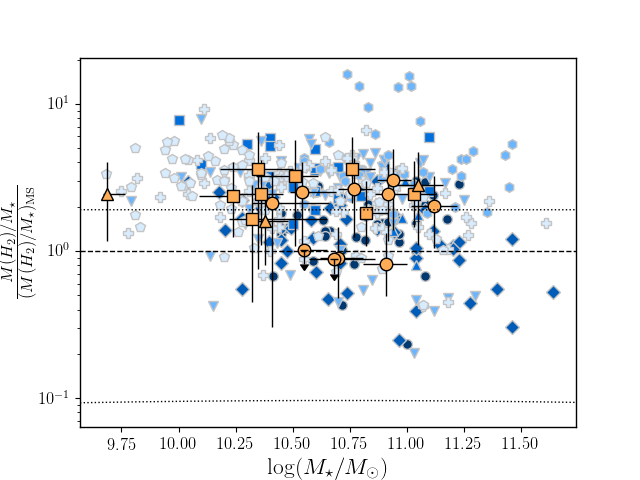}}
\subfloat[]{\includegraphics[trim={0cm 0cm 0cm 0cm},clip,width=0.5\textwidth,clip=true]{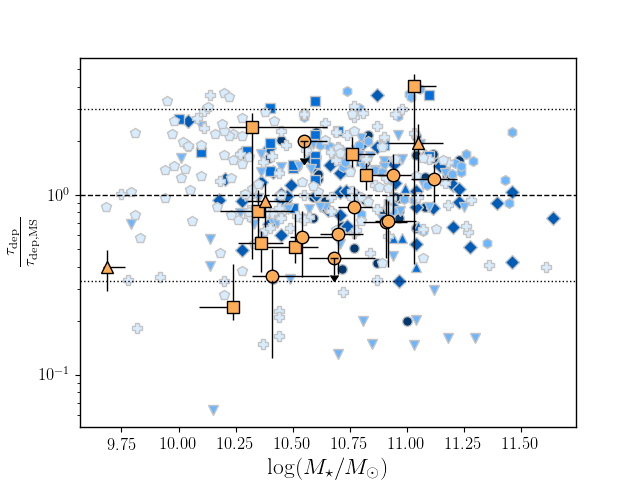}}\\
\caption{(a) Star formation rate as a function of stellar mass, the MS relation at different redshifts from \citet{Speagle2014} is shown as dotted green lines. (b,c,d) The SFR, molecular gas to stellar mass ratio, and depletion time, respectively, all normalized by the corresponding MS values \citep{Speagle2014,Tacconi2018}, as a function of the stellar mass. { The color code for the data points is analogous to Fig.~\ref{fig:our_targets_2}.} 
{In the panels (b,c,d) the horizontal dashed line refers to the MS value, that is, the y-axis value equal to unity, while the horizontal dotted lines show the range of y-axis values corresponding to MS galaxies.}}
\label{fig:our_targets}
\end{figure*}

{{In Fig.~\ref{fig:our_targets_2} we report, as a function of redshift, the SFRs, stellar masses, and molecular gas to stellar mass ratios of the LIRGs in our sample and those of the galaxies in the comparison samples. }
As seen in Fig.~\ref{fig:our_targets_2}a, our sources have SFRs in the range
$\simeq(4-50)~M_\odot/{\rm yr}$, with values typical of intermediate-redshift MS galaxies within the same mass range ($\log(M_\star/M_\odot)\simeq10-11$), as  displayed in Fig.~\ref{fig:our_targets_2}b, {which shows the galaxy stellar mass as a function of redshift.}

At fixed stellar mass, our LIRGs, however, have lower SFRs than the more distant $1.0<z<1.6$ sources
\citep{Daddi2010,Tacconi2013} in the comparison sample. This is the consequence
of the increment of the SFR at the MS with redshift, as shown in
Fig.~\ref{fig:our_targets}a, {where the SFR versus stellar mass scatter plot is shown, along with the MS relation at different redshifts.}

Once SFRs are normalized to {the corresponding MS values (Fig.~\ref{fig:our_targets}b)}, our LIRG sample populates the same {region in the }
${\rm SFR}/{\rm SFR}_{\rm MS}$ versus $\log(M_\star/M_\odot)$ {diagram}  as the comparison sample sources,  irrespective of their SFR, from normal star-forming galaxies to LIRGs, {as well as distant ULIRGs \citep{Daddi2010}.}
We also note that the comparison sources preferentially lie on the upper part of the MS. {This is most likely an observational bias and is linked to the fact that CO observations for ${\rm SFR}<{\rm SFR}_{\rm MS}$ sources are more uncertain and require a longer integration time.}

Fig.~\ref{fig:our_targets}b shows a significant scatter in SFR/SFR$_{\rm{MS}}$, {or equivalently in sSFR/sSFR$_{\rm{MS}}$}, at a given log($M_\star/M_\odot$). Most of our LIRGs are distributed within the {fiducial}   $|\log({\rm SFR/SFR}_{\rm MS})|< \log(3)= 0.48$   {scatter of} the MS, with the exception of {8 LIRGs}  that have a higher level of star formation activity {(see also Table~\ref{tab:galaxies_SEDproperties}).} {No strong trend is observed }
between SFR and stellar mass, as indeed low SFRs are seen both at the low- and the high-mass ends of our sample, although the most active of our systems have
log($M_\star$)$\le$10.5.  {This last aspect is reflected} in a tentative anticorrelation, with a significance of 2.5$\sigma$ (p-value = 0.013), between ${\rm SFR/SFR}_{\rm MS}$ and $\log(M_\star/M_\odot)$, that we find with the Spearman test.


{Furthermore, despite the LIRGs in our sample {having} been primarily selected on the basis of their FIR luminosity, they span a relatively wide range (an order of magnitude) in SFR, which is because both FIR and UV luminosities are useful to constrain the SFR. This can be appreciated from the comparison between the SEDs of A1763-1 and A2219-1 in Fig.~\ref{fig:SED}. While A1763-1 is brighter in the FIR, it also has a lower SFR than A2219-1. This is ultimately because A2219-1 has a stronger UV excess \citep[GALEX,][]{Morrissey2007}, as seen in the SEDs. 
We also stress that using the total IR as a SFR tracer may lead to an overestimation of the SFR, unless the contribution by the diffuse interstellar medium to the total IR luminosity is properly taken into account \citep[see also discussion in][]{Kennicutt2009,daCunha2012,Hayward2014,Hunt2019}. For the LIRGs in our sample, we find that by using Eq.~\ref{eq:LIR_to_SFR}, with $L_{\rm IR}$ replaced by $L_{\rm dust}$ provided by MAGPHYS, the SFR is $2.0^{+5.9}_{-0.7}$ higher\footnote{In this work we report the median value and the 1$\sigma$ uncertainties, corresponding to the 68.27\% confidence region.} than that estimated by MAGPHYS (Table~\ref{tab:galaxies_SEDproperties}) using the full FIR-to-UV SED. }


\subsection{Molecular gas and depletion timescale}\label{sec:results_molecular_gas_tdep}


{ In Figs.~\ref{fig:our_targets_2}cd we show} the redshift evolution
of the {molecular gas to stellar mass ratio} $\mu =M(H_2)/M_\star$ and illustrate {the impact on $\mu$ when choosing different $\alpha_{\rm CO}$ prescriptions.} 
{The majority (13) of the LIRGs have SFR/SFR$_{\rm MS}\lesssim3$ and are therefore formally consistent with being on the MS. Among the remaining eight LIRGs, half of them have ${\rm SFR}/{\rm SFR}_{\rm MS}\simeq(3-4)$ and the other half have ${\rm SFR}/{\rm SFR}_{\rm MS}\simeq(6-10)$. To account for such a broad range in SFR, from the MS up to larger values, in Sect.~\ref{sec:molecular_gas_masses} we have introduced a heuristic dependence of $\alpha_{\rm CO}$ on the SFR, denoted as $\alpha_{\rm CO}({\rm   SFR})$ prescription. In panel~(c) of Fig.~\ref{fig:our_targets_2} the values of $\mu$ are estimated using this prescription, while in panel~(d) a Galactic H$_2$-to-CO conversion factor $\alpha_{\rm CO}=4.36~M_\odot\,({\rm K~km~s}^{-1}~{\rm   pc}^2)^{-1}$ is used.} 

Large differences in $\mu$ between the two prescriptions, up to a
factor of $\sim5$ in molecular gas masses are seen, but only in the case
of very active and gas rich systems with $\mu\gtrsim1$.  {However, the discrepancy in $\mu$}  for 
normal star-forming galaxies and LIRGs is limited and does not
exceed a factor of $\sim2$.

Our sample of LIRGs covers a factor $\sim$30 in SFR and shows a
small but clear increase of star formation activity from $(9.3^{+1.2}_{-1.6})~M_\odot$/yr at $z\sim0.25$ to $(20.7^{+13.1}_{-4.9})~M_\odot$/yr at $z\sim0.5$ following standard relations for MS galaxies
\citep[e.g.,][]{Speagle2014}.\footnote{Here and throughout this Sect.~\ref{sec:results_molecular_gas_tdep}, when reporting values with associated uncertainties, we refer to the median and the $68.27\%$ confidence region (1$\sigma$) uncertainties.}}

{On the other hand, the LIRGs span a lower factor $\sim7$ in $\mu$, with an increment going from $0.27^{+0.02}_{-0.06}$ at $z\sim0.25$ to $0.42^{+0.09}_{-0.10}$ at $z\sim0.5$, following the general trend driven by standard relations for MS galaxies  \citep{Tacconi2018}.}
Furthermore, while the SFRs cover the full
range of SFRs reported so far in the literature at similar redshifts (Fig.~\ref{fig:our_targets_2}a),
the {molecular gas to stellar mass ratios} of our LIRGs seem to lie in the high tail of the distribution (Fig.~\ref{fig:our_targets_2}c). { However, by applying standard nonparametric tests (Mann-Whitney-Wilcoxon, Kolmogorov-Smirnov) we did not find any statistically significant difference in the distributions of $\mu$ between our LIRGs and sources in the comparison sample at similar redshifts $0.2<z<0.6$ than the LIRGs.}




In Figs.~\ref{fig:our_targets}cd we compare the ratios $\mu$ and the depletion timescale $\tau_{\rm dep}=M(H_2)/{\rm SFR}$, both normalized to their values at the MS, of the LIRGs in our sample with those of the sources in the comparison samples. The normalized $\mu$ and $\tau_{\rm dep}$ are plotted against the stellar mass on the x-axis.

Our LIRG sample shows normalized values of $\mu$ equal to $2.4^{+0.1}_{-0.3}$ and  therefore covers the upper part of {the range typically associated with MS galaxies \citep{Tacconi2018}.} {However, our sample of LIRGs} does not contain any extremely gaseous system, with normalized values of $\mu$ in the range  $\sim(0.8-3.6)$, that is, a factor of $\sim$4 dispersion.   

{As shown in Fig.~\ref{fig:our_targets}b a much larger dispersion  is observed for the normalized SFR, SFR/SFR$_{\rm MS}=(2.1^{+1.8}_{-0.6})$. The normalized SFR spans indeed the range between $\sim(0.5-9.6)$, corresponding to a factor of $\sim20$ dispersion, which is much higher than that found for the normalized $\mu$. }


As seen in Fig.~\ref{fig:our_targets}d and Table~\ref{tab:galaxy_properties_mol_gas} depletion times within $\tau_{\rm dep}=(0.3-5)$~Gyr are found for the targeted
LIRGs, {with values scattering around those associated with the MS and equal to $(0.8^{+0.4}_{-0.2})\times\tau_{\rm dep,MS}$. Similarly to what has been found for the SFR (Sect.~\ref{sec:results_Mstar_SFR}), by applying the Spearman test a hint for a correlation between the normalized  $\tau_{\rm dep}$ and $\log(M/M_\star)$ is found at 2.5$\sigma$ (p-value~=~0.014).}
The location of the LIRGs in the normalized $\tau_{\rm dep}$  versus $\log(M_\star/M_\odot)$ plane
reflects the trend observed in the ${\rm SFR}/{\rm SFR}_{\rm MS}$ versus $\log(M_\star/M_\odot)$ plot in Fig. \ref{fig:our_targets}b. This is a result 
of the  flat behavior of the normalized $\mu$ with $M_\star$ and its relatively small dispersion. Therefore, the small timescales $\tau_{\rm dep}<\tau_{\rm dep,MS}$ observed for a large fraction of our LIRGs (i.e.,13 out of 21) are the result of stronger activity rather than exhaustion of gas.

{Overall, the results presented in Figs.~\ref{fig:our_targets_2} and  \ref{fig:our_targets} show that the present study is one of the first to probe statistically large molecular gas reservoirs in the still overlooked population of star-forming { cluster} galaxies at intermediate redshifts. Our results thus complement, in terms of  redshift and environment, those found for field galaxies, and also those of star-forming cluster or field galaxies in the local Universe, of which the LIRGs in our sample are the higher-$z$ counterparts.}

\subsection{Galaxy properties versus cluster-centric distance}\label{sec:results_prop_vs_radius}

In the following we focus on the properties of our sample of intermediate-redshift LIRGs in Abell~1763, Abell~2219, Abell~697, Abell~963, Cl~0926+1242,
Cl~1416+4446, and MACS~J0717.5+3745 (Tables \ref{tab:galaxy_properties_general},  \ref{tab:galaxies_SEDproperties}, and \ref{tab:galaxy_properties_mol_gas}).

\begin{figure*}\centering
\subfloat{\hspace{0.2cm}\includegraphics[trim={0cm 0cm 0cm 0cm},clip,width=0.5\textwidth,clip=true]{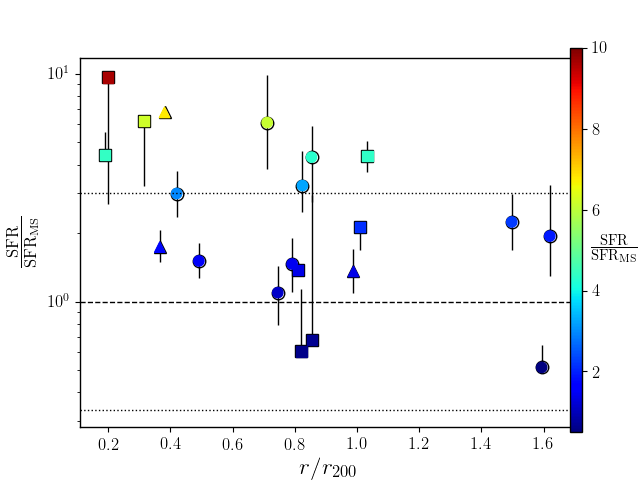}}
\subfloat{\hspace{0.2cm}\includegraphics[trim={0cm 0cm 0cm 0cm},clip,width=0.5\textwidth,clip=true]{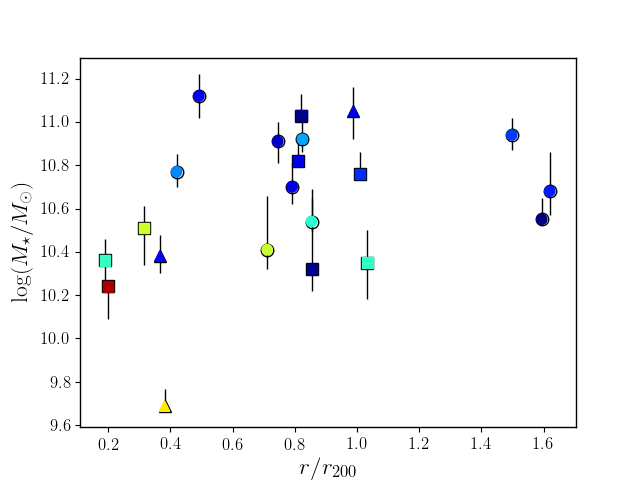}}
\caption{Star formation rate to SFR$_{\rm MS}$ (left panel) and stellar mass (right panel), as a function of $r/r_{200}$ of the LIRGs. Symbols are shown as  in Fig.~\ref{fig:our_targets_2}b. { Blue, green, yellow, and red data points correspond to increasing SFR/SFR$_{\rm MS}$ values, as illustrated in the color bar (left).} In the left panel the horizontal dashed line refers to the MS value, while the horizontal dotted lines show the $\pm0.48$~dex scatter corresponding to MS galaxies.}
\label{fig:SFR_Mstar_vs_radius}
\end{figure*}

Figure~\ref{fig:SFR_Mstar_vs_radius} {(left)  suggests} that the 
SFR/SFR$_{\rm MS}=({\rm SFR}/M_\star)/({\rm SFR}_{\rm MS}/M_\star)=$~sSFR/sSFR$_{\rm{MS}}$ values scatter within the MS range. However, while the inner regions of clusters {(i.e., $r/r_{200}<0.6$)} contain galaxies with normal star formation activity, all LIRGs formally above the MS, with ${\rm SFR}>3\times~{\rm SFR}_{\rm MS}$, are found within the cluster virial radius, irrespective of the cluster redshift. 
{Among the 16 LIRGs in our sample with projected cluster-centric distances $<r_{200}$, 7 ($44\%\pm12\%$)\footnote{The fraction and the uncertainties are estimated using the binomial distribution} have enhanced ${\rm SFR}>3\times{\rm SFR}_{\rm MS}$.}

As seen in the right panel of Fig.~\ref{fig:SFR_Mstar_vs_radius} these {cluster-core, star-forming} galaxies tend to have lower stellar masses, $\log(M_\star/M_\odot)<10.6$, than the rest of the LIRGs.  However the stellar mass alone is not a sufficient criterion {to explain the high star formation activity in the cluster-core LIRGs. For example, both GAL0926+1242-A  and A2219-2 are on the MS and have low stellar masses, $\log(M_\star/M_\odot)\simeq(10.3-10.4)$.} Similarly, the two MACS~J0717.5+3745 LIRGs
at the very upper edge of the MS have $\log(M_\star/M_\odot)\gtrsim10.8$.{We statistically quantified our results. By applying the Spearman test we found a tentative  anticorrelation that has a significance of 2.2$\sigma$ (p-value=0.03), between SFR/SFR$_{\rm MS}$ and $r/r_{200}$, which is ultimately due to the presence of LIRGs with ${\rm SFR}>3\times{\rm SFR}_{\rm MS}$ in the cluster inner regions.
However, we stress that additional observations in CO of LIRGs in the cores of intermediate-redshift clusters are needed to strengthen our results.}

\begin{figure}[!ht]\centering
\subfloat{\includegraphics[trim={0.6cm 0cm 0cm 1cm},clip,
width=0.5\textwidth,clip=true]{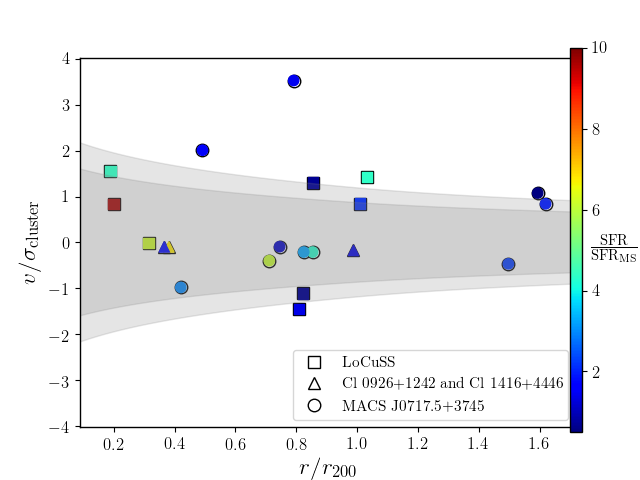}}
\caption{Phase space diagram for all cluster galaxies in our sample. In the x-axis we plot the projected cluster centric radius, $r$, normalized to $r_{200}$, while in the y-axis we plot the line-of-sight velocity, normalized to the cluster velocity dispersion $\sigma_{\rm cluster}$. The dark and light gray areas show the range of virialized regions defined by \citep{Jaffe2015} for the clusters in our sample.}
\label{fig:phase_space_diagram}
\end{figure}

Figure~\ref{fig:phase_space_diagram} places our sample in the cluster phase-space (line-of-sight velocity versus cluster-centric radius) diagram.
The LIRGs are color coded according to their ${\rm SFR}/{\rm SFR}_{\rm MS}$
ratios. We also highlight the
smallest and largest virialized regions in dark and light gray, respectively, accounting for the different cluster masses and
concentrations of the clusters considered.  The virialized regions were derived with the analytical
model of \citet{Jaffe2015}. Our targets span a broad range of cluster-centric
distances, from the cluster cores to their infall regions, out to $\sim$1.6
r$_{200}$.  Most of the sample LIRGs are overall located within the cluster
virialized region; the line-of-sight velocity is not greater than $\sim 2$ times
the cluster velocity dispersion.

Interestingly, two LIRGs in MACS~J0717.5+3745, which
have projected cluster-centric distances below $r_{200}$ but normal SF
activity (SFR/SFR$_{\rm MS}<3$)  have relative velocities larger than 2$\sigma_{\rm{cluster}}$.  
{ Large velocities are often found for cluster-core members; this also applies to  MACS~J0717.5+3745 ($\sigma_{\rm cluster}\simeq1660$~km/s; Table~\ref{tab:cluster_properties}). Considering these aspects, the location of the two sources with respect to the cluster center does not significantly differ from that of the LIRGs in our sample with lower values of $v/\sigma_{\rm cluster}$, but higher cluster centric distances $r>r_{200}$. Thus the two LIRGs likely belong to the infall regions of MACS~J0717.5+3745 and we have not discarded them, although these two LIRGs are formally outside the virialized region of the cluster (Fig.~\ref{fig:phase_space_diagram}).}





\begin{figure*}[!ht]\centering
\subfloat{\hspace{0.2cm}\includegraphics[trim={0cm 0cm 0cm 0cm},clip, page=5,width=0.5\textwidth,clip=true]{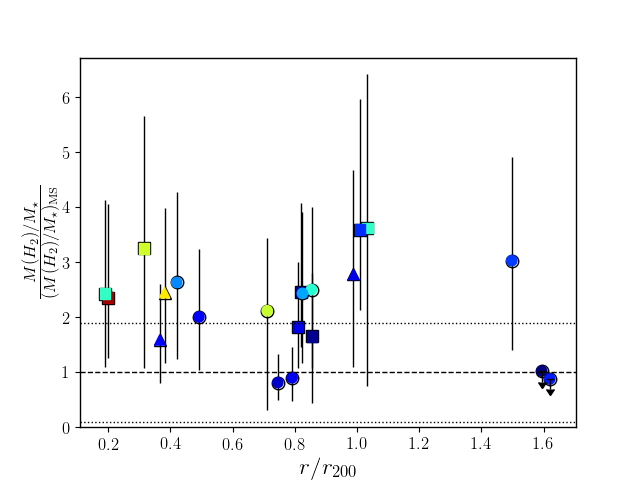}}
\subfloat{\hspace{0.2cm}\includegraphics[trim={0cm 0cm 0cm 0cm},clip,page=5,width=0.5\textwidth,clip=true]{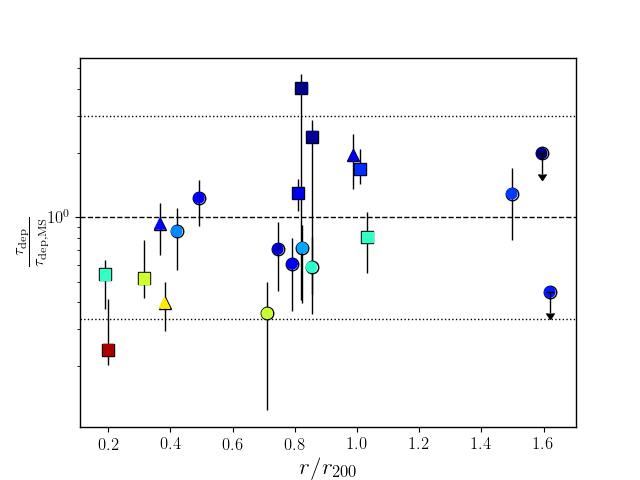}}

\caption{Molecular gas to stellar mass ratio (left) and depletion time (right), normalized to the corresponding MS values, as a function of $r/r_{200}$. The data points have different symbols and are color-coded according to the corresponding SFR/SFR$_{\rm MS}$, as in Fig.~\ref{fig:phase_space_diagram}. In each panel, the horizontal dashed line refers to the MS value, while the horizontal dotted lines show the range of y-axis values corresponding to MS galaxies.}\label{fig:prop_vs_radius}
\end{figure*}

The left panel of Fig.~\ref{fig:prop_vs_radius} shows that the factor $\sim$4 dispersion in the normalized $\mu$ that was
noted in Sect.~\ref{sec:results_molecular_gas_tdep} from Fig.~\ref{fig:our_targets}c has no clear link with the spatial position of the galaxy. 
{A similar behavior is found for the H$_2$-to-dust mass ratio, as a function of $r/r_{200}$. On average we find $M({{\rm H}_2})/M_{\rm dust} = 74^{+25}_{-20}$, which is consistent with the typical ratio $\sim100$ found for star-forming galaxies \citep{Scoville2014,Scoville2016,Berta2016}.}
However, interestingly, as seen in Fig.~\ref{fig:prop_vs_radius} (right), when both molecular gas and star formation activity are considered together via the gas depletion timescale, $\tau_{\rm dep}$, {a deficit of sources with $\tau_{\rm dep}\gtrsim\tau_{\rm dep, MS}$ is observed, as we move from the outskirts $r\gtrsim r_{200}$ down to the cluster inner regions.}

  
  This radial trend stands out more clearly than {in the case of SFR/SFR$_{\rm MS}$ versus $r/r_{200}$ (Fig.\ref{fig:SFR_Mstar_vs_radius}, left panel).} The Spearman test
{ suggests a possible} correlation between $\tau_{\rm dep}/\tau_{\rm dep, MS}$ and $r/r_{200}$, which we find at a significance of 2.8$\sigma$ (p-value = 0.005) {despite the significant scatter of $\tau_{\rm dep}/\tau_{\rm dep, MS}$ at a given cluster-centric distance.}



\section{Discussion}\label{sec:discussion}


\subsection{Star formation enhancement and gas enrichment}
For this work we selected a sample of LIRGs based on their FIR luminosities. However, once we modeled their FIR-to-UV SEDs the LIRGs show a wide range in SFR$\simeq(4-50)~M_\odot/{\rm yr}$, which is a result of the combination of both obscured (in the FIR) and unobscured (in the UV) ongoing star formation activity.

{The presence of enhanced star formation activity observed in a large fraction of the LIRGs within the virial radius suggests that} environmental processing mechanisms such as ram-pressure stripping, galaxy harassment, and starvation may have not been sufficiently effective in suppressing the star formation and then quenching the LIRGs. Recent simulations of dwarf galaxies by \citet{Hausammann2019} show that, at variance with the hot gas, cold gas may not be efficiently removed by ram pressure. 

{Our results suggest that the enhancement of  star formation for the cluster LIRGs, toward the cluster cores, is the main responsible for the corresponding decrease of depletion timescales (i.e., $\tau_{\rm dep}\lesssim\tau_{\rm dep,MS}$).}  {The short depletion times imply that the star formation in the LIRGs is shortly suppressed, which ultimately quenches them,} meanwhile increasing the fraction of passive galaxies in clusters \citep[e.g., SFR  versus density and morphology versus density relations,][]{Dressler1980,Lewis2002,Peng2010,Andreon2006,Raichoor_Andreon2012}.
{The short depletion times observed for a large fraction of LIRGs within the virial radius also imply} that molecular gas is consumed { rapidly} in cluster LIRGs. This may {help to} explain the fast exponential decrease with time, since $z\sim0.8$ of the fraction of LIRGs in clusters found in previous studies  \citep{Finn2010,Saintonge2008,Popesso2012}.

{To further investigate this scenario {it is interesting to look for possible differences} among the LIRGs, in terms of denser molecular gas reservoirs, from those probed by the low ${\rm J}=2\rightarrow1$ and ${\rm J}=1\rightarrow0$ transitions of CO considered in this work.}


Visual inspection of the images of our LIRGs (Figs.~\ref{fig:HSTimages},\ref{fig:SUBARUimages}) also suggests the presence of compact or clumpy optical morphologies, which seem to be common in distant star-forming and {\it Herschel}-selected galaxies. Consistent with these results, using ALMA observations \citet{Puglisi2019} recently found that a significant fraction, that is, $\sim50\%$, of their sample comprising 93 {\it Herschel}-selected galaxies at $1.1\leq z\leq 1.7$ and at the MS show a compact morphology. The authors interpret their sources as  early post-starburst galaxies. 



\subsection{Gas compression and environmental processing}
Similar to the targeted LIRGs of this work, star-forming galaxies have been occasionally found in the cores of nearby clusters in previous studies. \citet{Miller_Owen2001} found an excess of star-forming galaxies with enhanced radio emission, possibly not due to active nuclei, in the cores of nearby Abell clusters.   \citet{Bressan2002} proposed that this excess could be associated with a population of starburst galaxies. \citet{Miller_Owen2001} proposed instead that the excess is due to compression of the galactic magnetic field by thermal pressure of the intra-cluster medium, while \citet{Gavazzi_Jaffe1986} suggested that the excess is due to the ram pressure, which ultimately strengthens the magnetic field of the galaxy, as it moves through the intra-cluster medium.
Overall, the compression of gas is invoked in these studies to explain the excess of star-forming galaxies in the cores of nearby clusters. {Consistent with these studies, simulations by \citet{Bekki2014} found that the ram pressure can compress the interstellar medium gas and ultimately enhance the star formation of cluster and group galaxies.} {In the following we investigate the possibility that gas compression is also responsible for the strong star formation activity in some of the targeted LIRGs.}


A2219-1 is close to the shock front of A2219, which {might have favored the gas compression and an enhancement of star formation in the galaxy. The galaxy has a high} ${\rm SFR}=21.2~M_\odot$/yr, corresponding to ${\rm SFR}\simeq 18\,{\rm  SFR}_{\rm MS}$, {which is strengthened by} its UV excess (Fig.~\ref{fig:SED}). {Similarly, for A2744 at $z=0.308$, which is part of the HLS project, \citet{Rawle2014} found that cluster sources with elevated SFR could be associated with a large-scale shock front, suggesting gas compression as the cause.}

A large fraction $\sim50\%$ (10/21) of the LIRGs in our sample belong to MACS~J0717.5+3745. Seven among the ten LIRGs are formally located within the $r_{200}$ radius, {in projection,} while HLS071760+373709, HLS071805+373805, and HLS071814+374117 are found at larger cluster-centric distances (see Table~\ref{tab:galaxy_properties_general}). {As shown in Fig.~\ref{fig:M0717field} the LIRGs follow the pattern in dark matter defined by the weak-lensing analysis by \citet{Jauzac2012}. In particular, the three LIRGs at projected cluster-centric distances $>r_{200}$ fall within the filament extending toward the southeast from the cluster core.}

{We did not find any statistically significance difference concerning the molecular gas content of the MACS~J0717.5+374 LIRGs from the cluster inner regions out to the outskirts; see Fig.~\ref{fig:prop_vs_radius} (left). This result suggests that there is no strong evidence of preprocessing of molecular gas in these LIRGs by the large-scale dense environment (cluster and filament). However, the  MACS~J0717.5+374 LIRGs contribute to the observed increase of $\tau_{\rm dep}$, normalized to its MS, with increasing cluster centric distances (Fig.~\ref{fig:prop_vs_radius}, right), which has been discussed in Sect.~\ref{sec:results_prop_vs_radius} and is primarily due to the increase of the normalized SFR toward the cluster core.}

{Interestingly, we found four MACS~J0717.5+3745 LIRGs, namely  HLS071754+374303, HLS071731+374250, HLS071740+374755, and HLS071814+374117 at the edge of both X-ray and radio extended emissions associated with the cluster \citep[see Fig.~1 of][]{Bonafede2018}. HLS071754+374303 is also close to an X-ray infalling group to the southeast of the cluster, along the filament. The four sources have SFR$\gtrsim30~M_\odot$/yr and are therefore among those in our sample with the strongest star formation activity.
Furthermore, while HLS071740+374755 is located to the northeast of the cluster, the other three LIRGs are all found along the filament direction and also formally above the MS, having ${\rm SFR}\gtrsim 3\,{\rm  SFR}_{\rm MS}$.}

To explain such high (normalized) SFRs a possible scenario may be that a large-scale shock in MACS~J0717.5+374 has compressed the gas reservoirs in some of the LIRGs along the filament direction thus ultimately enhancing their star formation activity.
Interestingly, previous studies suggested that the $\sim(0.7-0.8)$~Mpc radio relic found in MACS~J0717.5+374 \citep{Bonafede2018}  traces  a  large-scale  shock  wave  propagating through the cluster and  originated   from either the  merger  events \citep{vanWeeren2009} or an accretion shock related to the SE  filament \citep{Bonafede2009}.










{Nevertheless, we note that  HLS071708+374557 is highly star forming and has  SFR$\simeq40~M_\odot/{\rm yr}\simeq6\,{\rm  SFR}_{\rm MS}$. However, it is not found in correspondence of the X-ray or radio emission of the cluster. As further outlined below it is thus likely that shock-induced gas compression is not the only mechanism possibly responsible for the high star formation activity observed in the LIRGs of MACS~J0717.5+374.}

{Interestingly, the only two sources with only upper limits to the CO flux have a normal MS star formation activity and are both located at the periphery ($r/r_{200}\simeq1.6$) of MACS~J0717.5+374. These two sources might be backsplash galaxies that have already experienced the processing by the cluster environment and are thus depleted in CO, similar to what was found in simulations by \citet{Bahe2013}.} 



By summing up the SFRs of the seven MACS~J0717.5+374 LIRGs {with projected cluster-centric distances $<r_{200}$} we derive lower limits to the total SFR of the cluster, ${\rm SFR}\gtrsim227~M_\odot/{\rm yr}$, and to the SFR normalized to the cluster mass, ${\rm SFR}/M_{200}\gtrsim8\times10^{-14}~{\rm yr}^{-1}$. 
The latter estimate makes MACS~J0717.5+3745 among the clusters at intermediate redshift with  the highest SFR, normalized to the cluster mass \citep[e.g.,][]{Finn2005,Geach2006,Haines2009b,Chung2010}

Similar to what was proposed by \citet{Chung2010} for the Bullet cluster and by \citet{Fadda2008} for Abell~1763 the significant normalized star formation activity of MACS~J0717.5+3745 might be explained if its LIRGs, or a fraction of them, belong to a population of infalling galaxies, possibly associated with the filament, that have not been quenched yet. 
If this is the case, the accretion timescale is not greater than a few hundred million years in order to be shorter than the depletion timescale $\tau_{\rm dep}\simeq(0.3-1.3)$~Gyr needed to quench the selected LIRGs of MACS~J0717.5+3745.

{For our environmental study we considered the location of the LIRGs in the cluster, parameterized by means of their $r/r_{200}$ (e.g., Figs.~\ref{fig:SFR_Mstar_vs_radius}, \ref{fig:phase_space_diagram}, \ref{fig:prop_vs_radius}).  While it would have been more appropriate to parameterize the effect of the environment by means of the local density \citep[e.g.,][]{Dressler1980,Peng2010}, to account for cluster substructures, this analysis requires a complete spectroscopic census of the cluster members, which is however very difficult to obtain, from the core out to the periphery of distant clusters.}

\section{Conclusions}\label{sec:conclusions}

We investigated the role of dense Mpc-scale environments in processing the molecular gas in galaxies as part of a larger search for CO in distant cluster galaxies.  To this aim we considered a sample of 17 cluster galaxies at intermediate redshifts $z\sim0.2-0.5$ with available FIR-to-UV photometry, as well {as high-resolution images from} Subaru and {\it HST}. The galaxies were selected for having total IR luminosities $\gtrsim10^{11}~L_\odot$, which classify the sources as LIRGs. The sources belong to well-studied clusters (Abell~697, 963, 1763, 2219, and MACS~J0717.5+3745) from the LoCuSS and HLS surveys.
The LIRGs also span a broad range in cluster centric distances, out to $\sim1.6\times r_{200}$, and are therefore an optimal sample to study the preprocessing of gas as the galaxies in the outskirts of clusters fall into their cores.

We observed the LIRGs in CO(1$\rightarrow$0) or CO(2$\rightarrow$1)  with several observational programs carried out between 2012 and 2017 with the IRAM PdBI and its successor NOEMA. The sample of 17 LIRGs has been complemented with 4 additional cluster LIRGs, 3 of which have already been observed in CO by \citet{Jablonka2013}, and the fourth by \citet{Cybulski2016}.  Accurate multiwavelength SED modeling from the FIR to UV was performed for all 21 LIRGs, which has allowed a homogeneous analysis of the full LIRG sample. 

We compared the SFRs, molecular gas to stellar mass ratios $M(H_2)/M_\star$, and  depletion times of the LIRGs with, first, those estimated using empirical relations for MS field galaxies and, second, those of a compilation of galaxies from the literature, out to $z\simeq1.6$. Our analysis suggests that the targeted LIRGs have {SFRs, molecular gas contents, and depletion times} that are consistent with those of both MS field galaxies and star-forming galaxies from the comparison sample, independent of the stellar mass considered, within $\log(M_\star/M_\odot)\simeq10-11$.

However, a 2.8$\sigma$ correlation between the depletion time, normalized to the value at the MS, and the projected cluster-centric distance, in units of $r_{200}$, is found for the sample of LIRGs. The correlation is ultimately due to an increase of SFR from a few $M_\odot/{\rm yr}$ in the cluster outskirts up to $\sim50~M_\odot/{\rm yr}$ in the inner regions $r\lesssim0.4\,r_{200}$ of the clusters. 
{Indeed, a large fraction of our cluster LIRGs, 7 out of the 16 LIRGs in our sample with projected cluster-centric distances $<r_{200}$, that is, $44\%\pm12\%$,  have enhanced ${\rm SFR}>3\times{\rm SFR}_{\rm MS}$.}
On the other hand, the ratio $M(H_2)/M_\star\simeq(0.2-1)$ scatters around the MS value with no observed trend as a function of the cluster-centric distance, {neither of the stellar mass.}

We discussed possible scenarios to explain the presence of significant large reservoirs of molecular gas in the LIRGs from the outskirts down to the cluster cores as well as the rising of the star formation with decreasing cluster-centric distance.
{The presence of enhanced star formation activity observed in a large fraction of the LIRGs within the virial radius suggests that environmental processing mechanisms such as ram-pressure stripping, galaxy harassment, and starvation may have not been sufficiently effective in suppressing the star formation and then quenching the LIRGs, consistently with recent results found in simulations \citep{Hausammann2019}. On the other hand, this work shows that
the enhancement of star formation we observe in the LIRGs toward the cluster cores, together with the decrease of the depletion time, implies that the molecular gas reservoirs feeding the star formation exhaust rapidly. {Therefore,  star formation is rapidly suppressed in the LIRGs, which are ultimately quenched.}}
{The rapid consumption of molecular gas observed in the cluster LIRGs may also explain the fast exponential decrease with time, since $z\sim0.8$, of the fraction of LIRGs in clusters, already found in previous studies  \citep{Finn2010,Saintonge2008,Popesso2012}. }



We separately discussed the most distant cluster in our sample, MACS~J0717.5+3745 at $z=0.546$, {since $\sim50\%$ of the LIRGs in our sample belong to this cluster.}
{ We did not find any statistically significant difference concerning the molecular gas content of the MACS~J0717.5+374 LIRGs from the cluster inner regions out to the outskirts, in the southeast filament. This suggests that the dense and cold molecular gas is not strongly preprocessed by the large-scale dense environment (cluster and filament). However, the  MACS~J0717.5+374 LIRGs contribute to the observed increase of $\tau_{\rm dep}$, normalized to its MS, with increasing cluster-centric distances.}

To explain the low depletion timescales and high SFRs observed in some MACS~J0717.5+374 LIRGs, we suggest a possible scenario in which a large-scale shock in MACS~J0717.5+374 has compressed the gas reservoirs in some LIRGs along the filament direction, thus ultimately enhancing their star formation activity. {Similarly, the high SFR of  A2219-1 may be due to shock-induced gas compression. This scenario is consistent with previous studies, which found increased SFR associated with shock fronts \citep{Rawle2014,Stroe2014}.}


We also discussed the other possible scenario in which the LIRGs of MACS~J0717.5+3745, or a fraction of these sources, belong to a population of infalling galaxies. If this is the case, the accretion timescale is not greater than a few hundred million years in order to be shorter than the depletion timescale $\tau_{\rm dep}\simeq(0.3-1.3)$~Gyr needed to quench the selected LIRGs.

{ Larger samples of cluster LIRGs with observations in CO, including those at higher spatial resolution and/or at higher J transitions, to probe denser molecular gas, will help to provide further insights into the physical processes involved in processing molecular gas of the LIRGs, as they fall into the cluster cores. }

\begin{acknowledgements}
{ We thank the anonymous referee for helpful comments which contributed to improve the paper.}
This work is based on observations carried out with the IRAM NOEMA Interferometer. IRAM is supported by INSU/CNRS (France), MPG (Germany) and IGN (Spain).
GC acknowledges financial support from the Swiss National Science Foundation (SNSF).
MJ is supported by the United Kingdom Research and Innovation (UKRI) Future Leaders Fellowship 'Using Cosmic Beasts to uncover the Nature of Dark Matter' [grant number MR/S017216/1] and by the Science and Technology Facilities Council [grant number ST/L00075X/1]. FB acknowledges funding from the European Research Council (ERC) under the European Union’s Horizon 2020 research and innovation program (grant agreement No. 726384).\end{acknowledgements}

\begin{appendix}
\section{NOEMA detections and images}\label{app:NOEMAdetections_images}
We report the images of the targeted LIRGs and the associated
NOEMA detections below.

\begin{figure*}[!ht]\centering
\captionsetup[subfigure]{labelformat=empty}
\subfloat[]{\includegraphics[trim={0.cm 0.cm 0.cm 0cm},clip,width=0.45\textwidth,clip=true]{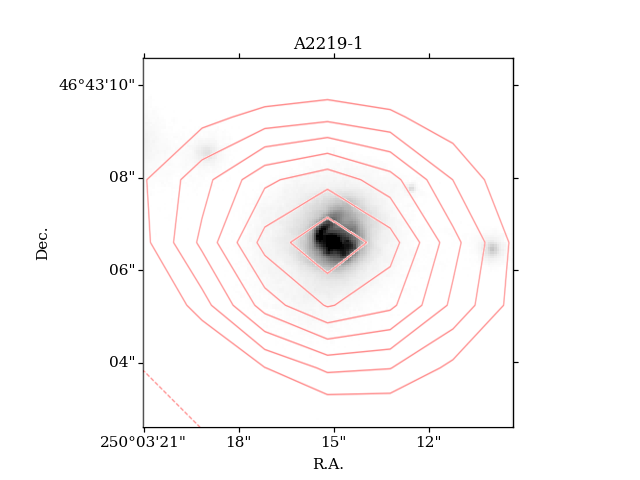}}
\subfloat[]{\includegraphics[trim={0cm 0cm 0cm 0cm},clip,width=0.45\textwidth,clip=true]{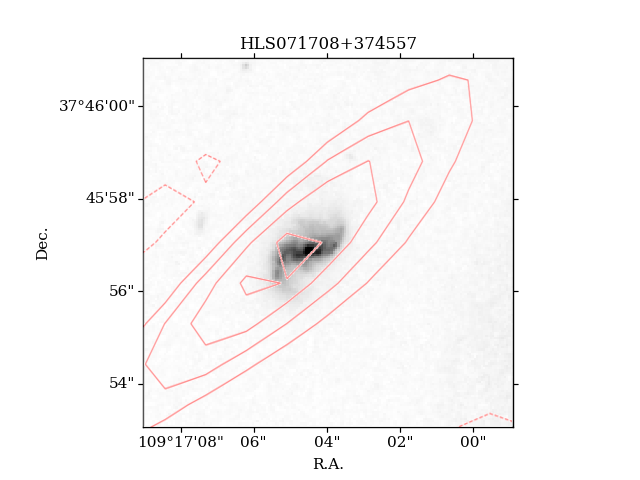}}\\
\subfloat[]{\includegraphics[trim={0.cm 0.cm 0.cm 0cm},clip,width=0.45\textwidth,clip=true]{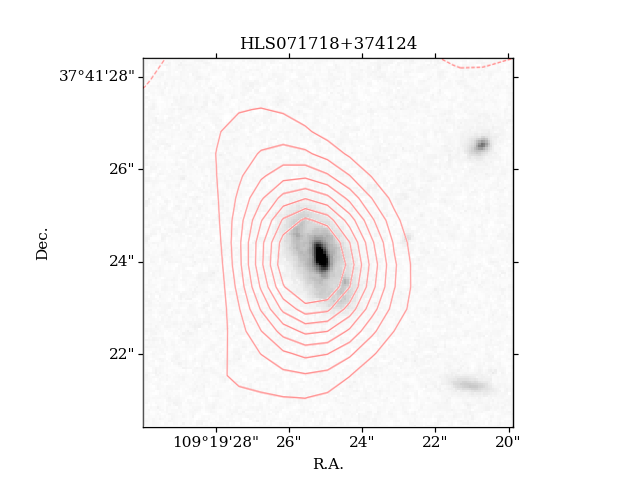}}
\subfloat[]{\includegraphics[trim={0cm 0cm 0cm 0cm},clip,width=0.45\textwidth,clip=true]{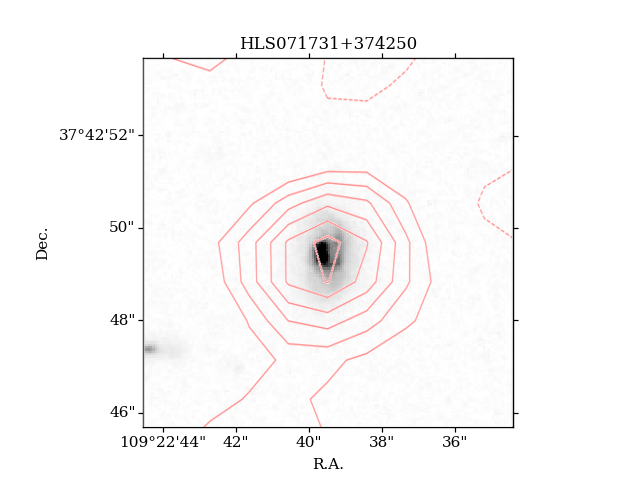}}\\
\subfloat[]{\includegraphics[trim={0.cm 0.cm 0.cm 0cm},clip,width=0.45\textwidth,clip=true]{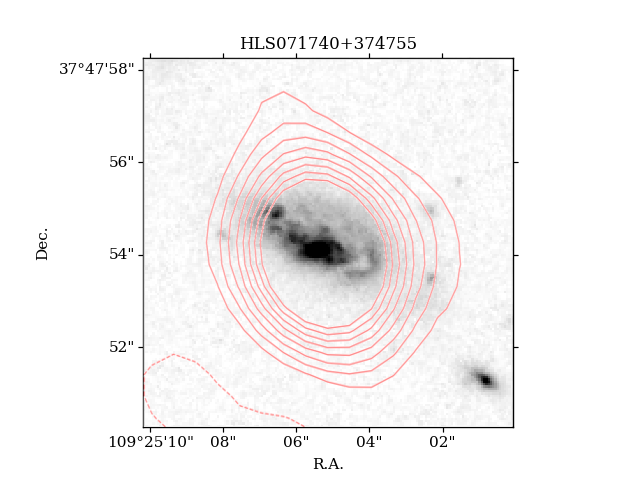}}
\subfloat[]{\includegraphics[trim={0cm 0cm 0cm 0cm},clip,width=0.45\textwidth,clip=true]{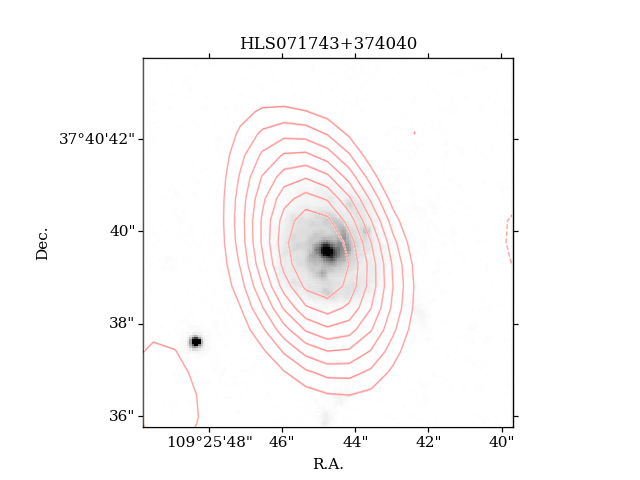}}\\
\caption{{\it HST}-ACS images ($8''\times8''$ size) centered at the coordinates of some of our targets, as shown at the top of each image. North is up, east is left. As
in Fig.~\ref{fig:NOEMAdetections}, NOEMA contours  are shown in red, {in case of CO detections}. The archival {\it HST} image of A2219-1 is taken with the F850LP filter, while those of HLS sources are taken with the F818W filter. Small shifts $\leq0.6$~arcsec are applied to the images, consistent with GCS catalog uncertainties in the absolute astrometry used by {\it HST}.}\label{fig:HSTimages}
\end{figure*}
\begin{figure*} \centering
\captionsetup[subfigure]{labelformat=empty}
\ContinuedFloat
\subfloat[]{\includegraphics[trim={0.cm 0.cm 0.cm 0cm},clip,width=0.45\textwidth,clip=true]{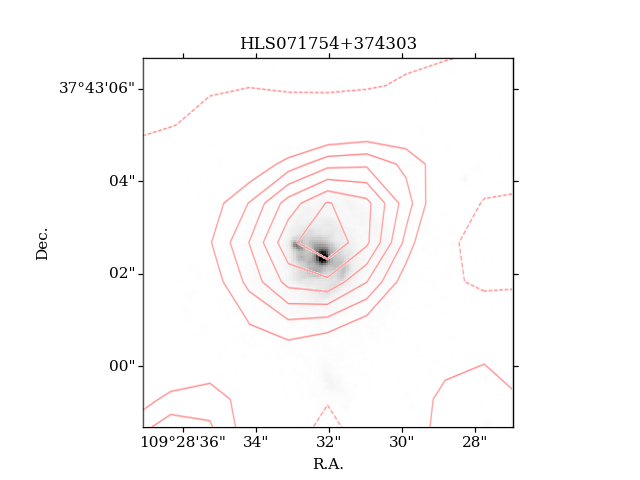}}
\subfloat[]{\includegraphics[trim={0cm 0cm 0cm 0cm},clip,width=0.45\textwidth,clip=true]{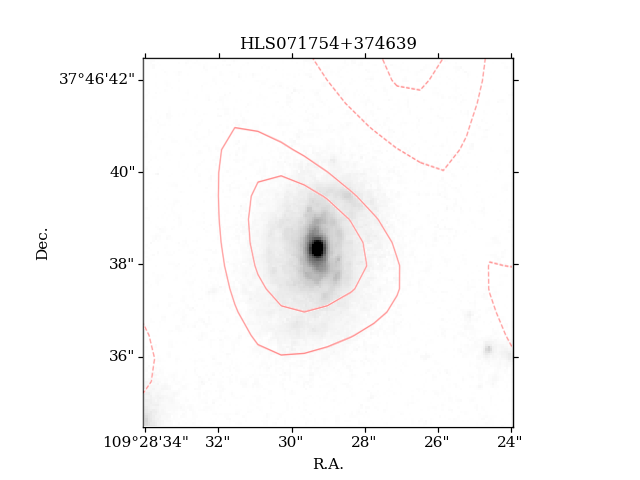}}\\
\subfloat[]{\includegraphics[trim={0.cm 0.cm 0.cm 0cm},clip,width=0.45\textwidth,clip=true]{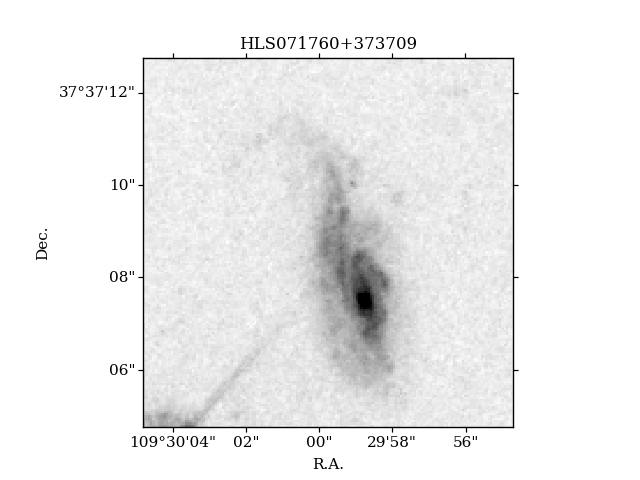}}
\subfloat[]{\includegraphics[trim={0cm 0cm 0cm 0cm},clip,width=0.45\textwidth,clip=true]{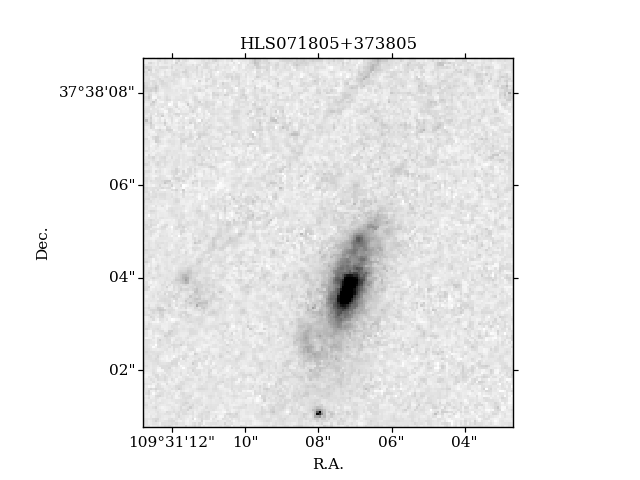}}\\
\subfloat[]{\includegraphics[trim={0.cm 0.cm 0.cm 0cm},clip,width=0.45\textwidth,clip=true]{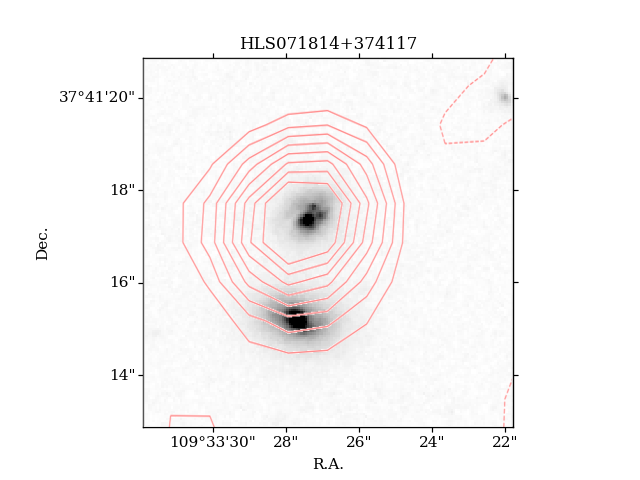}}\\
\caption{Continued.}
\end{figure*}

\begin{figure*}\centering
\captionsetup[subfigure]{labelformat=empty}
\subfloat[]{\includegraphics[trim={0.cm 0.cm 0.cm 0cm},clip,width=0.45\textwidth,clip=true]{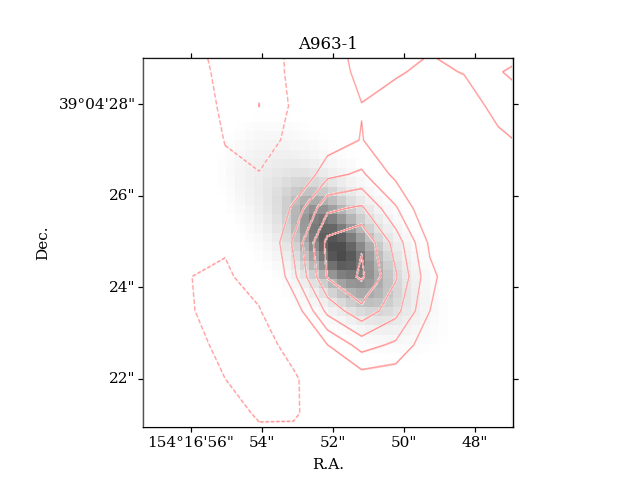}}
\subfloat[]{\includegraphics[trim={0cm 0cm 0cm 0cm},clip,width=0.45\textwidth,clip=true]{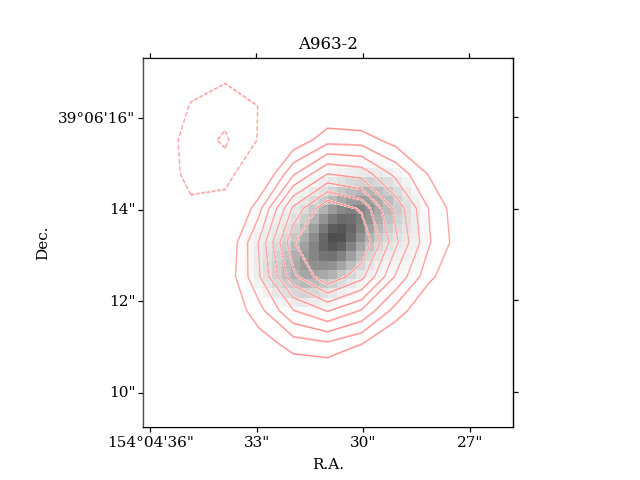}}\\
\subfloat[]{\includegraphics[trim={0cm 0cm 0cm 0cm},clip,width=0.45\textwidth,clip=true]{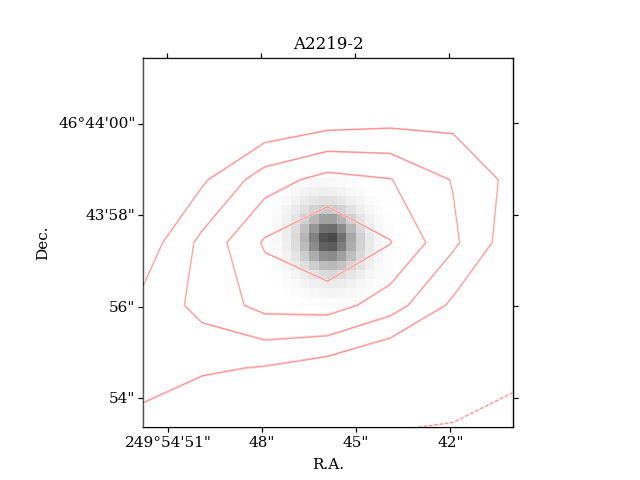}}
\subfloat[]{\includegraphics[trim={0cm 0cm 0cm 0cm},clip,width=0.45\textwidth,clip=true]{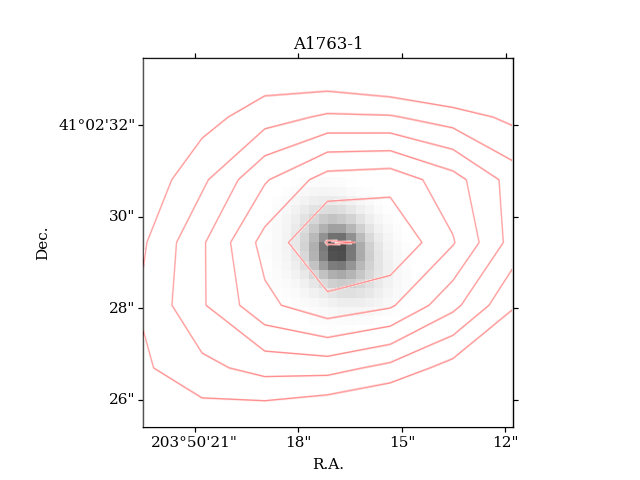}} \\
\subfloat[]{\includegraphics[trim={0.cm 0.cm 0.cm 0cm},clip,width=0.45\textwidth,clip=true]{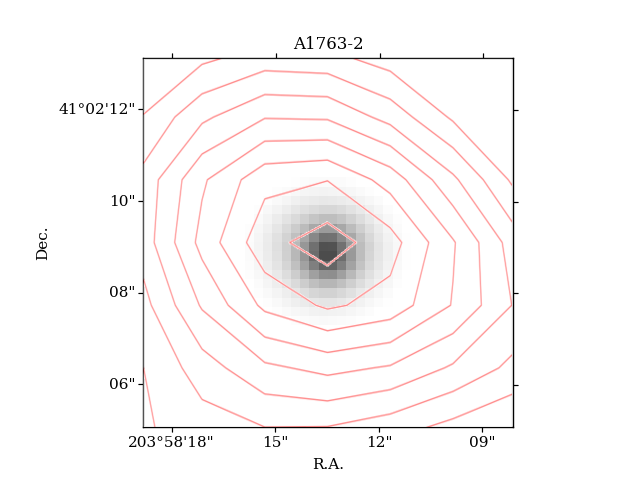}} 
\subfloat[]{\includegraphics[trim={0.cm 0.cm 0.cm 0cm},clip,width=0.45\textwidth,clip=true]{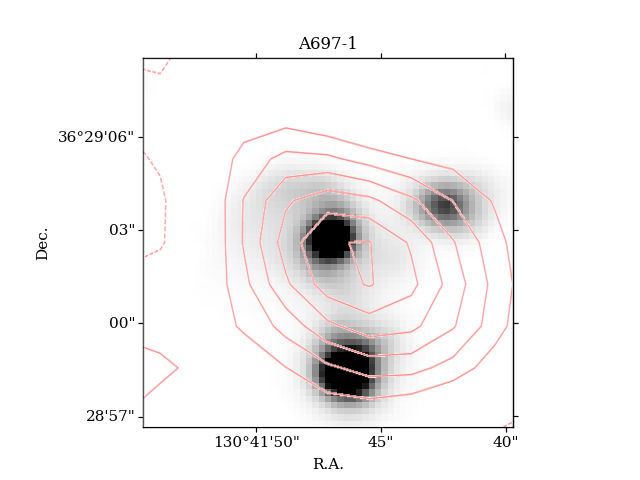}} \\
\caption{Subaru images ($8''\times8''$ size), {with the exception of A697-1 ($12''\times12''$ size),} centered at the coordinates of some of our targets, as shown at the top of each image. The \textsf{i}$^+$-band is considered for all sources except for A963-1 and A963-1, for which the \textsf{I$_{\textsf c}$}-filter is used.
The sources do not have available {\it HST} observations and are therefore not reported in Fig.~\ref{fig:HSTimages}.
North is up, east is left. As in Fig.~\ref{fig:NOEMAdetections},
NOEMA contours are shown in red. Small shifts $\leq0.34$~arcsec are applied to the images, consistent with the absolute Subaru astrometric uncertainties.}\label{fig:SUBARUimages}
\end{figure*}

\begin{figure*}[]\centering
\captionsetup[subfigure]{labelformat=empty}
\subfloat[A963-1]{\includegraphics[trim={0.5cm -0.5cm 0.5cm 1.1cm},clip,
page=13,width=0.5\textwidth,clip=true]{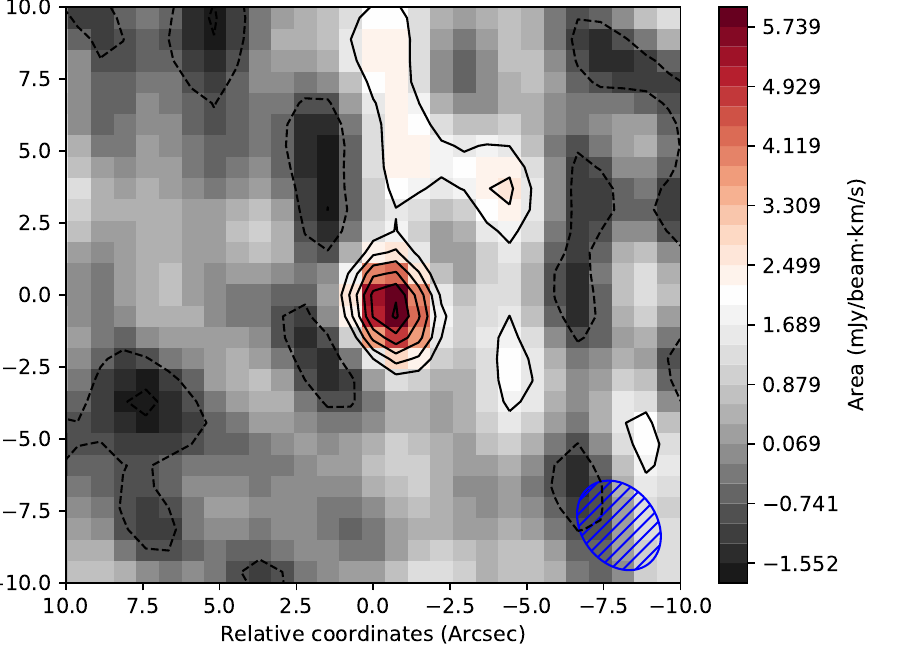}}
\subfloat[A963-1]{\includegraphics[trim={0cm 0cm 0cm 1.5cm},clip,
page=14,width=0.6\textwidth,clip=true]{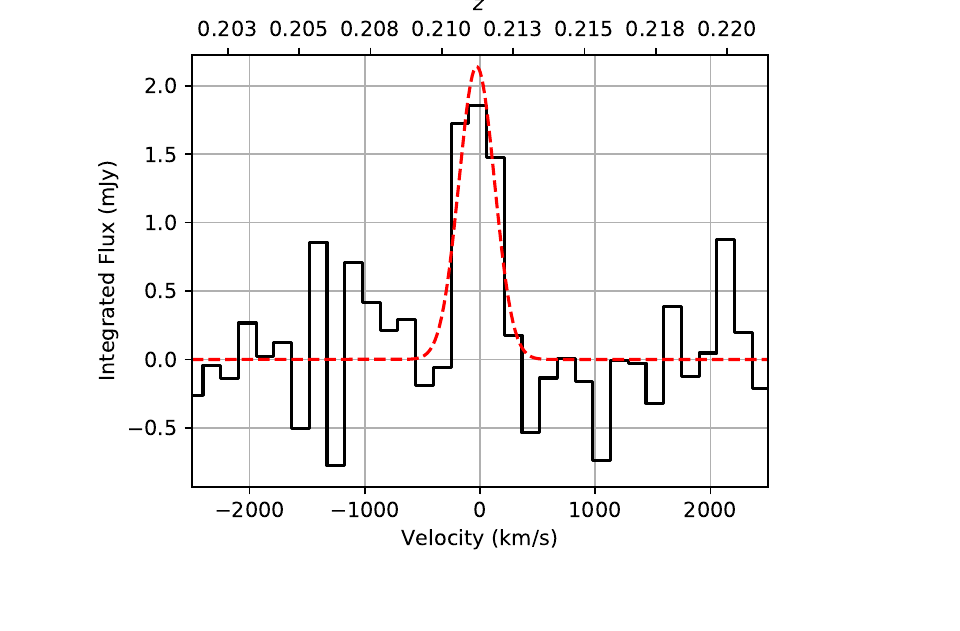}}\\
\subfloat[A963-2]{\includegraphics[trim={0.5cm -0.5cm 0.5cm 1.1cm},clip,
page=15,width=0.5\textwidth,clip=true]{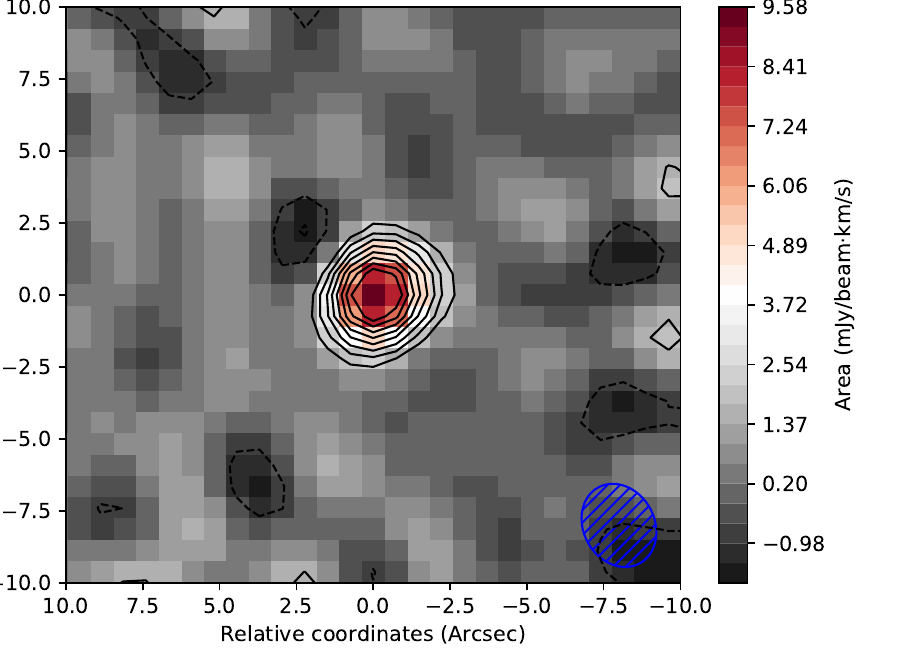}}
\subfloat[A963-2]{\includegraphics[trim={0cm 0cm 0cm 1.5cm},clip,
page=16,width=0.6\textwidth,clip=true]{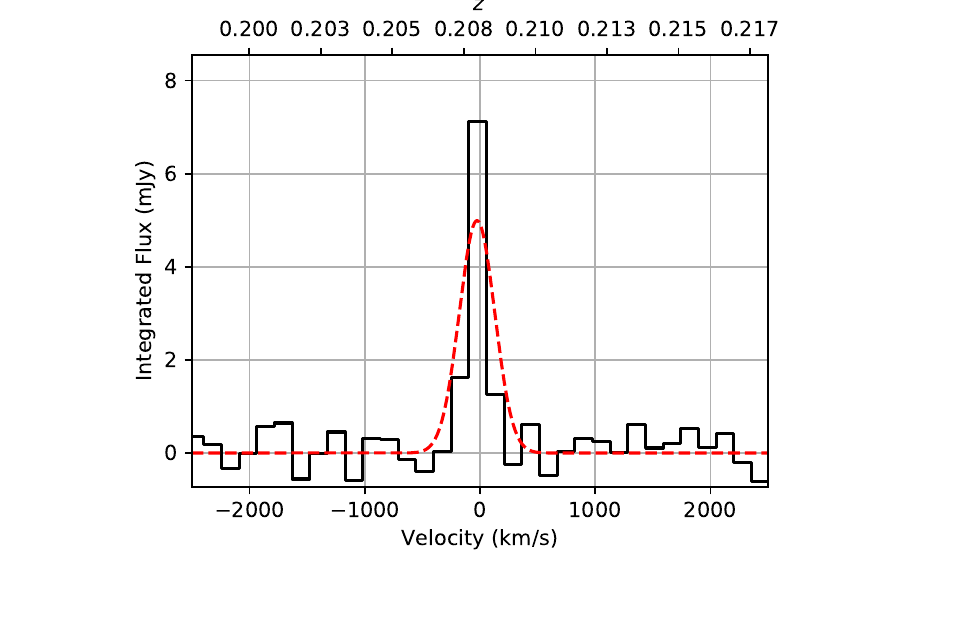}}\\
\caption{Left: clean intensity maps showing the CO detections obtained with NOEMA. Coordinates are reported as angular separations from the target sources. For each map, the velocity range considered corresponds to the velocity support associated with the corresponding CO emission line, see right panels. The solid and dashed contour levels
are superimposed and correspond to positive and negative fluxes, starting from +2$\sigma$ and -1$\sigma$ levels, respectively. Consecutive levels correspond to an absolute increment of 1$\sigma$ in significance.  The dashed ellipses (bottom right) show the beam size.  Right: spectrum (black solid line) obtained with NOEMA within an aperture corresponding to the beam size at the location of
the detection, as reported in the left panels. The best fits, baseline subtracted, are reported
(red dashed line).}
\label{fig:NOEMAdetections}
\end{figure*}
\begin{figure*} \centering
\captionsetup[subfigure]{labelformat=empty}
\ContinuedFloat
\subfloat[A2219-1]{\includegraphics[trim={0.5cm -0.5cm 0.5cm 1.1cm},clip,
page=21,width=0.5\textwidth,clip=true]{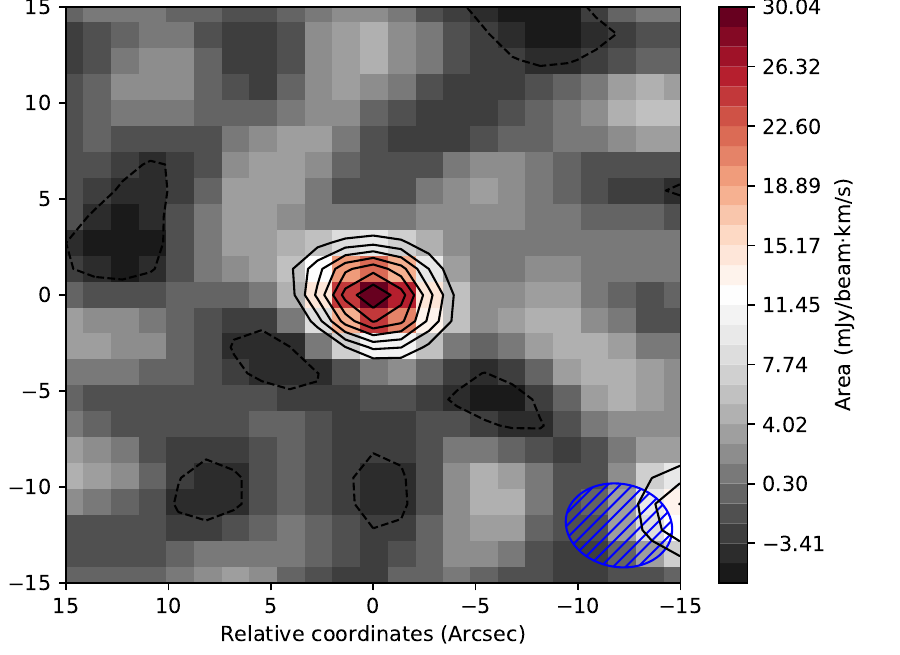}}
\subfloat[A2219-1]{\includegraphics[trim={0cm 0cm 0cm 1.5cm},clip,
page=22,width=0.6\textwidth,clip=true]{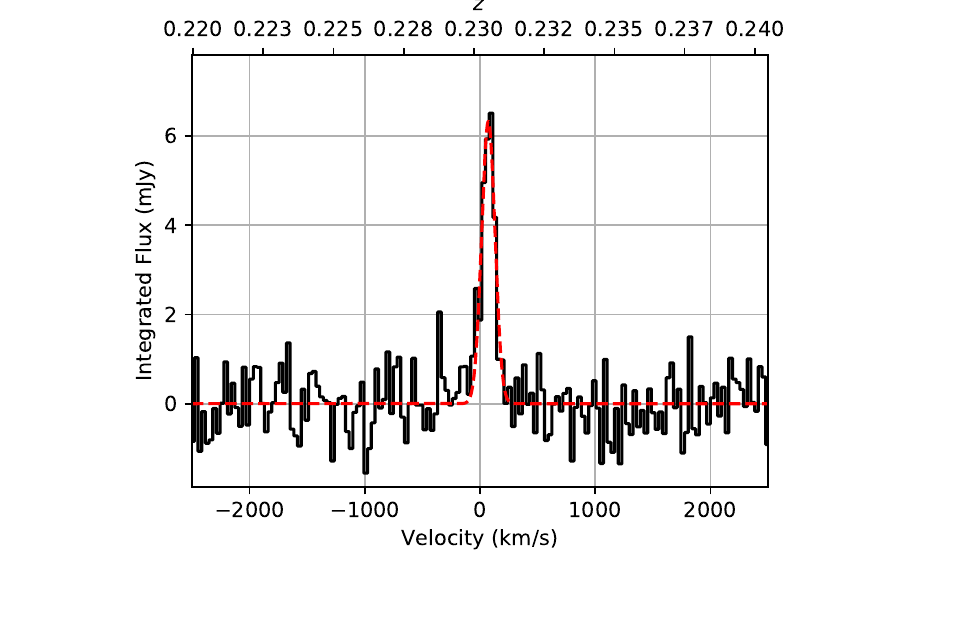}}\\
\subfloat[A2219-2]{\includegraphics[trim={0.5cm -0.5cm 0.5cm 1.1cm},clip,
page=23,width=0.5\textwidth,clip=true]{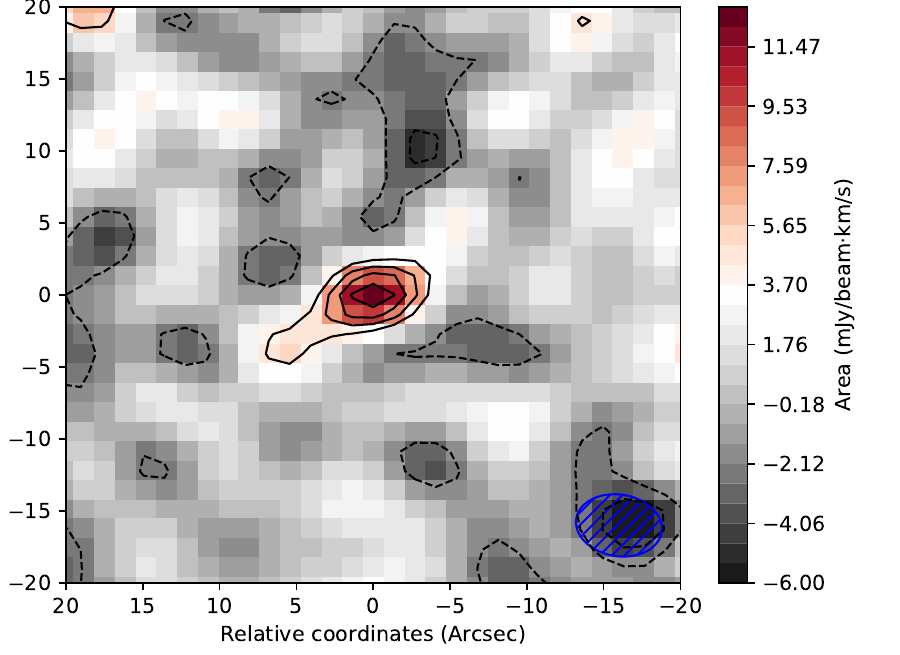}}
\subfloat[A2219-2]{\includegraphics[trim={0cm 0cm 0cm 1.5cm},clip,
page=24,width=0.6\textwidth,clip=true]{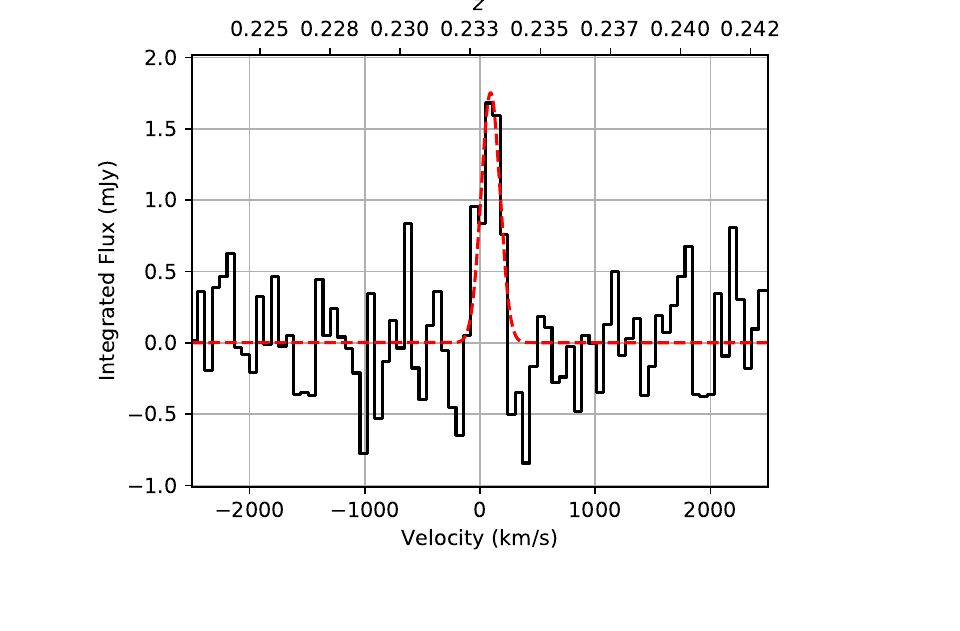}}\\
\subfloat[A1763-1]{\includegraphics[trim={0.5cm -0.5cm 0.5cm 1.1cm},clip,page=25,width=0.5\textwidth,clip=true]{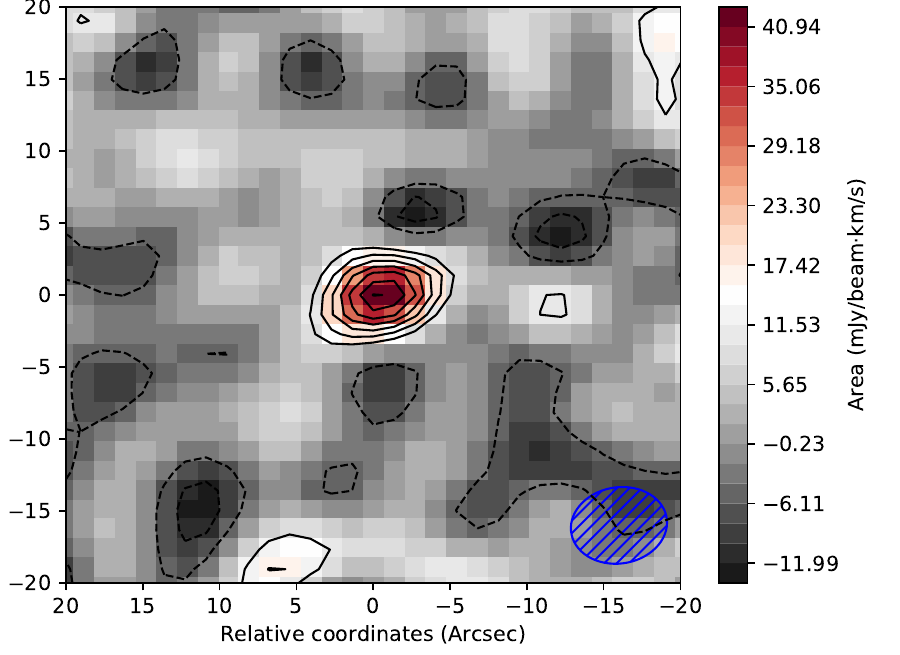}}
\subfloat[A1763-1]{\includegraphics[trim={0cm 0cm 0cm 1.5cm},clip,
page=26,width=0.6\textwidth,clip=true]{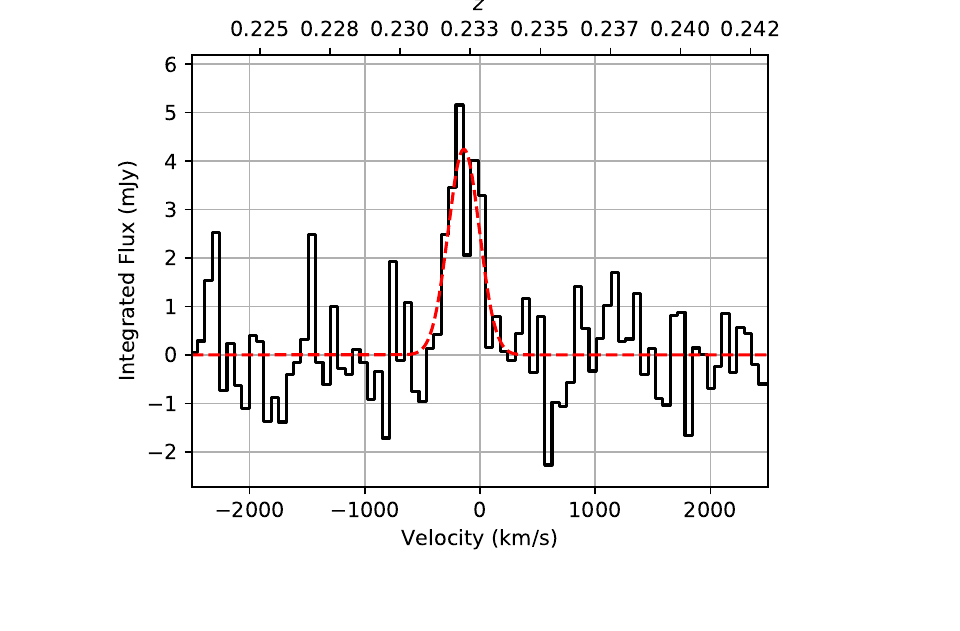}}\\
\caption{Continued.}
\end{figure*}
\begin{figure*} \centering
\captionsetup[subfigure]{labelformat=empty}
\ContinuedFloat
\subfloat[A1763-2]{\includegraphics[trim={0.5cm -0.5cm 0.5cm 1.1cm},clip,
page=27,width=0.5\textwidth,clip=true]{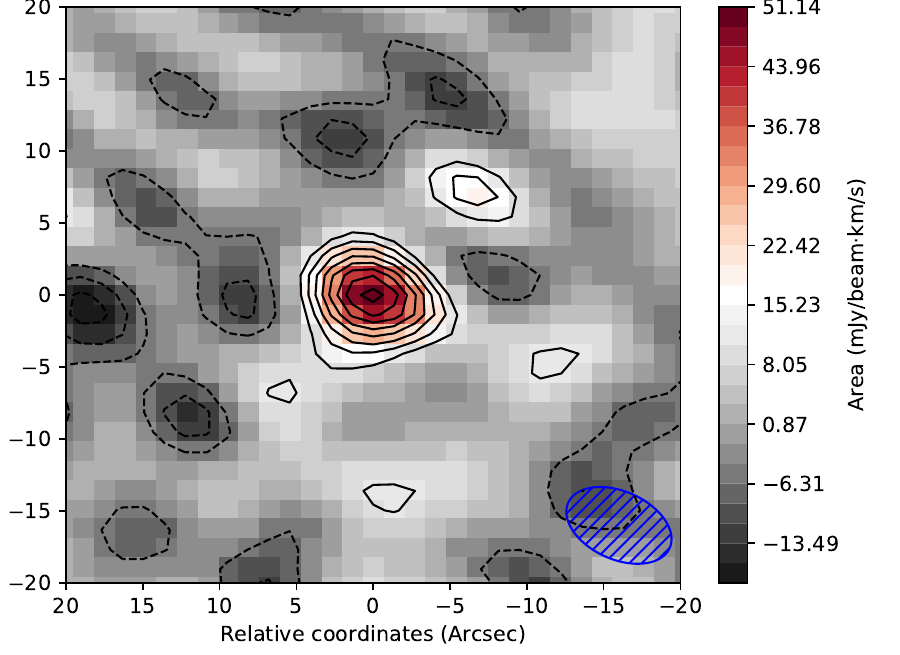}}
\subfloat[A1763-2]{\includegraphics[trim={0cm 0cm 0cm 1.5cm},clip,
page=28,width=0.6\textwidth,clip=true]{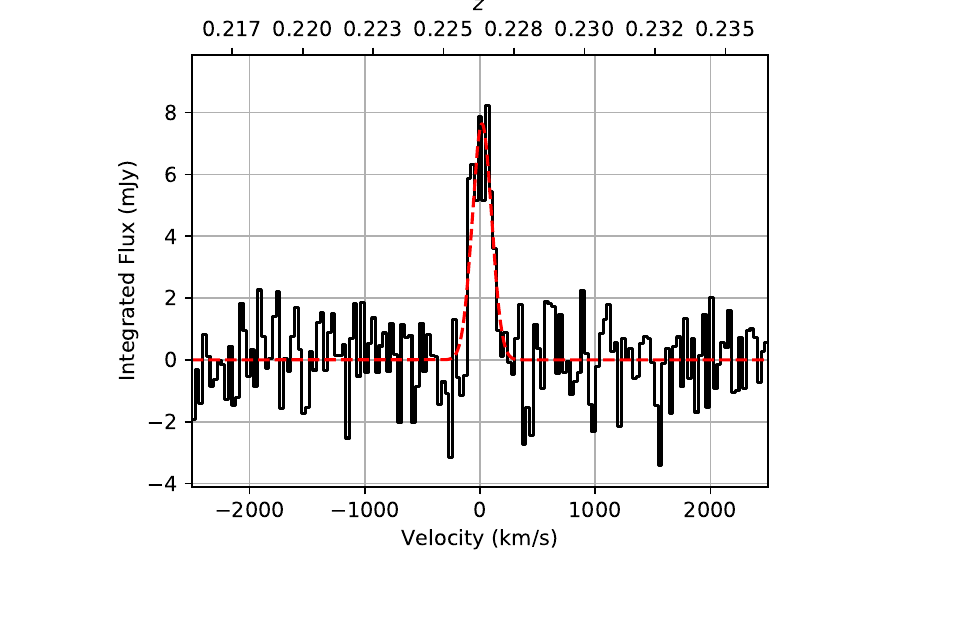}}\\
\subfloat[A697-1]{\includegraphics[trim={0.5cm -0.5cm 0.5cm 1.1cm},clip,
page=29,width=0.5\textwidth,clip=true]{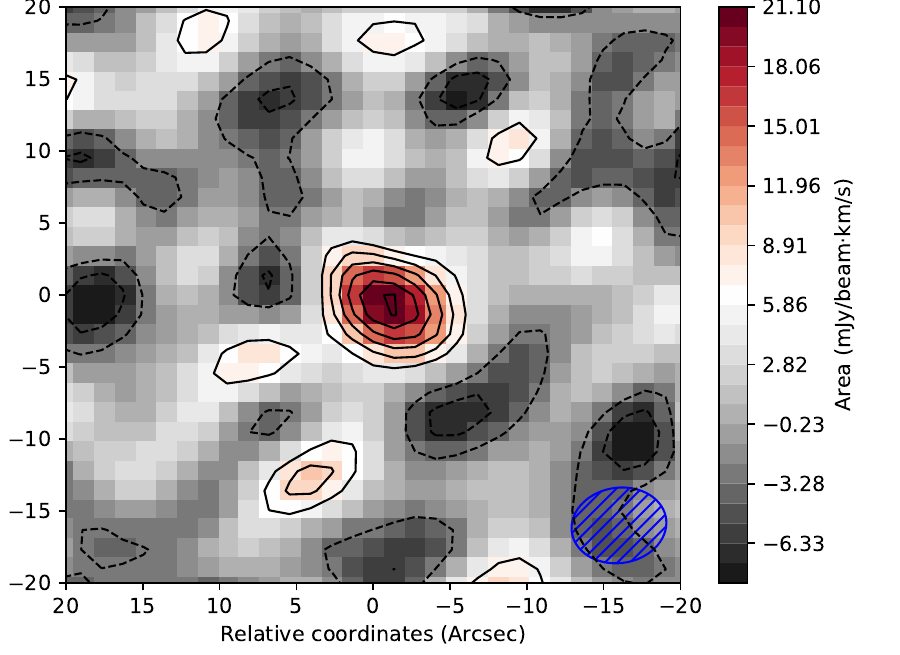}}
\subfloat[A697-1]{\includegraphics[trim={0cm 0cm 0cm 1.5cm},clip,
page=30,width=0.6\textwidth,clip=true]{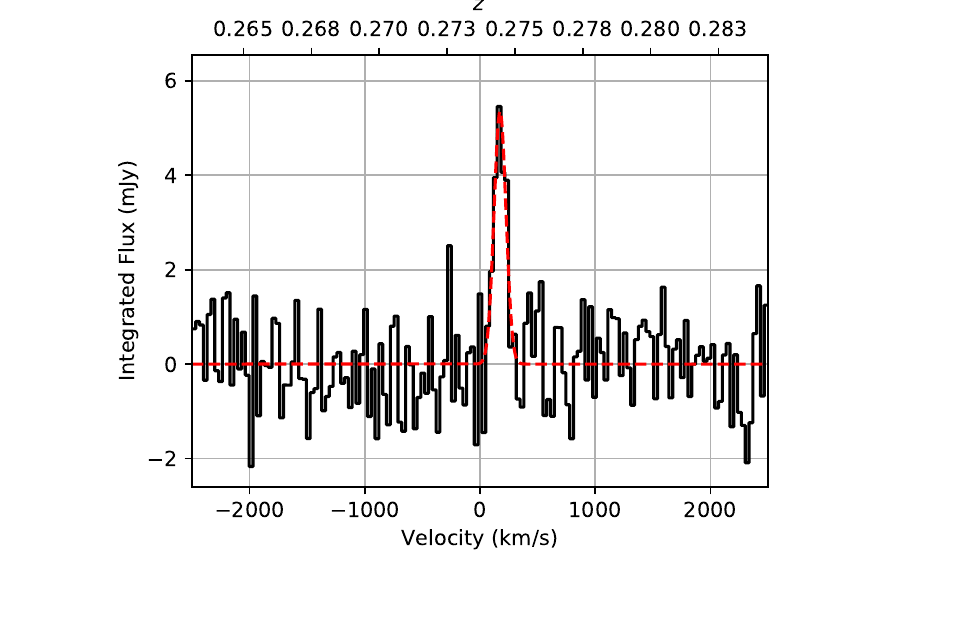}}\\
\subfloat[HLS071708+374557]{\includegraphics[trim={0.5cm -0.5cm 0.5cm 1.1cm},clip,
page=19,width=0.5\textwidth,clip=true]{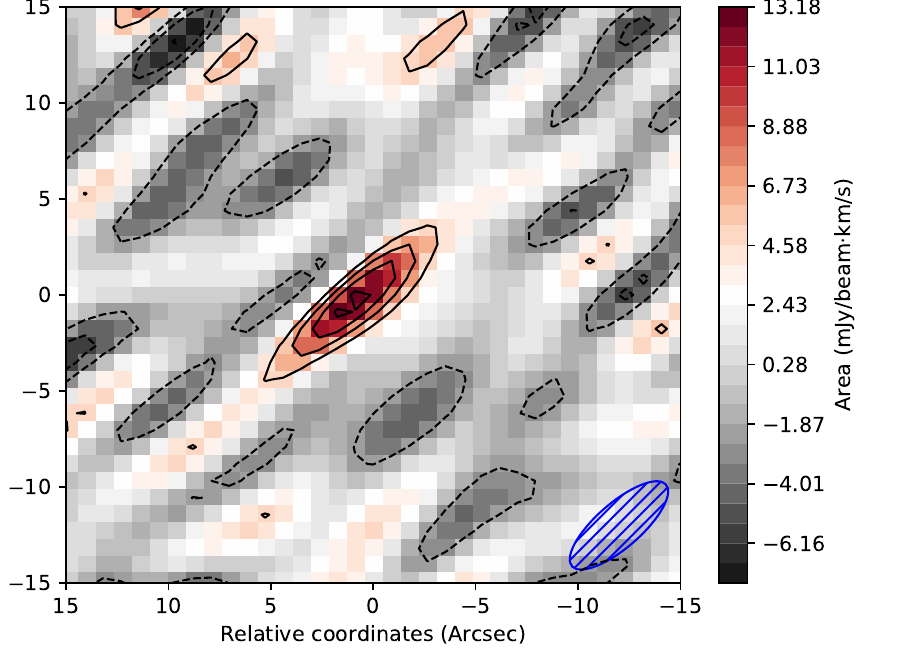}}
\subfloat[HLS071708+374557]{\includegraphics[trim={0cm 0cm 0cm 1.5cm},clip,
page=20,width=0.6\textwidth,clip=true]{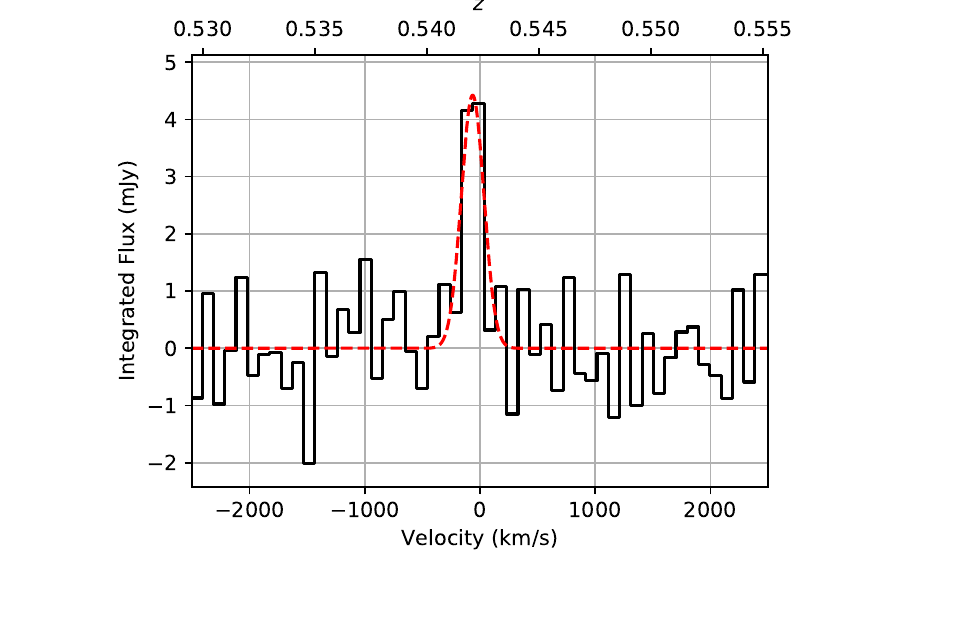}}\\
\caption{Continued.}
\end{figure*}
\begin{figure*} \centering
\captionsetup[subfigure]{labelformat=empty}
\ContinuedFloat
\subfloat[HLS071718+374124]{\includegraphics[trim={0.5cm -0.5cm 0.5cm 1.1cm},clip,
page=43,width=0.5\textwidth,clip=true]{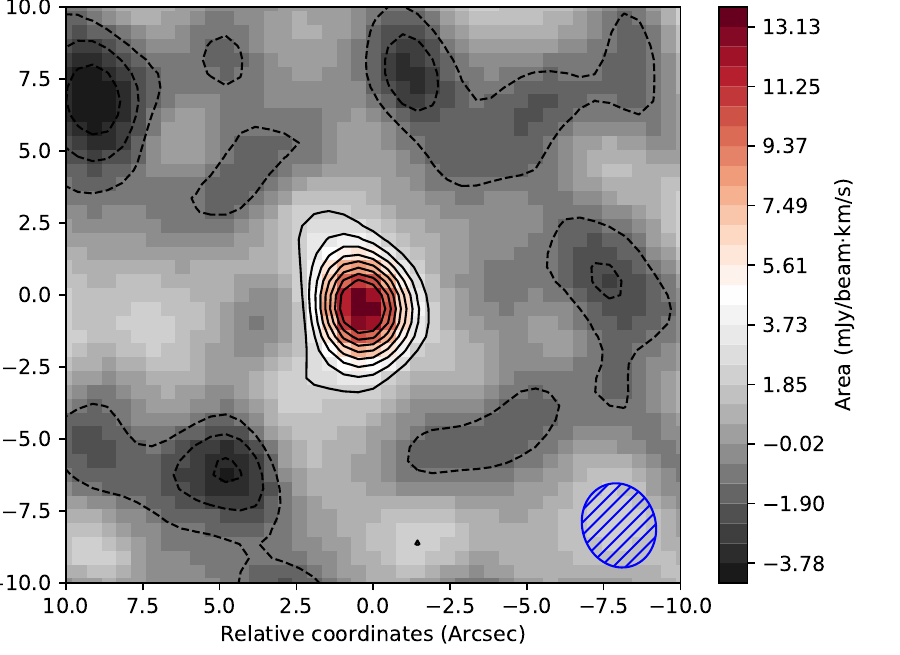}}
\subfloat[HLS071718+374124]{\includegraphics[trim={0cm 0cm 0cm 1.5cm},clip,
page=44,width=0.6\textwidth,clip=true]{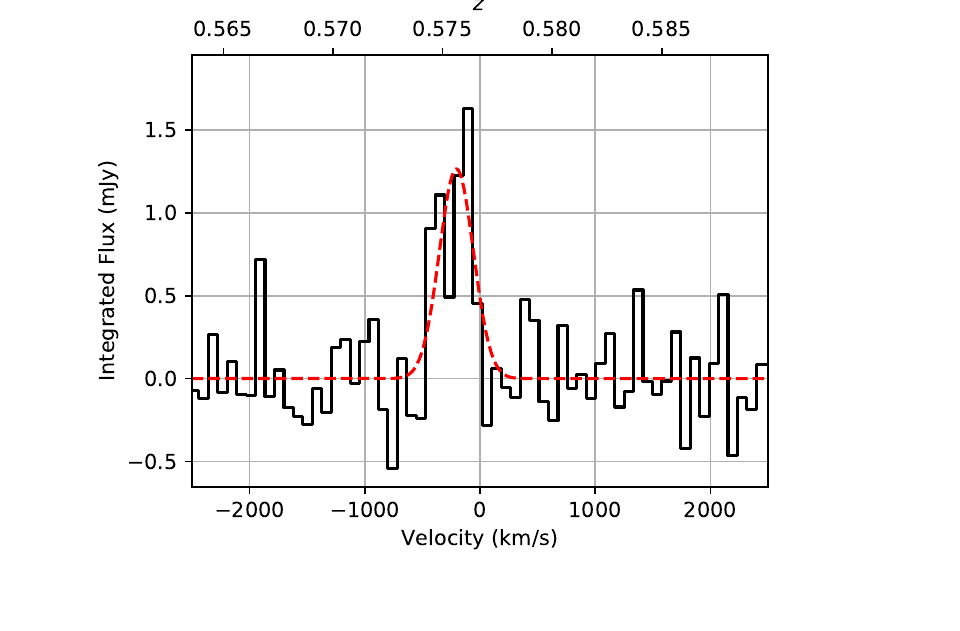}}\\
\subfloat[HLS071731+374250]{\includegraphics[trim={0.5cm -0.5cm 0.5cm 1.1cm},clip,
page=5,width=0.5\textwidth,clip=true]{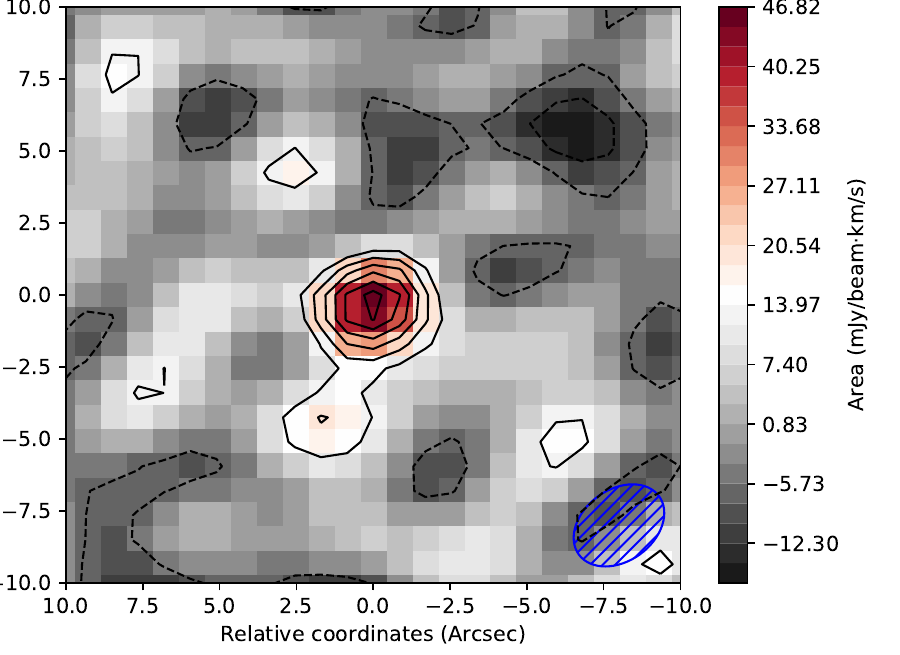}}
\subfloat[HLS071731+374250]{\includegraphics[trim={0cm 0cm 0cm 1.5cm},clip,
page=6,width=0.6\textwidth,clip=true]{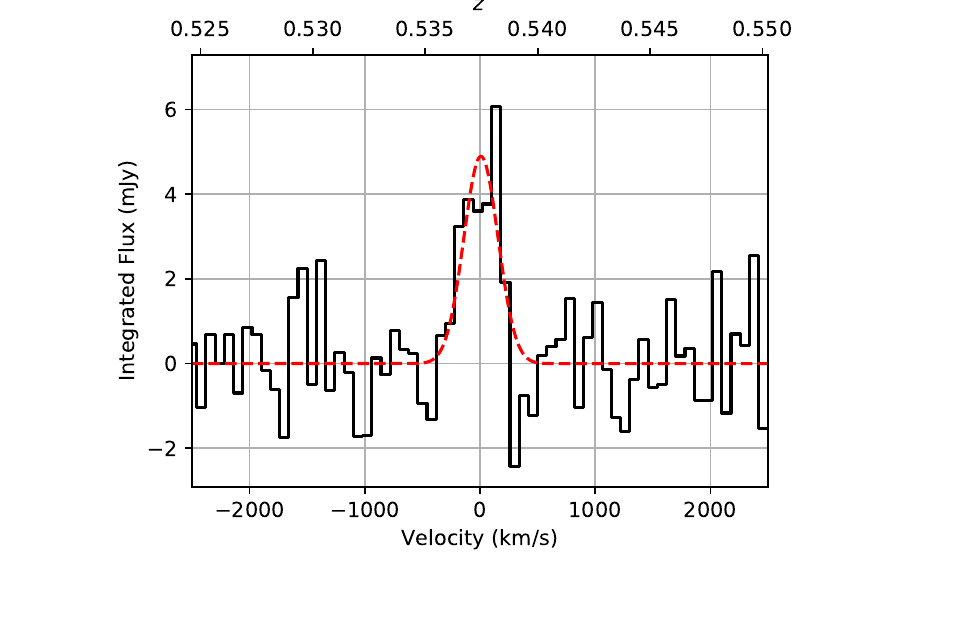}}\\
\subfloat[HLS071740+374755]{\includegraphics[trim={0.5cm -0.5cm 0.5cm 1.1cm},clip,
page=41,width=0.5\textwidth,clip=true]{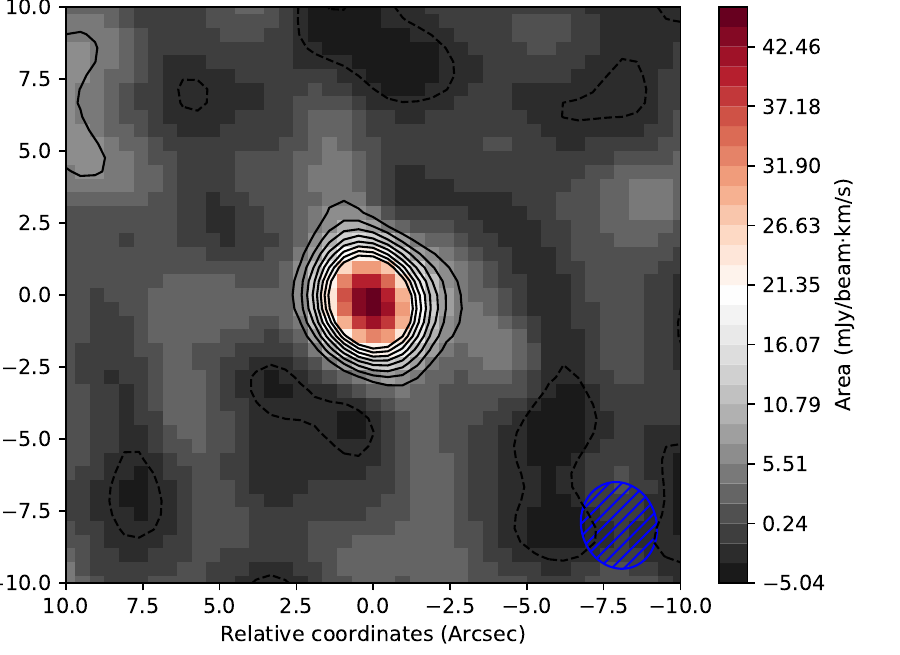}}
\subfloat[HLS071740+374755]{\includegraphics[trim={0cm 0cm 0cm 1.5cm},clip,
page=42,width=0.6\textwidth,clip=true]{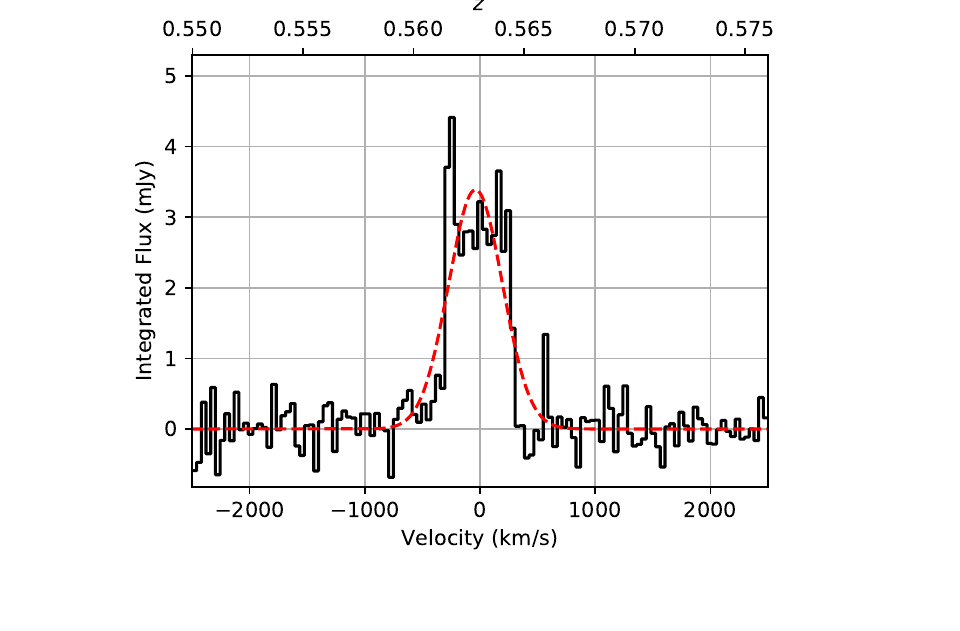}}\\
\caption{Continued.}
\end{figure*}
\begin{figure*} \centering
\captionsetup[subfigure]{labelformat=empty}
\ContinuedFloat
\subfloat[HLS071743+374040]{\includegraphics[trim={0.5cm -0.5cm 0.5cm 1.1cm},clip,
page=33,width=0.5\textwidth,clip=true]{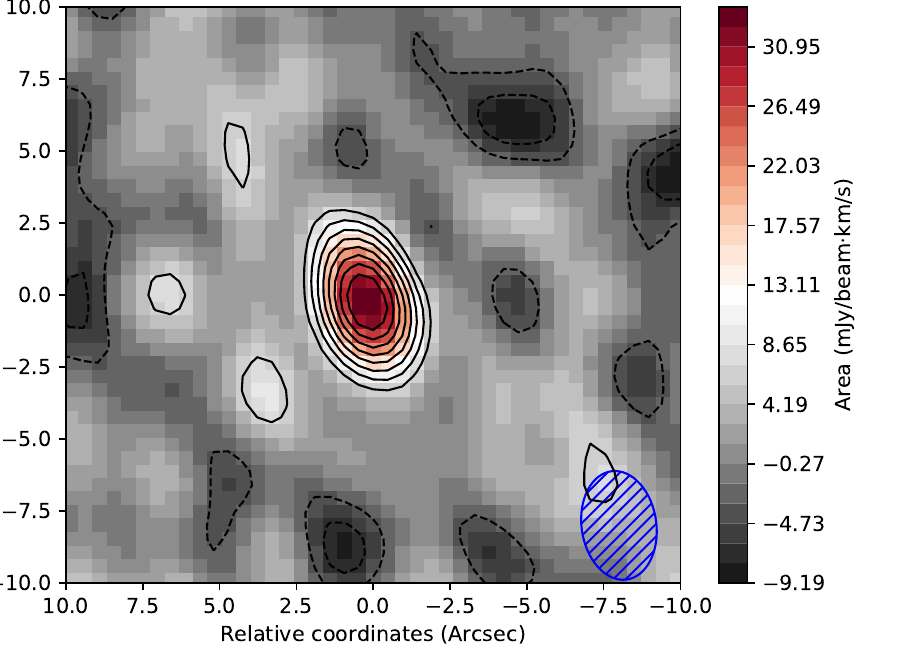}}
\subfloat[HLS071743+374040]{\includegraphics[trim={0cm 0cm 0cm 1.5cm},clip,
page=34,width=0.6\textwidth,clip=true]{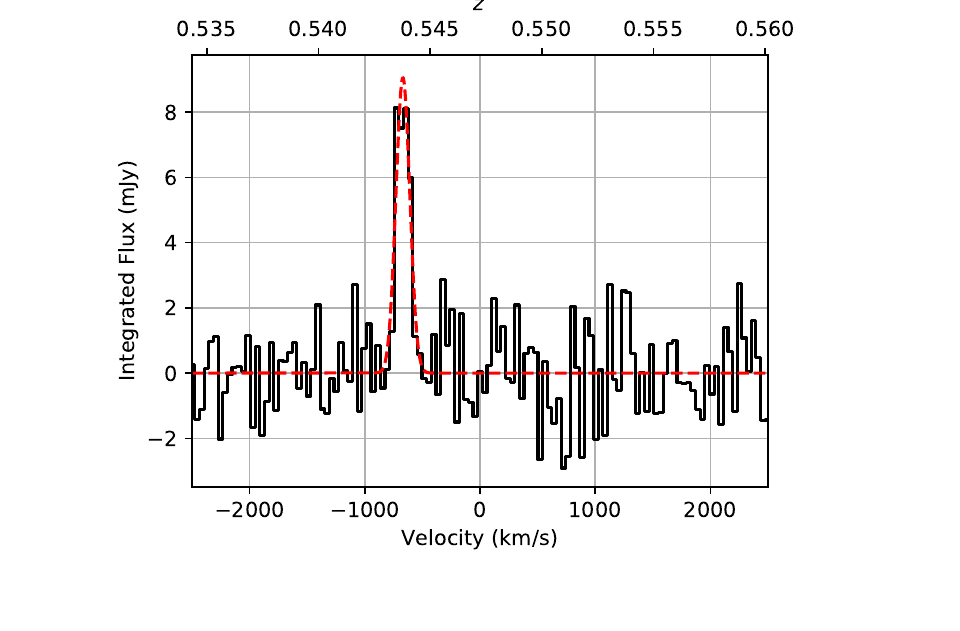}}\\
\subfloat[HLS071754+374303]{\includegraphics[trim={0.5cm -0.5cm 0.5cm 1.1cm},clip,
page=3,width=0.5\textwidth,clip=true]{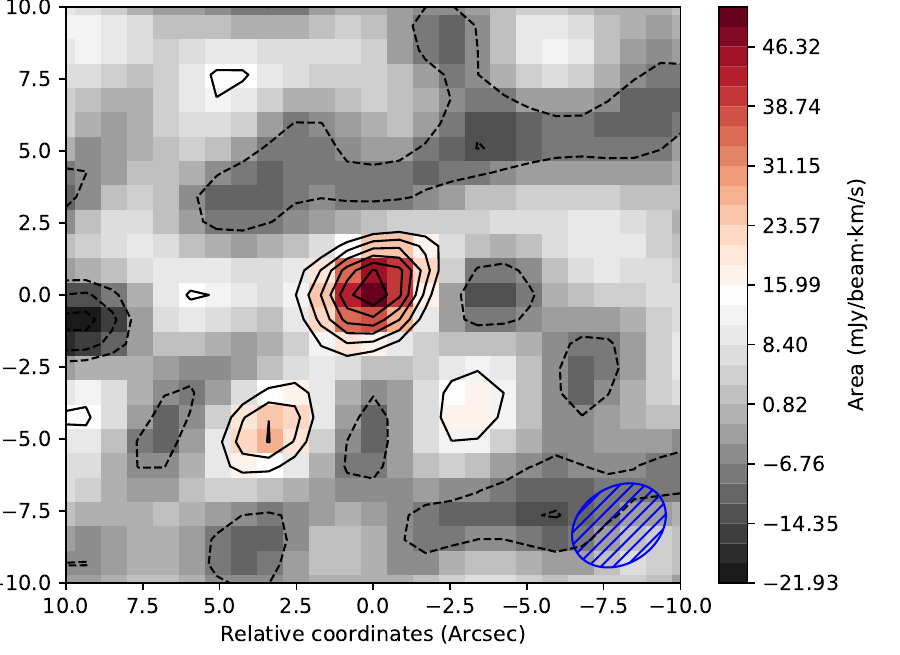}}
\subfloat[HLS071754+374303]{\includegraphics[trim={0cm 0cm 0cm 1.5cm},clip,
page=4,width=0.6\textwidth,clip=true]{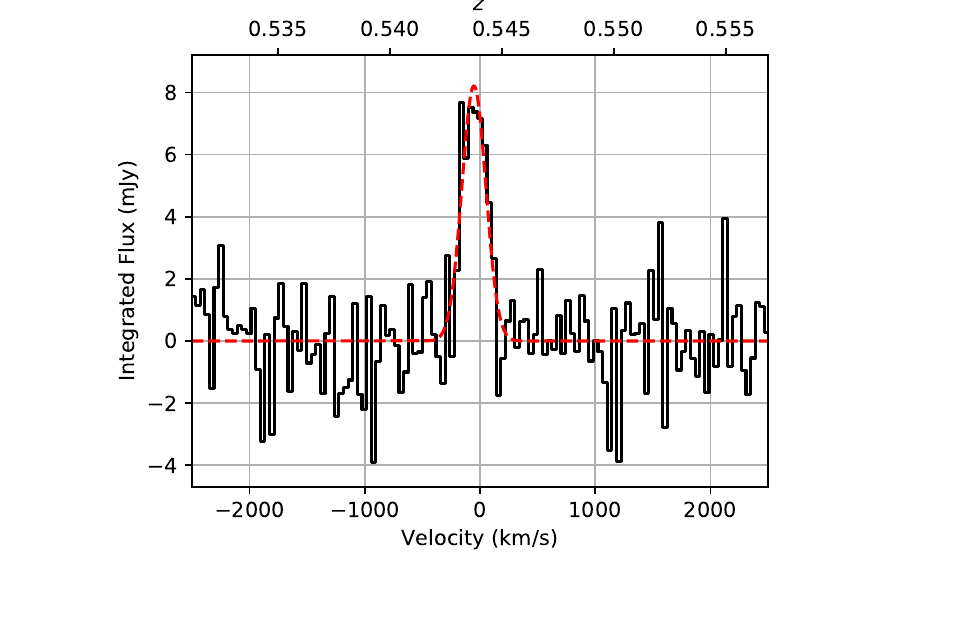}}\\
\subfloat[HLS071754+374639]{\includegraphics[trim={0.5cm -0.5cm 0.5cm 1.1cm},clip,
page=39,width=0.5\textwidth,clip=true]{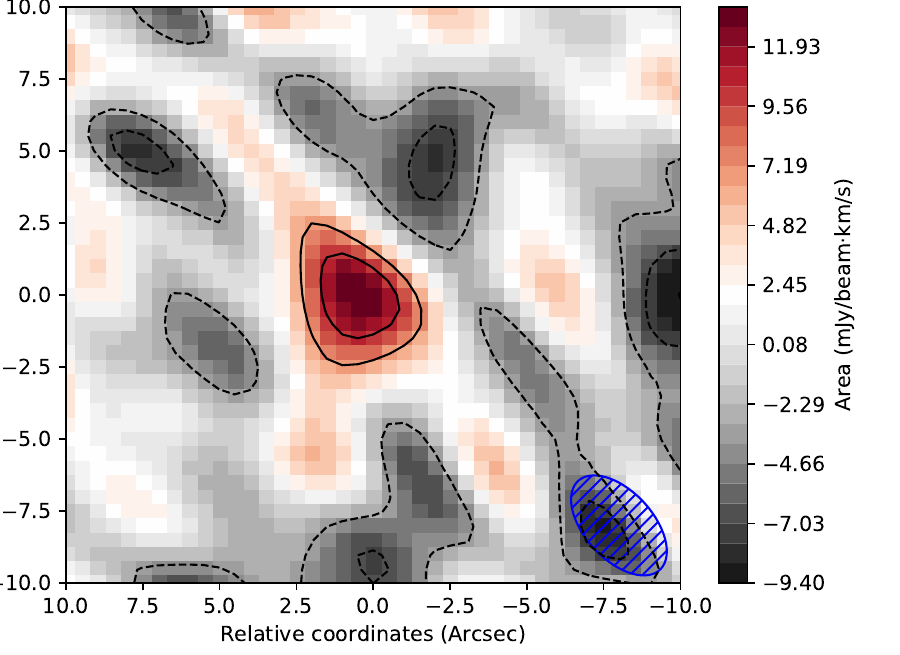}}
\subfloat[HLS071754+374639]{\includegraphics[trim={0cm 0cm 0cm 1.5cm},clip,
page=40,width=0.6\textwidth,clip=true]{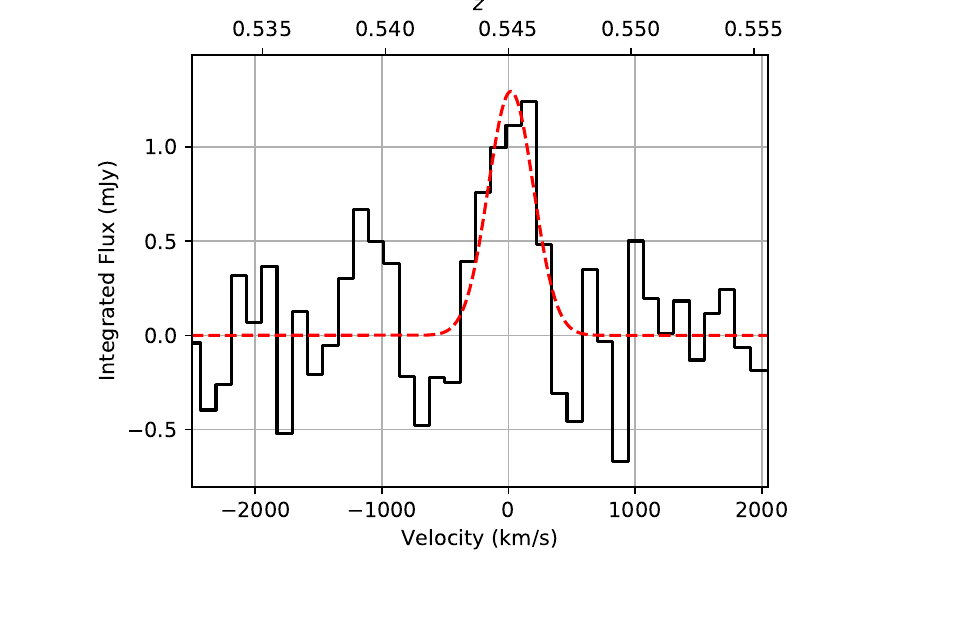}}\\
\caption{Continued.}
\end{figure*}

\begin{figure*} \centering
\captionsetup[subfigure]{labelformat=empty}
\ContinuedFloat
\subfloat[HLS071814+374117]{\includegraphics[trim={0.5cm -0.5cm 0.5cm 1.1cm},clip,
page=17,width=0.5\textwidth,clip=true]{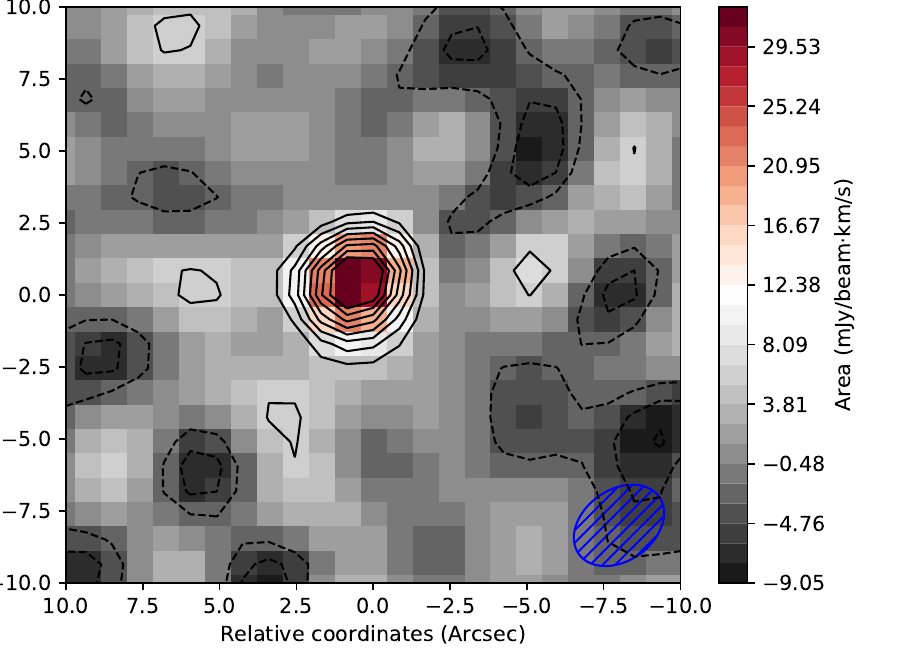}}
\subfloat[HLS071814+374117]{\includegraphics[trim={0cm 0cm 0cm 1.5cm},clip,
page=18,width=0.6\textwidth,clip=true]{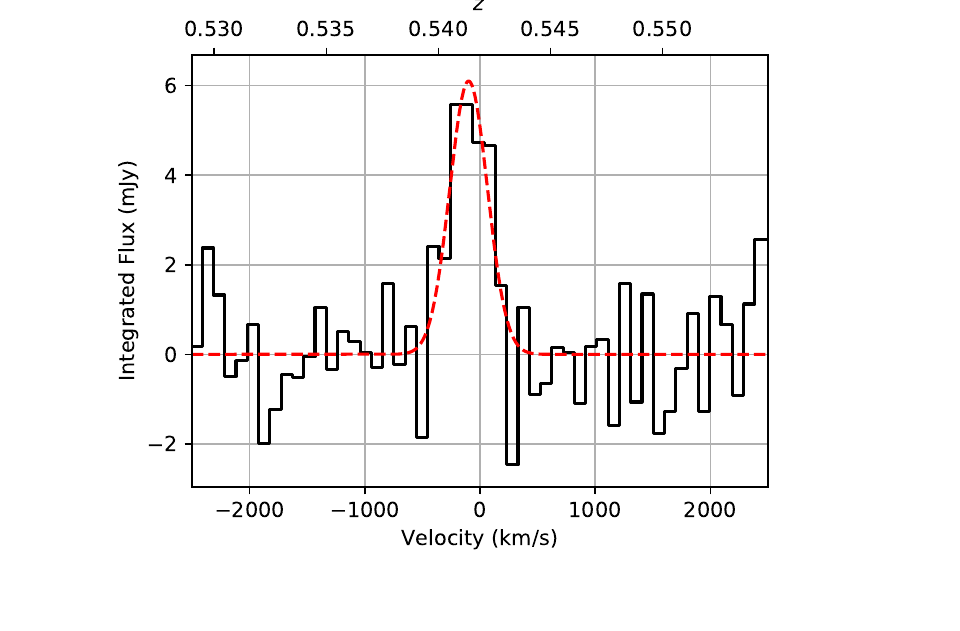}}\\
\caption{Continued.}
\end{figure*}
\end{appendix}
\end{document}